\documentclass[%
 preprint,
 amsmath,amssymb,
 aps,
]{revtex4-2}

\usepackage[english]{babel}
\usepackage{amsmath,amssymb,amsfonts}
\usepackage{graphicx}
\usepackage[colorlinks=True,linkcolor=red,citecolor=blue,urlcolor=blue]{hyperref}
\usepackage{bookmark}
\usepackage[dvipsnames]{xcolor}
\usepackage{braket}
\usepackage{nicefrac}
\usepackage{bm}
\usepackage{bbm}
\usepackage{braket}
\usepackage{slashed}
\usepackage{float}
\usepackage[normalem]{ulem}
\usepackage{array}
\usepackage{makecell}
\usepackage{subfigure}

\begin{document}


\title{Concentrated subradiant modes in one-dimensional atomic array coupled with chiral waveguides}

\author{Mengjie Yang$^{1}$, Luojia Wang$^{1,*}$, Xiaoxiong Wu$^{1}$, Han Xiao$^{1}$, Danying Yu$^{1}$, Luqi Yuan$^{1,\dagger}$, Xianfeng Chen$^{1,2,3}$}

\affiliation{$^1$State Key Laboratory of Advanced Optical
Communication Systems and Networks, School of Physics and
Astronomy, Shanghai Jiao Tong University, Shanghai 200240, China\\
$^2$Collaborative Innovation Center of Light Manipulations and
Applications, Shandong Normal University, Jinan
250358, China\\
$^3$Shanghai Research Center for Quantum Sciences, Shanghai 201315, China\\
$^*$Corresponding author: ljwang@sjtu.edu.cn;\\
$^\dagger$Corresponding author: yuanluqi@sjtu.edu.cn}

\date{\today}

\begin{abstract}

Non-Hermitian systems have recently attracted broad interest and exhibited intriguing physical phenomena, including the non-Hermitian skin effect, which have been widely studied in various fermionic and bosonic systems. Here we propose a non-Hermitian atom-waveguide system composed of a tilted one-dimensional atomic array coupled with two identical waveguides with opposite chiralities. Such system creates an effective lattice model including nonreciprocal long-range hoppings through the chiral-waveguide photon-mediated interactions. We find the excitations of the collective atomic states concentrate in the middle interface associated with subradiant modes, while, on the contrary, superradiant modes exhibit extended features. Such unique feature in our proposed system is linked to the non-Hermitian skin effect. Simulation results present subradiant funneling effect, with robustness against small atomic position disorders. Our work underpins the fundamental comprehension towards the non-Hermitian skin effect in open quantum systems and also provide prospective paths to study non-Hermitian systems in the area of quantum optics.

\end{abstract}


\maketitle


\newpage

\section{Introduction}

Non-Hermitian Hamiltonian recently attracts great interest due to its describing the interactions between a physical system and the environment that are ubiquitous in nature. It has been extensively studied in versatile fields, including the measurement of dissipation in open quantum systems \cite{gorini1976completely,lindblad1976generators}, the dynamics of the nonlinear instabilities in soft matter and quantum fluids \cite{carusotto2013quantum,marchetti2013hydrodynamics}, theoretical and experimental implementations in nonreciprocal coupling strengths \cite{feng2011nonreciprocal,popa2014non,longhi2015non,brandenbourger2019non,ghatak2020observation,hofmann2020reciprocal,li2020critical,zou2021observation} and many others \cite{ma2016acoustic,cummer2016controlling,zhang2021acoustic,nelson1998non,amir2016non,cao2015dielectric,budich2019symmetry,lee2019anatomy,edvardsson2019non,ashida2020non,borgnia2020non,bergholtz2021exceptional}. The non-Hermitian skin effect (NHSE), in which the eigenstates are found to be concentrated near the interface \cite{lee2016anomalous,alvarez2018non,lee2019anatomy,lee2019hybrid,song2019non,okuma2020topological,weidemann2020topological}, is one of the most remarkable phenomena in non-Hermitian systems in the past decade, and leads to many appealing physics in quantum systems \cite{kunst2018biorthogonal,yao2018edge,song2019non,zhu2020photonic,xiao2020non,helbig2020generalized,weidemann2020topological,li2020critical,okuma2020topological,zhang2021observation,budich2019symmetry,lee2019anatomy,edvardsson2019non,torres2019perspective,ashida2020non,borgnia2020non,bergholtz2021exceptional}. For example, the non-Hermitian photonics mesh lattice with anisotropic couplings has been studied towards building up the light-harvesting platforms and optical sensors with enhanced sensitivity \cite{weidemann2020topological}. Moreover, NHSE is a key component in active arguments on the collapse of the conventional bulk-boundary correspondence, indicating the new perspective of exotic properties in non-Hermitian systems \cite{kunst2018biorthogonal,yao2018edge,zhu2020photonic,xiao2020non,helbig2020generalized}.

Exploring light-matter interactions and hence manipulating quantum states are critically fundamental in quantum optics, which have been profoundly studied in various systems \cite{haroche2006exploring}. Optical waveguides provide achievable platforms to control efficient light-atom interactions, including providing single atom-photon couplings towards the photon transport process \cite{le2005spontaneous,shen2005coherent,shen2005coherentspontaneous,zheng2010waveguide,yuan2015achieving,kockum2018decoherence,xiao2020frequency}, and long-range atom-atom interactions in atomic array through mediating waveguide-guided propagating photons \cite{le2005nanofiber,lalumiere2013input,shahmoon2013nonradiative,masson2020atomic,corzo2019waveguide}. In particular, an ensemble of atoms coupled with 1D waveguide systems have been studied to show fruitful properties such as sub- and superradiant states \cite{zhang2019theory,albrecht2019subradiant,goban2015superradiance,solano2017super}, the topologically enhanced photons absorption \cite{nie2021topology}, non-local optical nonlinearities \cite{shahmoon2016highly}, and the electromagnetically induced transparency \cite{roy2011two,song2017photon}, which exhibit significant applications potentially in the quantum states storage and the quantum information processing. As a natural non-Hermitian system, the atom-waveguide system presents an interesting platform to explore the NHSE, which, however, has not been studied to the best of our knowledge. Fortunately, by virtue of recent developments in chiral quantum optics fields \cite{lodahl2017chiral,feng2011nonreciprocal,rodriguez2013near,mitsch2014quantum,petersen2014chiral,sollner2015deterministic,coles2016chirality,javadi2018spin}, it is possible to implement promising non-reciprocal control of light-atom interactions by employing the chiral waveguides \cite{feng2011nonreciprocal,petersen2014chiral,Mirza2016,mahmoodian2020dynamics,wang2020single}, which provides the possibility for exploring the NHSE in atom-waveguide systems.

In this work, we study a tilted one-dimensional atomic array coupled with two chiral waveguides, which exhibits subradiant states concentrated at the middle interface. By placing atomic array in-between waveguides with a small angle, position-dependent dipole-dipole interactions have been built through mediating photons that propagate at opposite directions in two waveguides. Such system gives a non-Hermitian lattice model including long-range couplings. Opposite asymmetric hoppings between two atomic dipoles are supported in the lattice model, while the middle atom holds same decays into both waveguides and hence provides an artificial interface at the center of the atomic array. Our system exhibits concentrated states with subradiant decays but extended states with superradiant decays. Similar with the funneling effect of light in a photonic system that refers to any light field travelling toward an interface \cite{weidemann2020topological}, we find excitations at various positions in the atomic array evolve towards the center interface but hold the subradiant feature. During these processes, the superradiant states of the system dissipate fast. Such concentrated subradiant states show robust property against atomic position disorders. Our study therefore points towards the NHSE in the atom-waveguide system with possible applications in quantum-state harvesting and robust photon storage in the sense of quantum optics.

\section{Model}

\begin{figure}[!htp]
\centering
\includegraphics[width=1\columnwidth]{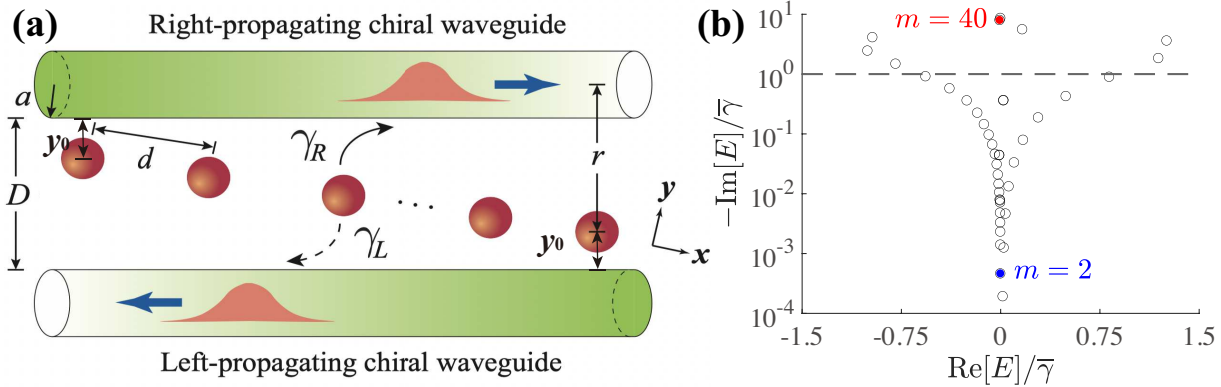}
\caption{(a) Schematic of a one-dimensional atomic array coupled with two chiral waveguides, with atoms uniformly arranged with a tilted small angle in-between waveguides. Atoms emit photons into the right- (left-) propagating chiral waveguide, and the corresponding spontaneous decay rate is $\gamma_{R}$ ($\gamma_{L}$), which depends on the distance $r-a$ (or $D-r+a$) from the atom to the top (or bottom) waveguide surface. (b) The complex eigen-energies $E$ of the non-Hermitian lattice model including 41 atoms. The order of the mode number $m$ is arranged descendingly by the imaginary part of $E$. Blue (Red) dot refers to the eigenvalue at $m=2$ ($m=40$).}\label{fig1}
\end{figure}

We study a one-dimensional atomic array coupled with two chiral waveguides, as schematically shown in Fig. \ref{fig1}(a). In particular, $N$ two-level atoms are aligned along the $x$ axis with equal spacing $d$, each of which is labeled by $j$ (for $j = -(N-1)/2,\ldots,0,\ldots,+(N-1)/2$, for odd $N$) at positions $x_{j}$. Two chiral waveguides with spacing $D$, placed at a small angle to the $x$-axis in $xy$ plane, are identical except that photons are allowed in propagating in opposite directions, i.e., the top one is right-propagating waveguide while the bottom one is left-propagating [see Fig. \ref{fig1}(a)]. Here we only consider the spontaneous decay rate of guided modes but ignore the effect caused by the environment, since the loss to the environment leads to a background dissipation and do not change the main feature of the findings in our proposed system (see Appendix \ref{A1} for details). The spontaneous decay rate, $\gamma_{Rj}$ (or $\gamma_{Lj}$), are dependent on the distance $r-a$ (or $D-r+a$) from the $j$-th atom to the top (or bottom) waveguide surface, where $r$ denotes the distance from each atom to the top waveguide cylinder axis and $a$ is the radius of each waveguide. We take $r > a$ throughout this paper.

We next discuss the effective model that supports dipole-dipole interactions from the mediated waveguide photon. To this purpose, we take $N = 41$ atoms into consideration, and the term $\overline{\gamma}$ used as a normalization factor throughout this paper refers to the average spontaneous decay rate of this specific $N = 41$ system, i.e., $\overline{\gamma} = \sum_{j=-20}^{+20} (\gamma_{Rj}+\gamma_{Lj})/(2N)$. We set $a=250$nm, $D=1000$nm, and $d=9073.8$nm ($d\gg \lambda$ for eliminating the effect caused by the environment in the inter-atomic interactions). $\lambda=852$nm is wavelength of waveguide photons. Atoms are uniformly arranged with a tilted angle $\theta \sim$ $0.002$ rad in-between waveguides (i.e., the $0$-th atom is placed in the middle of two waveguides).  $\gamma_{Rj}$ ($\gamma_{Lj}$) are calculated accordingly \cite{LeKien2017}. We notice that the atom experiences larger decay if it is close to one waveguide, and gives relatively very small decay rates into both waveguide if it is near the middle between two waveguides \cite{le2006scattering,Scheel2015,LeKien2017}. Moreover, there emerges a generalized interface in the center of the atomic array, which is critically important to the later demonstrations of the non-Hermitian skin effect and the funnel-like behavior.

By taking the Born-Markov approximation and neglecting the retardation caused by the finite propagation velocity of photons, one can write the chiral master equation for the evolution of the system density operator \cite{Pichler2015}

\begin{equation}
\begin{aligned}
\dot{\hat{\rho}}(t) &=-i\left[\hat{H}_{\mathrm{sys}}, \hat{\rho}(t)\right]+\sum_{\lambda=R, L} \sum_{j} \frac{\gamma_{\lambda j}}{2}\left(\left[\hat{\sigma}_{j}, \hat{\rho}(t) \hat{\sigma}_{j}^{\dagger}\right]-\left[\hat{\sigma}_{j}^{\dagger}, \hat{\sigma}_{j} \hat{\rho}(t)\right]\right) \\
&+\sum_{\substack{\lambda=R, L \\
k_{\lambda} x_{j}>k_{\lambda} x_{l}}} \sum_{\substack{j, l}} \sqrt{\gamma_{\lambda j} \gamma_{\lambda l}}\left(e^{-i k_{\lambda}\left(x_{j}-x_{l}\right)}\left[\hat{\sigma}_{j}, \hat{\rho}(t) \hat{\sigma}_{l}^{\dagger}\right]-e^{i k_{\lambda}\left(x_{j}-x_{l}\right)}\left[\hat{\sigma}_{j}^{\dagger}, \hat{\sigma}_{l} \hat{\rho}(t)\right]\right).
\end{aligned}\label{A7}
\end{equation}
where $\hat{\rho}(t)$ is the time-dependent system density operator. Here $\hbar=1$ for the simplicity. The Hamiltonian of two-level atoms system Hamiltonian reads $\hat{H}_{\mathrm{sys}} = \sum_{j} \Delta_{j}\hat{\sigma}^{\dagger}_{j}\hat{\sigma_{j}}$ with $j$ being the index of the $j$-th atom and the operator $\hat{\sigma}_{j}^{\dagger} = |e_{j}\rangle \langle g_{j}|$ being the operator representing the transition from the ground state $|g_{j}\rangle$ to the excited state $|e_{j}\rangle$. $k_{R} = -k_{L} = k$, where $k$ is the wave vector of the photon. $\hat{H}_{\mathrm{eff}}$ denotes the effective Hamiltonian.

To derive the effective Hamiltonian of our proposed system in Fig. \ref{fig1}(a), we re-write the chiral master equation in explicit Lindblad form as \cite{Pichler2015,Mirza2016}

\begin{equation}
    \dot{\hat{\rho}}(t)=-i\left[\hat{H}_{\mathrm{sys}}, \hat{\rho}(t)\right]-i\left(\hat{H}_{\mathrm{eff}} \hat{\rho}(t)-\hat{\rho}(t) \hat{H}_{\mathrm{eff}}^{\dagger}\right)+  \hat{c}_{L} \hat{\rho}(t) \hat{c}_{L}^{\dagger} + \hat{c}_{R} \hat{\rho}(t) \hat{c}_{R}^{\dagger}, \label{lindblad}
\end{equation}
where $\hat{c}_{L} = \sum_{j} \sqrt{\gamma_{Lj}} e^{ikx_{j}}\hat{\sigma}_{j}$ and $\hat{c}_{Rj} = \sum_{j} \sqrt{\gamma_{Rj}} e^{-ikx_{j}}\hat{\sigma}_{j}$. By comparing Eqs. \eqref{A7} and \eqref{lindblad}, we obtain the long-range effective Hamiltonian of our model as

\begin{equation}
    \hat{H}_{\mathrm{eff}}=-\frac{i}{2}\sum_{j}(\gamma_{Lj}+\gamma_{Rj})\hat{\sigma}_{j}^{\dagger}\hat{\sigma}_{j}-i\sum_{j>l}\sqrt{\gamma_{Ll}\gamma_{Lj}}\hat{\sigma}_{l}^{\dagger}\hat{\sigma}_{j} e^{ik(x_{j}-x_{l})}-i\sum_{j>l}\sqrt{\gamma_{Rl}\gamma_{Rj}}\hat{\sigma}_{j}^{\dagger}\hat{\sigma}_{l} e^{ik(x_{j}-x_{l})}.
    \label{Heff}
\end{equation}

The effective Hamiltonian in Eq. \eqref{Heff} is non-Hermitian, which is consistent with Hamiltonians in Refs. \cite{Pichler2015,yao2018edge,kunst2018biorthogonal,lee2019anatomy,lee2019hybrid,song2019non,ghatak2019new,weidemann2020topological,okuma2020topological,helbig2020generalized,xiao2020non,zhu2020photonic,zhang2021observation}, but includes non-uniform long-range dipole-dipole interactions where onsite decays in the first term and hopping coefficients in the second and third terms are dependent on positions of two atoms. In the following, we use the Hamiltonian \eqref{Heff} to explore the NHSE phenomena in the model of Fig. \ref{fig1}(a).

\section{Results}



The effective Hamiltonian in Eq. \eqref{Heff} includes $N=41$ atoms at different positions between waveguides, and hence has no translational symmetry. We therefore diagonalize Eq. \eqref{Heff} with the open boundary directly in the spatial space. The resulting band structure is plotted in Fig. \ref{fig1}(b), which has $41$ eigen-energies $E$ at complex values. We use $m=1,2,...,41$ to label the index of eigen-energies by increasingly sorting [$-$Im($E$)], the collective decay rate of the mode, from slow decay rate to fast decay rate. One can see that most of modes exhibit a collective decay rate smaller than the average spontaneous decay rate of the system, i.e., [$-$Im($E$)] $<\overline{\gamma}$, referring to subradiant modes, while the others having [$-$Im($E$)] $>\overline{\gamma}$, giving the superradiant modes of the system. Notably, there are two modes having [$-$Im($E$)] $
\sim 10^{-3}\overline{\gamma}$ ($m=1, 2$). The real parts of eigen-energies are around zero for most subradiant modes, but diverge towards $\sim \pm\overline{\gamma}$ for larger $m$ (larger collective decay rate).


\begin{figure}[!htp]
\centering
\includegraphics[width=1.0\columnwidth]{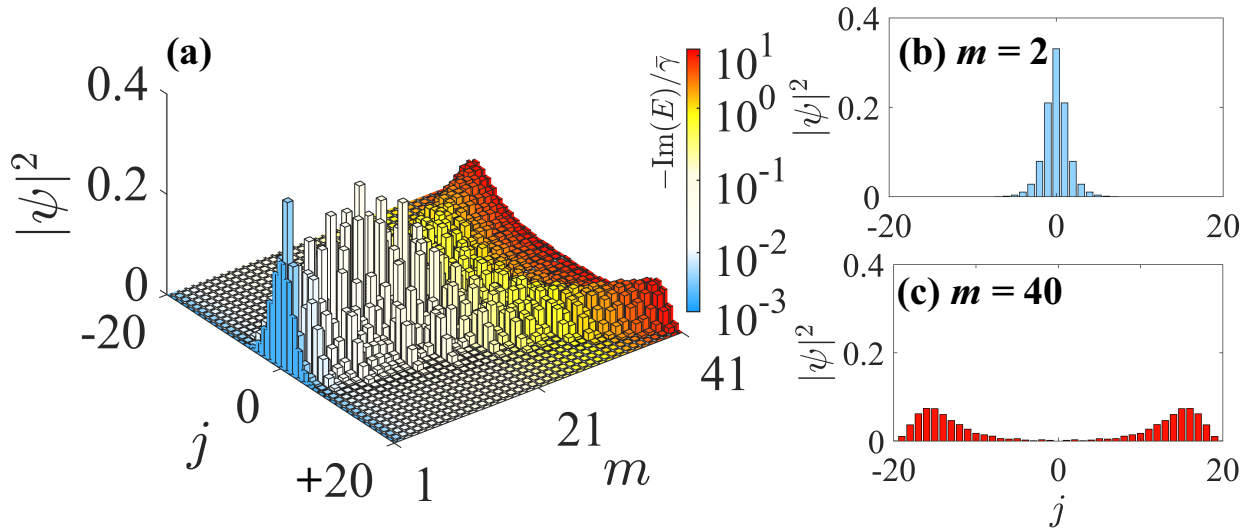}
\caption{(a) Intensity distributions versus atom site $j$ for all eigen-states with $m=1,2,\ldots,41$. Colors for each eigen-state indicate the collective decay rates. (b) and (c) The intensity distributiuons for eigen-state with $m=2$ and eigen-state with $m=40$, respectively.}\label{eigen}
\end{figure}

We further plot the intensity distribution versus atom site $j$ for all the eigen-states with $m=1,2,\ldots,41$, in Fig. \ref{eigen}(a). The remarkable feature is that the modes with very small collective decay rate (subradiant modes) exhibit concentrated intensity in the middle interface around the $0$-th atom. When the collective decay rate of modes increases, the concentration of intensity at the middle interface becomes weaker. Further increase of the collective decay rate to the superradiant regime results in the fact that the concentration disappears and the intensities of modes extended gradually towards two ends of the atomic array. We show two examples of very opposite cases in Fig. \ref{eigen}(b) for the eigen-state at $m=2$ and Fig. \ref{eigen}(c) for the eigen-state at $m=40$. One can see that, for the mode at $m=2$ in Fig. \ref{eigen}(b), the intensity distribution is concentrated mostly at the middle of the atomic array, with exponentially decaying into both sides. The corresponding collective decay rate is $\sim 1.4 \times 10^{-3} \overline{\gamma}$, which is a subradiant mode. However, for the mode at $m=40$ in Fig. \ref{eigen}(c), the collective decay rate gives a superradient decay with $\sim 8 \overline{\gamma}$, where the eigen-state exhibits the extended distributions with highest intensities at $j=\pm 17$-th atoms.

Next we conduct numerical simulations to study evolutions of the system under different conditions. We consider the wave function of the excited wavepacket as

\begin{equation}
    |\phi(t)\rangle = \sum_{j} v_{j}(t) \hat{\sigma}_{j}^{\dagger} |0\rangle,
    \label{wave function}
\end{equation}
where $v_{j}(t)$ is the amplitude of the wavepacket state at the $j$-th atom. By inserting the Eq. \eqref{wave function} into the Schr\"{o}dinger Eq. $H|\phi(t)\rangle = i\frac{d}{dt}|\phi(t)\rangle$ where $H$ is defined in Eq. \eqref{Heff}, we obtain the coupled time evolution equations at the position of $j$-th atom when exciting the $j_{s}$-th atom ($-20<j<+20$):

\begin{equation}
\dot{v}_{j}(t) = -\frac{\gamma_{Lj}+\gamma_{Rj}}{2}\ v_{j}(t) - \sum_{l = -20}^{j-1}\sqrt{\gamma_{Rl}\gamma_{Rj}}\ e^{ik\cdot (j-l)d}\ v_{l}(t) - \sum_{l = j+1}^{+20}\sqrt{\gamma_{Lj}\gamma_{Ll}}\ e^{ik\cdot (l-j)d}\ v_{l}(t) + s(t)\delta_{j_{s},j}.
\label{evolution equations}
\end{equation}

The $j_{s}$-th atom is excited by a temporal Gaussian-shape excitation source $s(t)$,

\begin{equation}
s(t) = e^{-\frac{(t-t_{s})^{2}}{2\tau^{2}}} \cdot e^{-i \omega_{s} t}, \label{source}
\end{equation}
where $t_{s}$ and $\tau$ give the temporal center and the width of the Gaussian excitation, respectively, and $\omega_{s}$ represents the excitation frequency.

\begin{figure}[!htp]
\centering
\includegraphics[width=1\columnwidth]{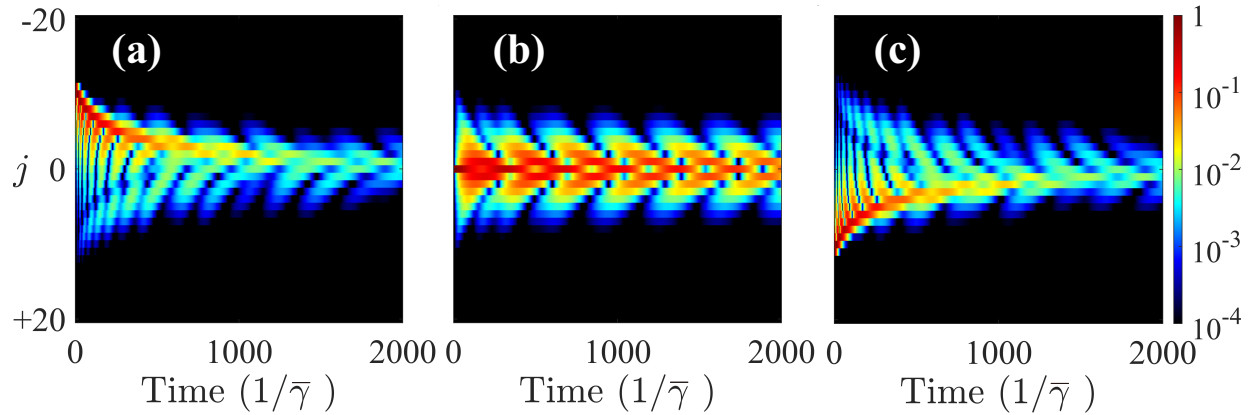}
\caption{Simulation results of the excited wavepacket evolutions with time, with three different initial excitation positions at (a) $j_{s} = -11$, (b) $j_{s} = 0$, and (c) $j_{s} = +11$, respectively.}\label{fig4}
\end{figure}

In simulations, we are aiming to excite the subradiant modes, which exhibits the concentration at the middle interface with a relatively long decay time [such as the mode at $m=2$ in Fig. \ref{eigen}(b)]. We therefore set $\omega_{s} = -0.0032 \overline{\gamma}$, $t_{s}$=$\overline{\gamma}^{-1}$, and $\tau$ = 2 $\overline{\gamma}^{-1}$. Note that the excitation frequency $\omega_s$ is chosen to be resonant with the eigen-frequency of the mode at $m=2$. However, as shown in Fig. \ref{fig1}(b), the real parts of eigen-states are around $0$, so the excitation source here may excites many modes. Fortunately, the subradiant modes not only share the same features of concentration at the middle interface, but also exhibits long decay time, which makes one possible to observe the corresponding phenomena in simulations. We also take three different excited positions at $j_{s} = -11, 0$, and $+11$, respectively, where simulation results showing the excited wavepacket evolutions with time are plotted in Fig. \ref{fig4}. One can see that, for all three cases, the normalized intensity of the excitation gradually focuses at the middle interface during evolutions, while the energy of the excitation decay exponentially. Such phenomena are consistent with the eigen-state distributions in Fig. \ref{eigen}(a). After all superradiant modes fastly decay, the subradiant modes exhibit a slower decay with the funnel-like behavior that the wavepacket propagates to the middle interface. Moreover, compared with excitations near two sides of the atomic array, the initial excitation at the middle [see Fig. \ref{fig4}(b)], shows the confined wavepacket with stronger intensities, because the initial excitation overlaps largely with the eigen-states of subradiant modes and hence more such states are excited.

Different from Ref. \cite{weidemann2020topological}, our model supports not only subradiant modes that concentrate the excited quantum states into the middle interface, but also includes superradiant modes that holds extended distributions but decays much faster. Therefore, in the funneling process in our proposed model, the total energy of the excitation decreases due to the energy loss not only from the background dissipation, but also mainly from these superradiant modes. Nevertheless, the concentrations of the wavepacket at the middle interface associated with subradiant modes reveal the manifestation of non-Hermitian skin effect in this atom-waveguide system.

\begin{figure}[!htp]
\centering
\includegraphics[width=1\columnwidth]{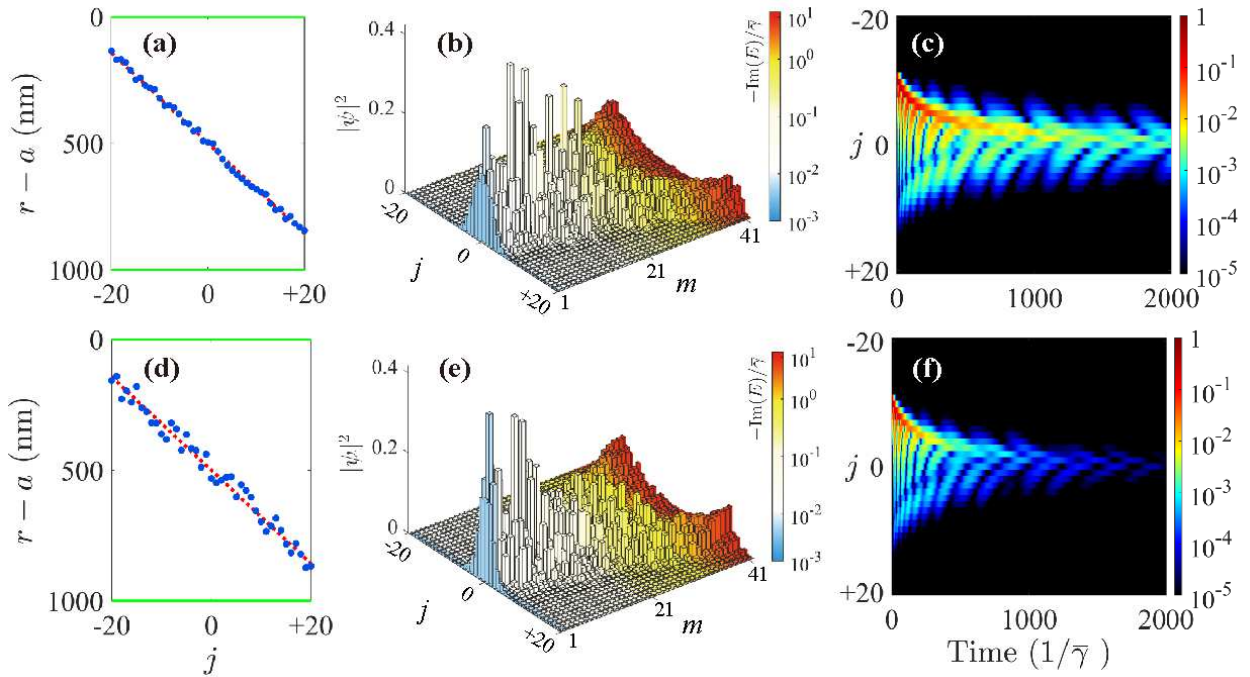}
\caption{Atomic positions with disorders at (a) $\delta = 2 \cdot \theta d$ and (d) $\delta = 6 \cdot \theta d$. Blue dots are positions arrangement of 41 atoms, red dashed line denotes the original atomic positions without disorders, and solid green lines indicate the two chiral waveguides surfaces. (b) and (e) The corresponding intensity distributions of all eigen-states, respectively. (c) and (f) The corresponding simulation results of the excited wavepacket evolutions with time, with initial excitation position at $j_{s} = -11$.}\label{fig5}
\end{figure}

To further explore the robustness of such funnel-like behavior in this non-Hermitian atom-waveguide system, we consider disorders of atomic positions, where each atom is deviated vertically from its original position by a factor of $\delta \cdot R$. Here $R$ is a random number chosen in a regime of $(-0.5,0.5)$ and $\delta$ is a constant reflecting the disorder of the system. In simulations, we take $\delta = 2 \cdot \theta d \approx 36$nm and $\delta = 
6 \cdot \theta d \approx 108$nm, which results in disordered atomic positions shown in Figs. \ref{fig5}(a) and \ref{fig5}(d). The corresponding intensity distributions of all eigen-states are presented in Figs. \ref{fig5}(b) and \ref{fig5}(e), respectively. One can see that, in both case, the subradiant modes are still concentrated near the interface, while the superradiant modes expand towards both sides of the atomic array. We also perform simulations at conditions similar as those in Fig. \ref{fig4}(a) with the excitation source applied on the $-11$-th atom. As seen in both Figs. \ref{fig5}(c) and \ref{fig5}(f), although larger disorders lead to faster decay of the collective excitation, excitations of the system are localized into the middle interface of the atomic array. Simulations with disorders presented here therefore demonstrate robust NHSE associated with the subradiant modes.

The concentrated subradiant states in our proposed model, are of different nature from localization effects under the topological protection \cite{ruostekoski2002particle,schomerus2013topologically,poli2015selective,zeuner2015observation,leykam2017edge,shen2018topological,zhao2018topological,parto2018edge,kawabata2019symmetry,li2020topological,okuma2020topological}, where only a few topological modes are selected to show the concentration. The subradiant states here are found to be associated with the NHSE, while the extended states are results from the competition between fast single-atom losses and the NHSE (see Appendix \ref{A3} for details). Moreover, the interesting phenomena shown in our work mainly result from the nonreciprocal couplings but not the loss profile, while the losses are indeed crucial for the extended states. We further explore our proposed model over a broad range of parameters in Appendices \ref{A3} and \ref{A5}, and conclude that the found existence of concentrated states with subradiant features persists for the scale of the system $N \rightarrow \infty$, which provides further evidence of connecting to NHSE \cite{yokomizo2021non}. Lastly, the occurrence of the found interesting phenomena mainly depends on the varying rate of $\gamma_{R(L)j}$ on atomic positions, which is a result of the geometry of the atomic array. In other words, if $\gamma_{R(L)j}$ varies largely, the distributions of eigenstates show the phenomena of concentrated subradiant states and extended superradiant states. However, when the varying rate of $\gamma_{R(L)j}$ decreases, a transition occurs and the system exhibits all bulk states (refer to Appendix \ref{A5}).


\section{Discussion and Conclusion}

Our proposal is potentially feasible for experimental realizations on optical platforms including nanofibers or nanophotonic waveguides coupled with atoms \cite{goban2012demonstration,reitz2013coherence,le2014propagation,goban2015superradiance,sayrin2015storage,tiecke2015efficient,solano2017super} and superconducting transmission lines with artificial atoms \cite{van2013photon,janvier2015coherent,kakuyanagi2016observation,wen2019large,fedorov2021photon}. Previous experiments have demonstrated that the nanofiber can effectively trap $\sim 2000$ atoms localized about $200$ nm around the nanofiber surface \cite{vetsch2010optical}, where the ratio between the guided mode decay rate ($\gamma$) and the vacuum decay rate of a single atom ($\gamma_0$) can reach $\gamma /$ $\gamma_0$ $\sim0.9\pm 0.1$ \cite{goban2015superradiance}. Moreover, it has been shown that one can further reduce the impact from the losses emitted into the environment \cite{van2013photon,yu2014nanowire,douglas2015quantum,gonzalez2015subwavelength,yu2019two}. For example, recent developments in experiments have greatly strengthened the guided mode coupling coefficients $\gamma/\gamma_0 \sim 50$ for transmon qubits coupled to a 1D coplanar microwave transmission line \cite{van2013photon}. On the other hand, chiral waveguides have been engineered in photonic nanostructures, where unidirectional transport of photons have been realized for chiral quantum optics \cite{feng2011nonreciprocal,rodriguez2013near,mitsch2014quantum,petersen2014chiral,sollner2015deterministic,coles2016chirality,javadi2018spin,orazbayev2018chiral}. All of these start-of-art technologies make the proposal of atoms localized in-between two chiral waveguides feasible in future experiments.

In summary, we have investigated a tilted one-dimensional atomic array coupled with two chiral waveguides, which supports a non-Hermitian lattice model with asymmetric long-range hoppings. Features of concentrated states with subradiant decays but extended states with superradiant decays exhibit. By exciting the system at different atoms, we numerically show the atomic funnel-like bahavior where the energy of the excitation is guided to the middle interface, with the robustness against small disorders. Our results reveal distinctive physics in a chiral atom-waveguide system, pointing to the NHSE associated with the subradiant modes, which paves a promising path to study intriguing non-Hermitian properties via quantum optics platforms. This theoretical proposal also shows potential applications towards manipulation of quantum states, which is of great significance in fields of the quantum storage and the quantum information processing.

\begin{acknowledgments}
This research is supported by National Natural Science Foundation
of China (12122407, 11974245) and Shanghai Municipal Science and
Technology Major Project (2019SHZDZX01-ZX06). L.Y. thanks the
sponsorship from Yangyang Development Fund and the support from
the Program for Professor of Special Appointment (Eastern Scholar)
at Shanghai Institutions of Higher Learning.
\end{acknowledgments}

\appendix

\section{The effect of the environment}\label{A1}

The total spontaneous decay rate for an single atom placed near a waveguide can be written as

\begin{equation}
    \gamma^{\text{(total)}} = \gamma^{(\text{g})} + \gamma^{(\text{r})},
    \label{eq1}
\end{equation}
where $\gamma^{(\text{g})}$ and $\gamma^{(\text{r})}$ refer to guided modes and radiation modes, respectively. The numerical results of $\gamma^{(\text{g})}$ can be obtained by the approach in Appendix \ref{calcu}.

From many well-known works in Refs. \cite{le2006scattering,Scheel2015,LeKien2017}, we know that as the distance increases, the decay rate of guided modes exponentially decrease to zero, while the decay rate of radiation modes exponentially decrease to the free-space decay $\gamma_{0}$. In our main text, we take all parameters in Ref. \cite{LeKien2017}, numerically calculate $\gamma^{(\text{g})}$ for a single waveguide and exhibit our result in Fig. B1, which is consistent with Fig. 2(a) in Ref. \cite{LeKien2017}. Since we take the same parameters and refer to their calculation methods, there is a comparability of radiation modes in our work. We notice that the decay rate of radiation modes stably reaches to free-space decay $\gamma_{0}$ at about $r-a = 125$nm. Therefore, we can take the decay to the environment as $\gamma_{0}$ throughout the main text as long as we make sure the nearest distance between an atom and the surface of the waveguide is bigger than 125nm. After making this approximation, the total spontaneous decay for a single atom is

\begin{equation}
    \gamma^{\text{(total)}} = \gamma^{(\text{g})} + \gamma_{0}.
    \label{eq2}
\end{equation}

Furthermore, we make our second reasonable approximation: the emission into the environment only provides independent single-atom decay, but not contributes to dipole-dipole interactions. The same approximation is used in Ref. \cite{asenjo2017atom} when dealing with the effect caused by the environment. This approximation is valid when the inter-atomic distance (labeled as $d$ in the main text) is much larger than the wavelength (calculated as $2\pi/k$ in the main text), i.e., $d >> \lambda = 2\pi/k = 852$nm. The reason is in the following: The waveguide mediates long-range interactions by photons, and thus the atoms interact with each other via the guided mode no matter what the atoms spacing is. In contrast, the field decays depending on distance as a series of power-laws in free space. Therefore, if the spacing of two atoms are too far, the cross-atom decay can be seen as zero. 

From what has been discussed above, we obtain a new effective Hamiltonian including the effect by the environment which describes this non-Hermitian system written as 

\begin{equation}
    \hat{H}_{\mathrm{eff}}=-\frac{i}{2}\sum_{j}(\gamma_{Lj}^{(\text{total})}+\gamma_{Rj}^{(\text{total})})\hat{\sigma}_{j}^{\dagger}\hat{\sigma}_{j}-i\sum_{j>l}\sqrt{\gamma_{Ll}^{(\text{g})}\gamma_{Lj}^{(\text{g})}}\hat{\sigma}_{l}^{\dagger}\hat{\sigma}_{j} e^{ik(x_{j}-x_{l})}-i\sum_{j>l}\sqrt{\gamma_{Rl}^{(\text{g})}\gamma_{Rj}^{(\text{g})}}\hat{\sigma}_{j}^{\dagger}\hat{\sigma}_{l} e^{ik(x_{j}-x_{l})}.
    \label{eq3}
\end{equation}

We set the same parameters as the main text. Fig. A1 shows the energy spectra for this new Hamiltonian in Eq. (\ref{eq3}).

\begin{figure}[!htp]
\centering
\subfigure[]{
\includegraphics[width=0.48\textwidth]{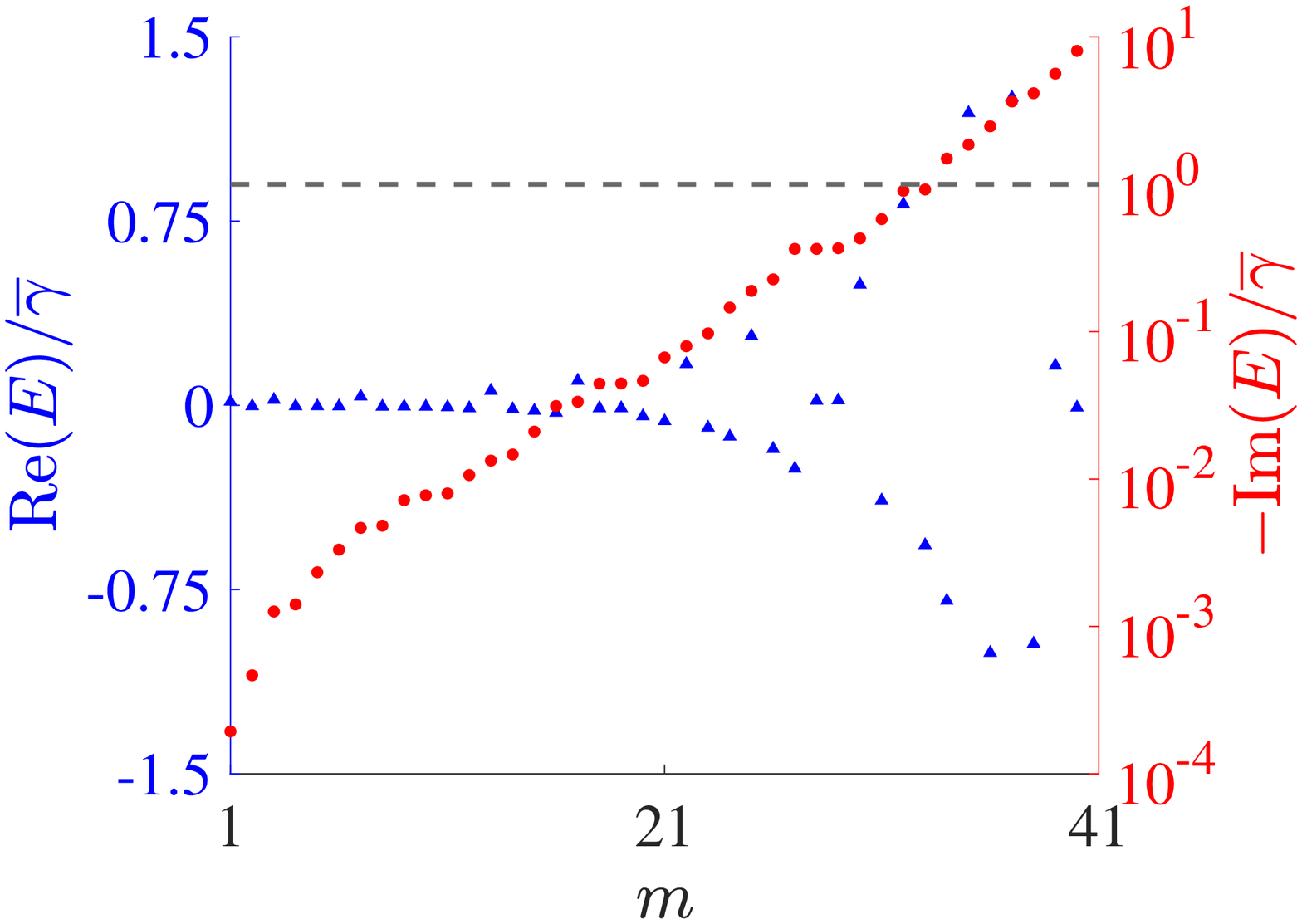}}
\subfigure[]{
\includegraphics[width=0.48\textwidth]{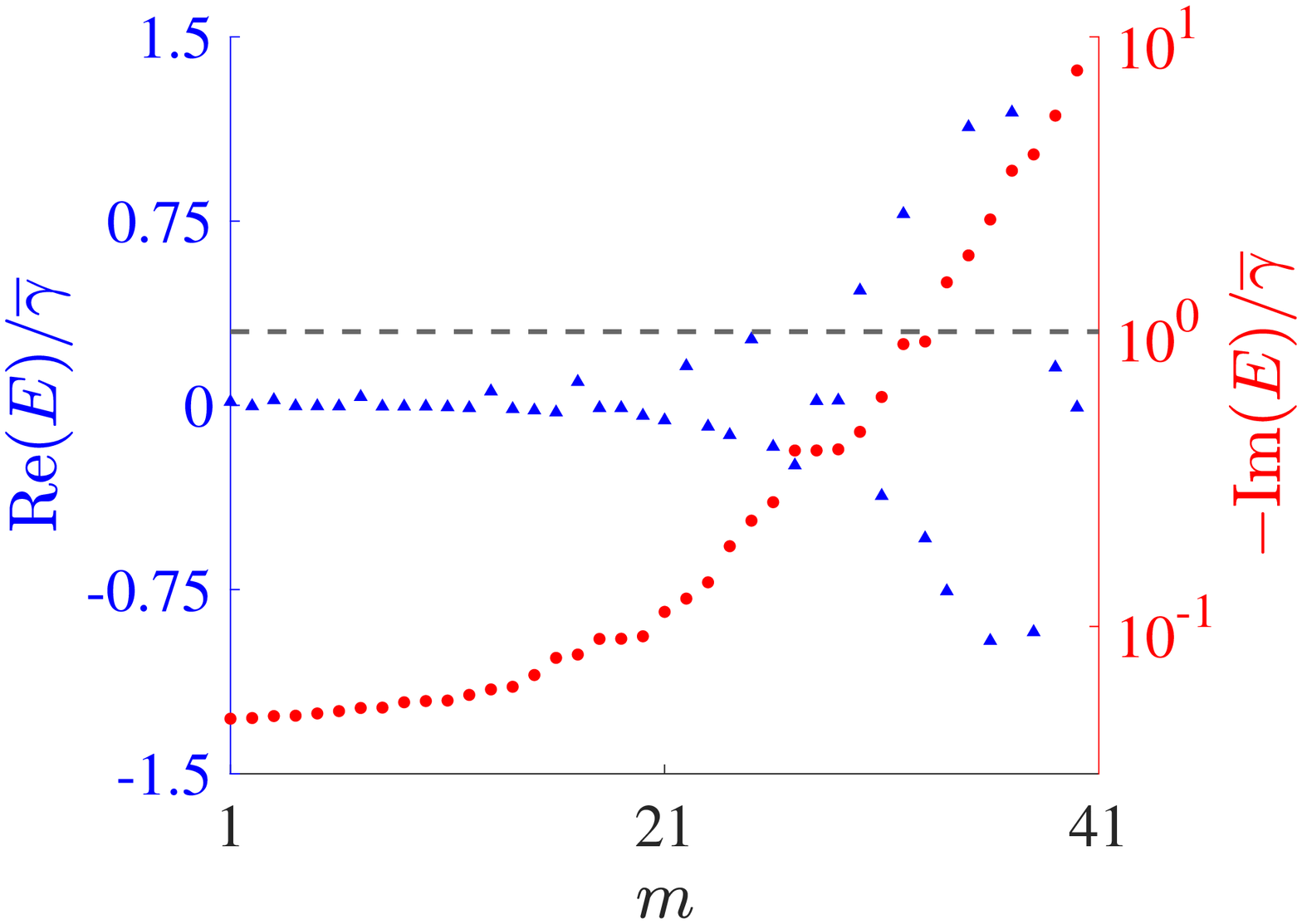}}\\
\small{Fig. A1: (a) The complex eigen-energies $E$ of the non-Hermitian lattice model including 41 atoms in the main text. (b) The complex eigen-energies $E$ of the non-Hermitian lattice model including 41 atoms which takes into account of the environment. The order of the mode number $m$ is arranged descendingly by the imaginary part of $E$.}
\end{figure}

According to Fig. A1, one can see that the real part of energies do not change, which indicates adding the impact from environment will not influence the exciting frequency in the later numerical simulation. In contrast, the imaginary part of energies changes as one can expect. The losses of subradiant states inevitably increase. Fortunately, the number of sub- and super-radiant states stays the same. However, the minimal loss is over $10^{-2} \overline{\gamma}^{\text{(total)}}$ ($\overline{\gamma}^{\text{(total)}} =\overline{\gamma}^{\text{(g)}}+\gamma_0$ due to linear algebra), where $\overline{\gamma}^{\text{(total)}}$ and $\overline{\gamma}^{\text{(g)}}$ refer to the average total and guided spontaneous decay rate of this specific $N = 41$ system, respectively.

\begin{figure}[!htp]
\centering
\subfigure{
\includegraphics[width=0.7\textwidth]{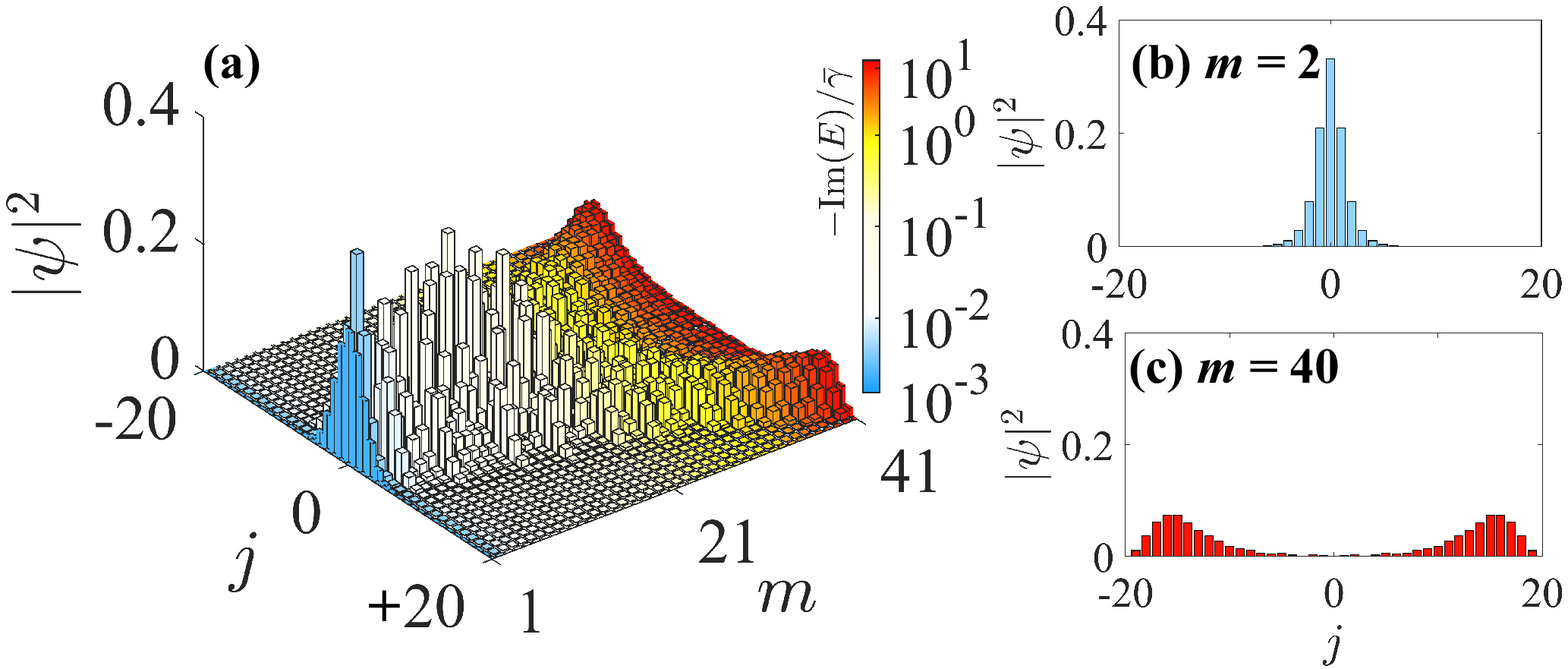}}
\subfigure{
\includegraphics[width=0.7\textwidth]{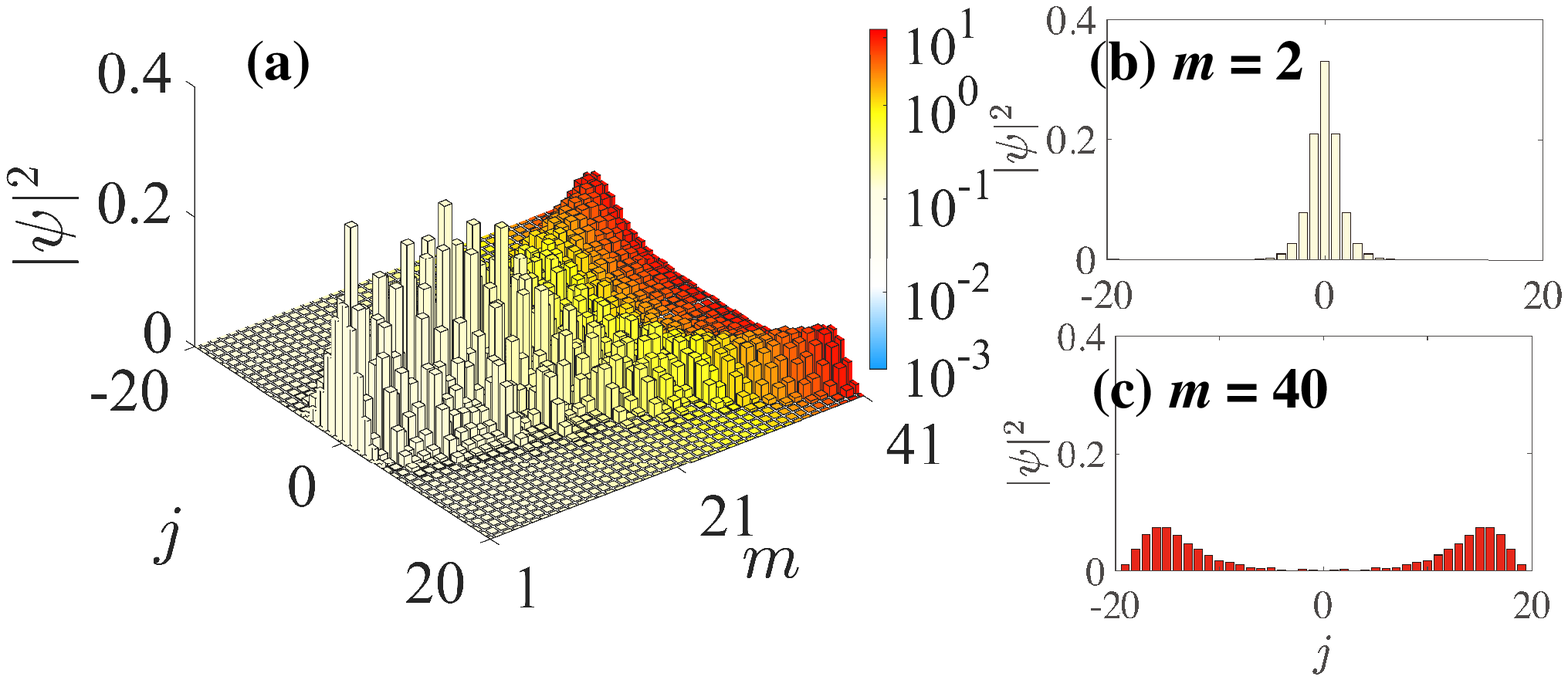}}\\
\small{Fig. A2: The localization without taking into account the environment (upper panel) versus the localization that takes into account the environment (down panel). (a) Intensity distributions versus atom site $j$ for all eigen-states with $m=1,2,\ldots,41$. Colors for each eigen-state indicate the collective decay rates. (b) and (c) The intensity distributiuons for eigen-state with $m=2$ and eigen-state with $m=40$, respectively.}
\end{figure}

As illustrated in Fig. A2, the intensity distributions still exhibit similar patterns after the environment is included. We can come to the similar conclusion as we have claimed in the main text: our system exhibits concentrated states with subradiant decays but extended states with superradiant decays.

In summary, by making two reasonable approximations, we find the environment critically important to exert impact on the energy spectra, especially the imaginary part, i.e., the losses. Nevertheless, the effect of environment does not change the main feature of our proposed system, i.e., the system has subradiant modes concentrated in the middle interface but also includes superradiant modes that hold extended distributions but decays much faster.

\section{The calculation of the spontaneous decay rate of guided modes for a single waveguide in the atom-waveguide system}\label{calcu}

In the main text, we use the spontaneous decay rate $\gamma$ of guided modes in the atom-waveguide system. Here we provide details in the following. We take the refractive index  of the waveguide $n_{1}$ and the surrounding environment has the refractive index $n_{2}$. The position-dependent rate of an atom decaying into the waveguide can be obtained by solving the time-dependent Schr\"{o}dinger equation and then performing a standard Wigner-Weisskopf treatment, which has been attentively discussed in Ref. \cite{Scheel2015}, and the resulting decay rate is

\begin{equation}
\gamma(r) = \frac{\Delta_{j}}{2 \epsilon_{0} \hbar}\left|\boldsymbol{\hat{d}}_{j} \cdot \boldsymbol{\hat{e}}(r)\right|^{2},
\label{eq4}
\end{equation}
where $\Delta_{j}$ denotes the atomic transition frequency; $\boldsymbol{\hat{d}}_{j}$ refers to the dipole moment for an atom, and $\boldsymbol{\hat{e}}(r)$ is the profile function for photon modes that propagate at the $z$ axis along the waveguide direction, which reads \cite{LeKien2017}

\begin{equation}
\begin{aligned}
&e_{r}=i C\left[(1-s) K_{0}(q r)+(1+s) K_{2}(q r)\right] ,\\
&e_{\varphi}=-C\left[(1-s) K_{0}(q r)-(1+s) K_{2}(q r)\right], \\
&e_{z}=C \frac{2 q}{k} K_{1}(q r) .
\end{aligned}
\label{eq5}
\end{equation}
where $K_{p}$ ($p = 0, 1, 2$) represents the modified Bessel functions of the second kind; $C$ satisfies the normalization condition that $\int_{0}^{2 \pi} d \varphi \int_{0}^{\infty} n_{\mathrm{2}}^{2}|\mathbf{\hat{e}}|^{2} r dr=1$. The parameters $q = \sqrt{k^{2}-n_{2}^{2}k_0^{2}}$ and $h = \sqrt{n_{1}^{2}k_0^{2}-k^{2}}$, and the propagation constant $k$ is the solution of the waveguide eigenvalue equation that reads \cite{le2004field,tong2004single}

\begin{equation}
\begin{aligned}
\frac{J_{0}(h a)}{h a J_{1}(h a)}=&-\frac{n_{1}^{2}+n_{2}^{2}}{2 n_{1}^{2}} \frac{K_{1}^{\prime}(q a)}{q a K_{1}(q a)}+\frac{1}{h^{2} a^{2}} \\
&-\left[\left(\frac{n_{1}^{2}-n_{2}^{2}}{2 n_{1}^{2}} \frac{K_{1}^{\prime}(q a)}{q a K_{1}(q a)}\right)^{2}\right.
\left.+\frac{k^{2}}{n_{1}^{2} k_0^{2}}\left(\frac{1}{q^{2} a^{2}}+\frac{1}{h^{2} a^{2}}\right)^{2}\right]^{1 / 2}
\end{aligned} \label{eq6}
\end{equation}

Here, in Eqs. (\ref{eq5}) and (\ref{eq6}) the parameter $s$ takes the form

\begin{equation}
s=\frac{1 / h^{2} a^{2}+1 / q^{2} a^{2}}{J_{1}^{\prime}(h a) / h a J_{1}(h a)+K_{1}^{\prime}(q a) / q a K_{1}(q a)}, \label{eq7}
\end{equation}
where $J_{1}$ is the Bessel functions of the first kind.

Once we have all the information above, we can plot the decay rate by taking Eqs. (\ref{eq4})--(\ref{eq7}) with parameters of the cesium atom and a silica cylinder nanofiber from Refs. \cite{Scheel2015,LeKien2017}, as illustrated in Fig. B1.

\begin{figure}[!htp]
    \centering
    \includegraphics[width = 0.5\textwidth]{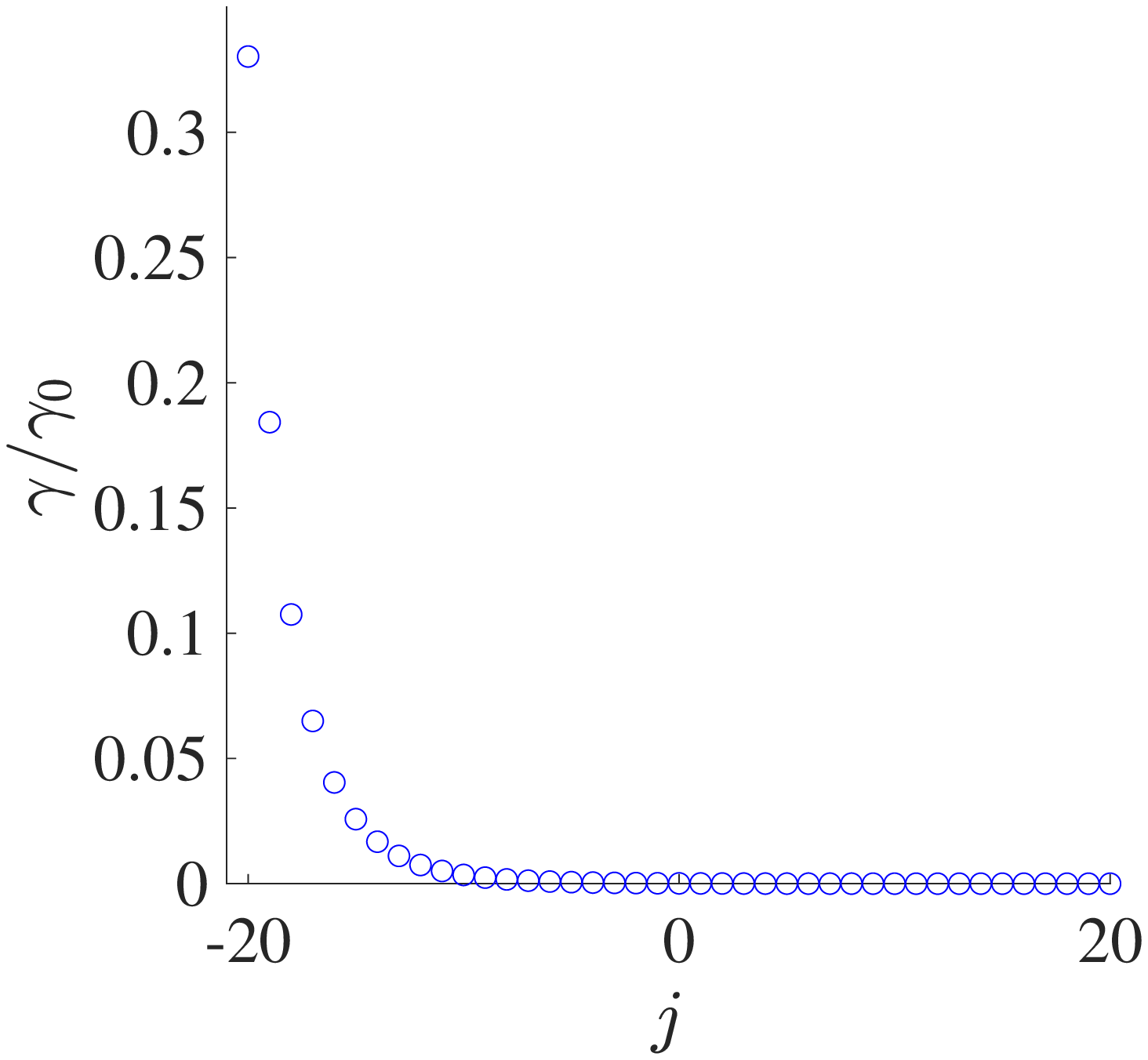}\\
    \small{Fig. B1: The numerical results of the spontaneous decay rate determined by the distance $r-a$ (or $D-r+a$) from the atom to the top (or bottom) waveguide surface at the position of each atom $j=-20,-19,\cdots,+19,+20$.}
\end{figure}

\section{Explore the physical connection between concentrated subradiant states and the non-Hermitian skin effect}\label{A3}

Let us further discuss a little bit more on the physical connection between the phenomena of the concentrated subradiant states and the extended superradiant states in our system and the non-Hermitian skin effect. In our proposed model, the extended modes are mainly results of the losses profile, without which our system is supposed to exhibit the exact non-Hermitian skin effect. To be specific, the exact non-Hermitian skin effect which is supposed to emerge in our system will be destroyed by the quantity and the position of the losses, therefore, superradiant states behave to be extended on both sides of the array, while the states in the middle of the array, the subradiant modes still behave like the skin effect.

To illustrate our argument, let us refer to the one-dimensional ring resonator system (see Fig. 1(e) in Ref. \cite{song2020two})  and consider our model by ignoring the loss profile and long-range couplings, but only taking the position-dependent nearest-neighbor couplings with real values, with the corresponding Hamiltonian as:

\begin{equation}
H=\left(\begin{array}{llllll}
0 & \sqrt{\gamma_{L_{-20}}\gamma_{L_{-19}}} &0 & 0 & 0 & 0\\
\sqrt{\gamma_{R_{-20}}\gamma_{R_{-19}}} & 0 & \sqrt{\gamma_{L_{-19}}\gamma_{L_{-18}}} & 0 & 0 & 0\\
0 & \sqrt{\gamma_{R_{-19}}\gamma_{R_{-18}}} & 0 & \ddots & 0 & 0\\
0 & 0 & \ddots & 0 & \sqrt{\gamma_{L_{+18}}\gamma_{L_{+19}}} & 0 \\
0 & 0 & 0 &  \sqrt{\gamma_{R_{+18}}\gamma_{R_{+19}}} & 0  & \sqrt{\gamma_{L_{+19}}\gamma_{L_{+20}}}\\
0 & 0 & 0 & 0 &  \sqrt{\gamma_{R_{+19}}\gamma_{R_{+20}}} & 0
\end{array}\right).    
\label{eqR1}
\end{equation}

By diagonalizing the Hamiltonian (\ref{eqR1}), we can plot the distribution of all eigenstates, as shown in Fig. C1. One can find that here all modes are localized at the middle interface, which is similar as the results in Ref. \cite{song2020two}. 

\begin{figure}[!htp]
    \centering
    \includegraphics[width=0.6\columnwidth]{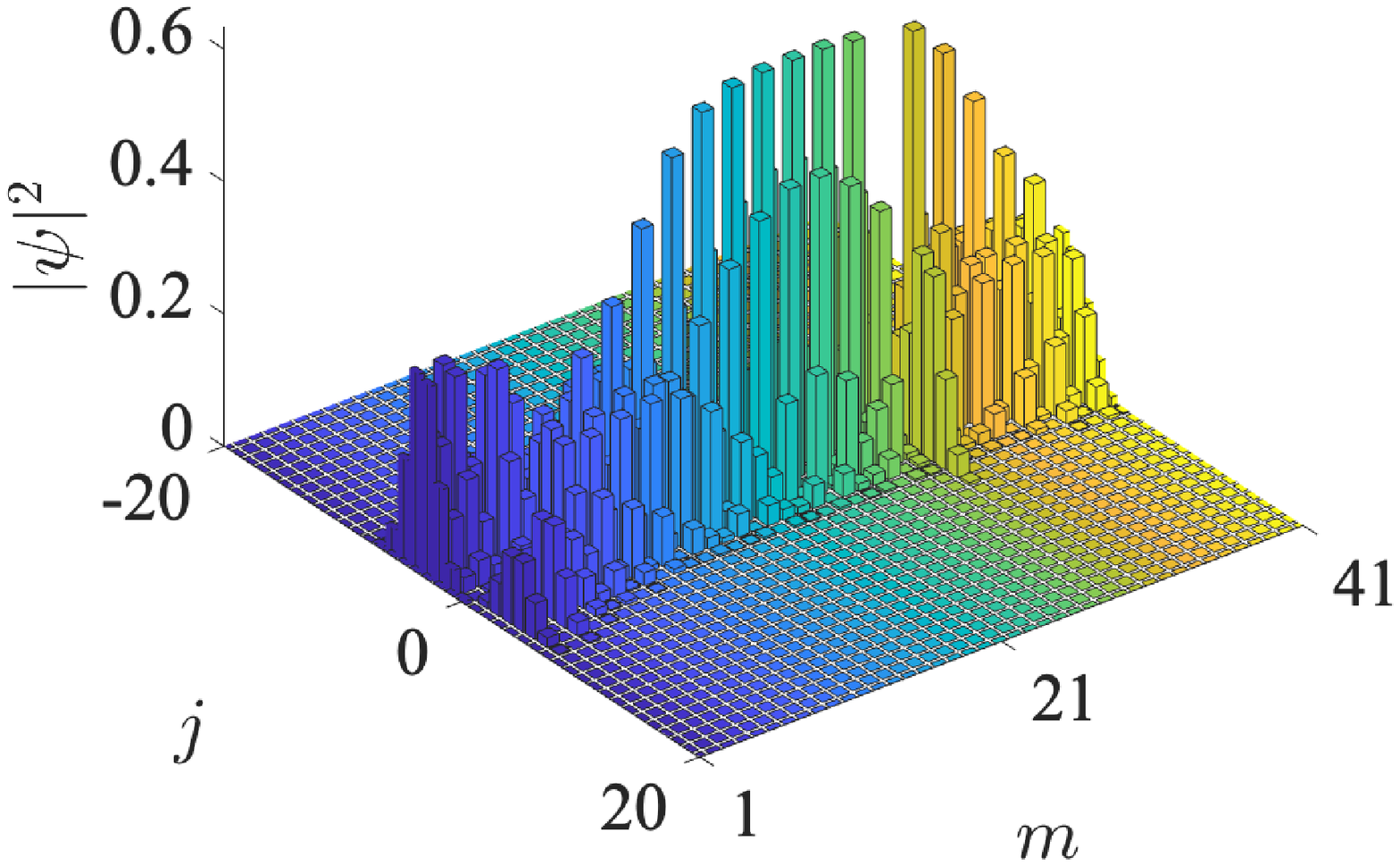}\\
    \small{Fig. C1: Intensity distributions versus atom site $j$ for all eigen-states with $m=1,2,\ldots,41$ from Eq. (\ref{eqR1}).}
\end{figure}

We further add single-atom losses terms into the toy in Eq. (\ref{eqR1}), and the new Hamiltonian is written as
\tiny{
\begin{equation}
H=\left(\begin{array}{llllll}
-\frac{i}{2}(\gamma_{L_{-20}}+\gamma_{R_{-20}}) & \sqrt{\gamma_{L_{-20}}\gamma_{L_{-19}}} &0 & 0 & 0 & 0\\
\sqrt{\gamma_{R_{-20}}\gamma_{R_{-19}}} & -\frac{i}{2}(\gamma_{L_{-19}}+\gamma_{R_{-19}}) & \sqrt{\gamma_{L_{-19}}\gamma_{L_{-18}}} & 0 & 0 & 0\\
0 & \sqrt{\gamma_{R_{-19}}\gamma_{R_{-18}}} & \ddots & \ddots & 0 & 0\\
0 & 0 & \ddots & \ddots & \sqrt{\gamma_{L_{+18}}\gamma_{L_{+19}}} & 0 \\
0 & 0 & 0 &  \sqrt{\gamma_{R_{+18}}\gamma_{R_{+19}}} & -\frac{i}{2}(\gamma_{L_{+19}}+\gamma_{R_{+19}})  & \sqrt{\gamma_{L_{+19}}\gamma_{L_{+20}}}\\
0 & 0 & 0 & 0 &  \sqrt{\gamma_{R_{+19}}\gamma_{R_{+20}}} & -\frac{i}{2}(\gamma_{L_{+20}}+\gamma_{R_{+20}})
\end{array}\right).    
\label{eqR2}
\end{equation}
}

\normalsize

\begin{figure}[!htp]
    \centering
    \includegraphics[width=0.6\columnwidth]{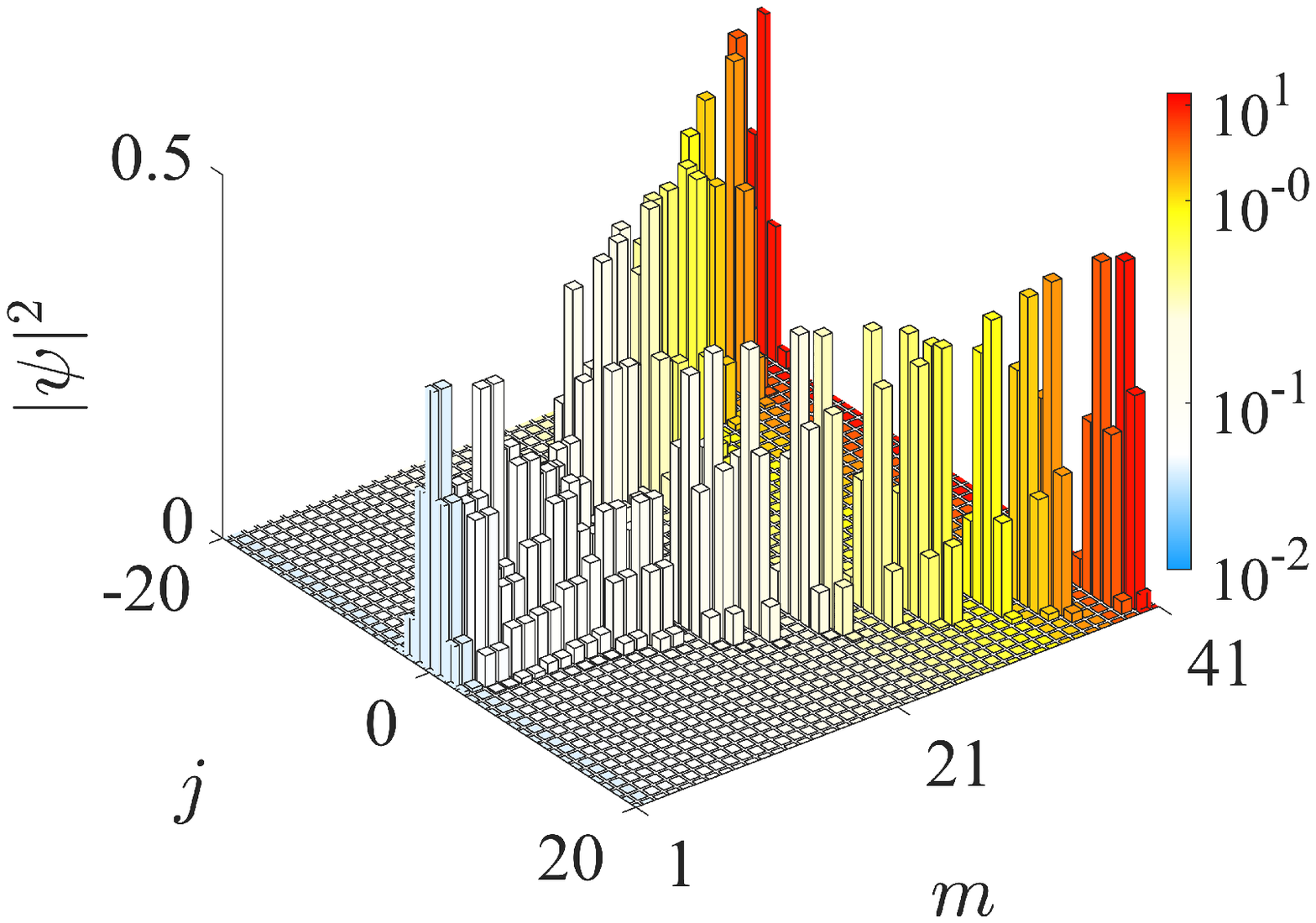}\\
    \small{Fig. C2: Intensity distributions versus atom site $j$ for all eigen-states with $m=1,2,\ldots,41$ from Eq. (\ref{eqR2}).}
\end{figure}

The resulting intensity distribution of eigen-states are shown in Fig. C2. One can notice that only when we add the single-atom losses, the extended states emerge, while the subradiant states remain concentrated in the vicinity of the middle interface as those in Fig. C1. By further adding back long-range terms towards the realistic model in Eq. (\ref{Heff}) in the main text, one can obtain the result shown in Fig. \ref{eigen} in the main text. From this perspective, we then conclude that the subradiant states are indeed associated with the non-Hermitian skin effect, and the extended states are results from the competition between fast single-atom losses and the non-Hermitian skin effect.

Last but not the least, it is worth noting that the non-Hermitian effect does not mean all eigenstates have to localize near the interface. In fact, it only requires partial eigenstates are localized near the interface. More specifically, it requires that the proportion of skin modes tends to be non-zero when the scale of the system $L\rightarrow\infty$ (see Fig. 2(b3) in Ref. \cite{yokomizo2021non}). We explore the impact from the system size, i.e., the atom number $N$ in Sec. \ref{para}, and we find that, with the size of the system being enlarged, the main feature of the concentrated subradiant states and the extended superradiant states found in the main text persists, as illustrated in Fig. E11.

\section{Exclusion of the effect caused by the loss profile}\label{A4}

Here we conduct more numerical simulation on an artificial reciprocal model, pointing out that the interesting phenomena shown in the main text mainly result from the nonreciprocal couplings but not the loss profile, while the losses are crucial for the extended states.

The Hamiltonian of our model is written as 
\begin{equation}
    \hat{H}_{\mathrm{eff}}=-\frac{i}{2}\sum_{j}(\gamma_{Lj}+\gamma_{Rj})\hat{\sigma}_{j}^{\dagger}\hat{\sigma}_{j}-i\sum_{j>l}\sqrt{\gamma_{Ll}\gamma_{Lj}}\hat{\sigma}_{l}^{\dagger}\hat{\sigma}_{j} e^{ik(x_{j}-x_{l})}-i\sum_{j>l}\sqrt{\gamma_{Rl}\gamma_{Rj}}\hat{\sigma}_{j}^{\dagger}\hat{\sigma}_{l} e^{ik(x_{j}-x_{l})}.
    \label{R4}
\end{equation}

To make a comparison, we devise a toy model which has the same local loss profile as our proposed model and the corresponding Hamiltonian is written as 
\begin{equation}
    \hat{\tilde{H}}_{\mathrm{eff}}=-\frac{i}{2}\sum_{j}(\gamma_{Lj}+\gamma_{Rj})\hat{\sigma}_{j}^{\dagger}\hat{\sigma}_{j}-i\sum_{j\neq l}t_{jl}\hat{\sigma}_{l}^{\dagger}\hat{\sigma}_{j} e^{ik|x_{j}-x_{l}|},
    \label{R5}
\end{equation}
where $t_{jl} =( \sqrt{\gamma_{Ll}\gamma_{Lj}}+\sqrt{\gamma_{Rl}\gamma_{Rj}})/2$ is the reciprocal long-range coupling between the $j$-th and $l$-th atom.

According to Eq. (\ref{R5}), we plot the localization of the reciprocal lattice in Fig. D1. One can see that the reciprocal toy model exhibits mostly bulk modes, while it is noticeable that the eigenstates with the biggest losses have very strong intensities localized on the both sides of the array. By comparing the Fig. \ref{eigen}(a) in the main text and Fig. D1, one can exclude the possibility that the subradiant states solely reflect the loss profile, but have a strong connection with the nonreciprocal couplings, while the high losses are crucial for the extended states.

\begin{figure}[!htp]
    \centering
    \includegraphics[width=0.7\textwidth]{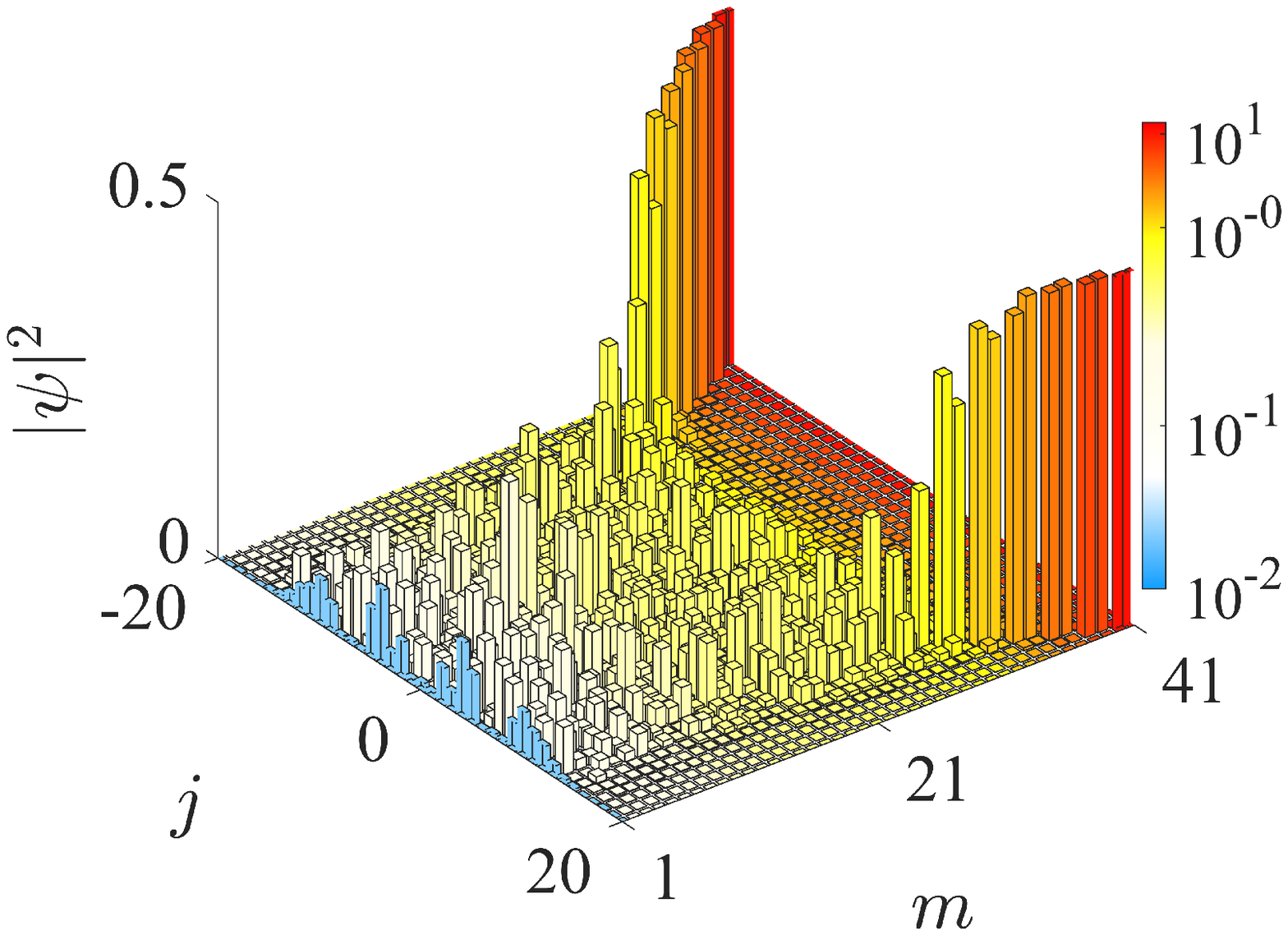}\\
    \small{Fig. D1: Intensity distributions versus atom site $j$ for all eigen-states with $m=1,2,\ldots,41$ of the to model from Eq. (\ref{R5}). Colors for each eigen-state indicate the collective decay rates.}
\end{figure}

\section{Study of the non-Hermitian model over a broad range of parameters}\label{A5}

Here we conduct more numerical calculations over a range of parameters, which are summarized in the following.

As labelled in Fig. E1, we consider four variables in our proposed model, i.e.,  $y_0, H=D-2y_0, d, N$, where $d$ is the atoms spacing. $H$ represents the width of the space that the atomic array takes. $y_0$ refers to the nearest distance from an atom to the vertical surface of the waveguide, namely the distance from the $j=-\frac{(N-1)}{2}$-th ($j=+\frac{(N-1)}{2}$-th) atom to the top (bottom) waveguide. $N$ is the atom number. 

\begin{figure}[!htp]
    \centering
    \includegraphics[width=0.8\textwidth]{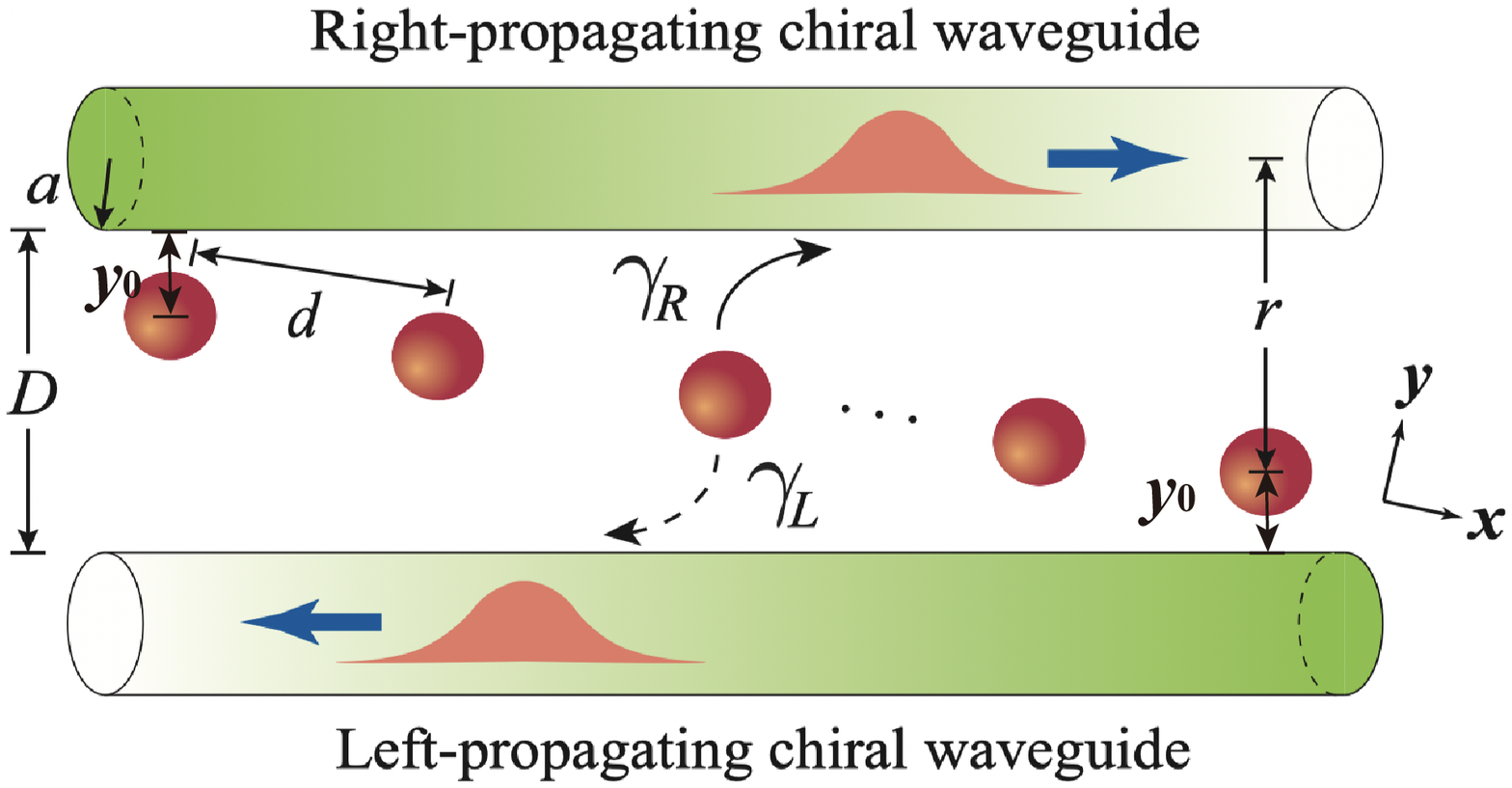}\\
    \small{Fig. E1: Schematic of a one-dimensional atomic array coupled with two chiral waveguides, with atoms uniformly arranged with a tilted small angle in-between waveguides. Atoms emit photons into the right- (left-) propagating chiral waveguide, and the corresponding spontaneous decay rate is $\gamma_{R}$ ($\gamma_{L}$), which depends on the distance $r-a$ (or $D-r+a$) from the atom to the top (or bottom) waveguide surface.}
\end{figure}

At the very first, let us introduce three important factors, which quantify the proposed system and thus make it more convenient for study over a range of parameters.

\begin{itemize}
    \item The average life time of all subradiant states, which reflects the average level of decay rate of all subradiant states, is labelled as $\left \langle \tau \right \rangle$.
    \item The FWHM of the most concentrated state, which confers the quantity of the concentration of the most localized state, is labelled as FWHM$_{\text{min}}$. The bigger the FWHM$_{\text{min}}$ is, the lower concentration the state is. We obtain this quantity by first calculating eigenstates, and then we conduct the Gaussian fitting. The unit is atom site number.
    \item The average FWHM of all subradiant states, which provides the relevant information of the concentration of all subradiant states, is labelled as $\left \langle \text{FWHM} \right \rangle$. This quantity describes the average level of concentration for this system. We obtain the $\left \langle \text{FWHM} \right \rangle$ by the following steps: We first numerically calculate the eigenstates, then conduct the Gaussian fitting on all subradiant states to get the FWHM for each subradiant state. At last we do the average operation. The unit is atom site number.
\end{itemize}

We then study the effects of  $y_0, H, d$ and $N$ on our proposed system, with details as follows:


\subsection*{E1. Effects of parameter $H$ and $y_0$}

\setcounter{subfigure}{0}
\begin{figure}[!htp]
\centering
\subfigure[]{
\includegraphics[width=0.3\textwidth]{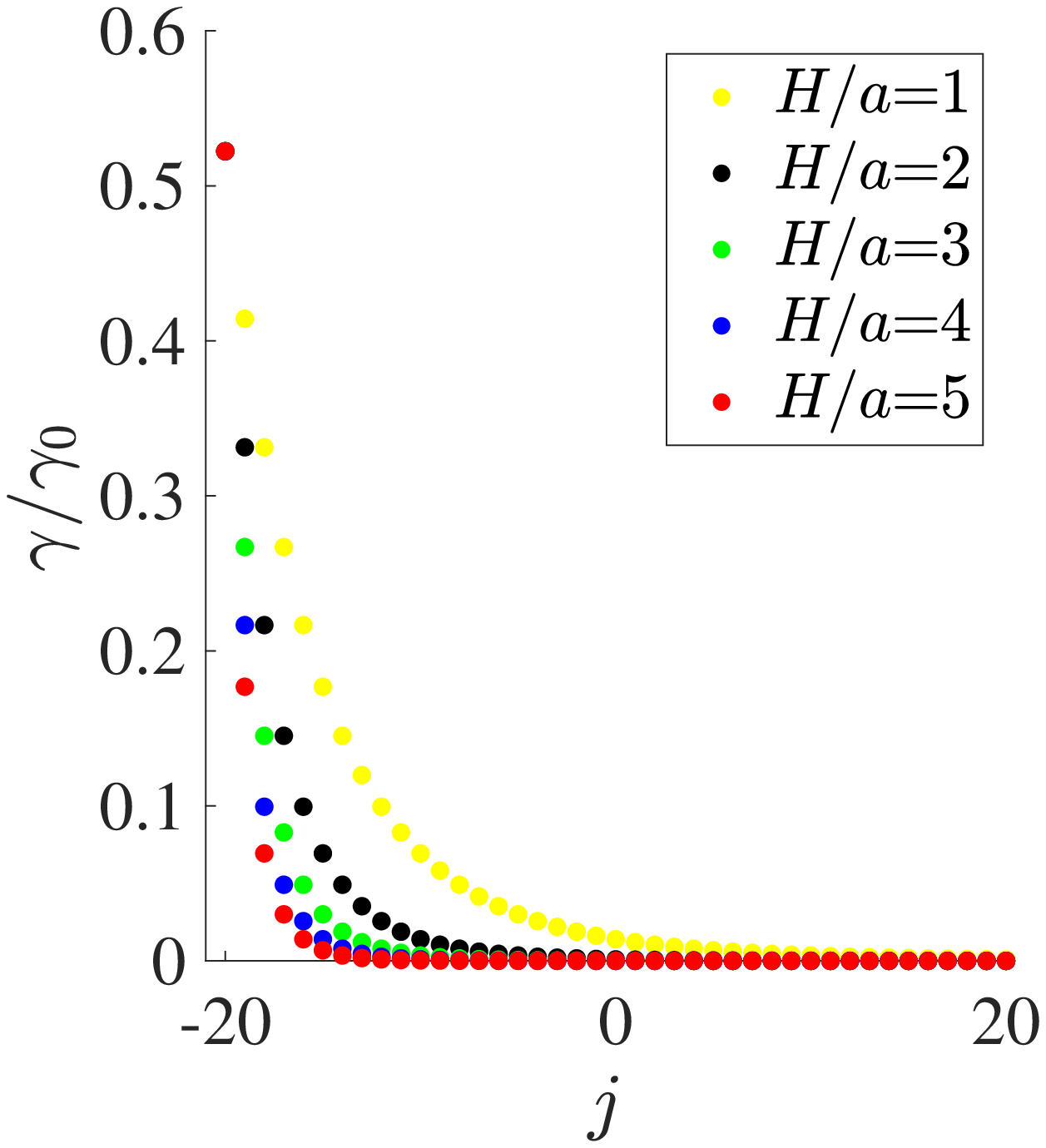}}
\subfigure[]{
\includegraphics[width=0.3\textwidth]{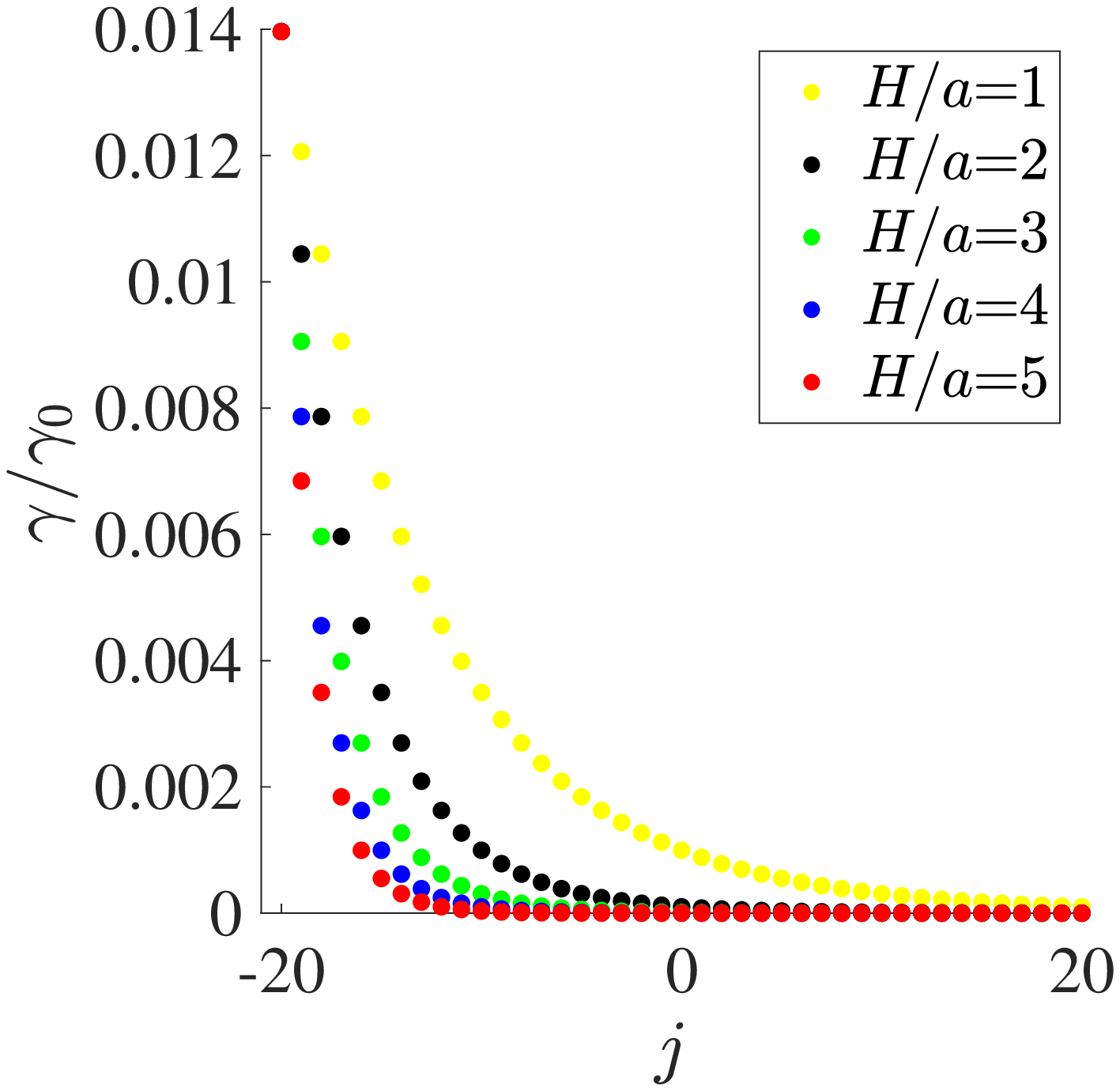}}
\subfigure[]{
\includegraphics[width=0.3\textwidth]{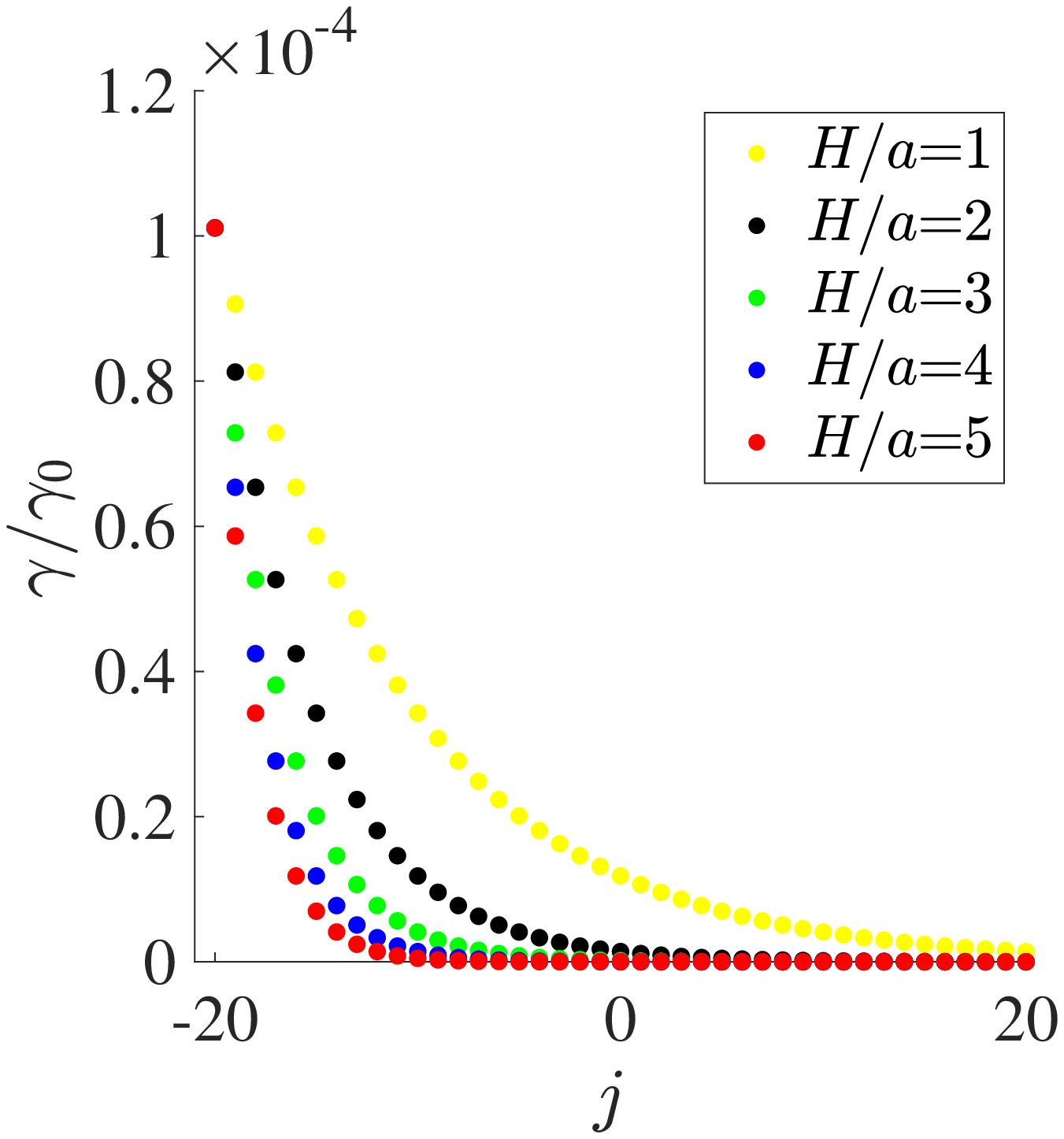}}
\subfigure[]{
\includegraphics[width=0.3\textwidth]{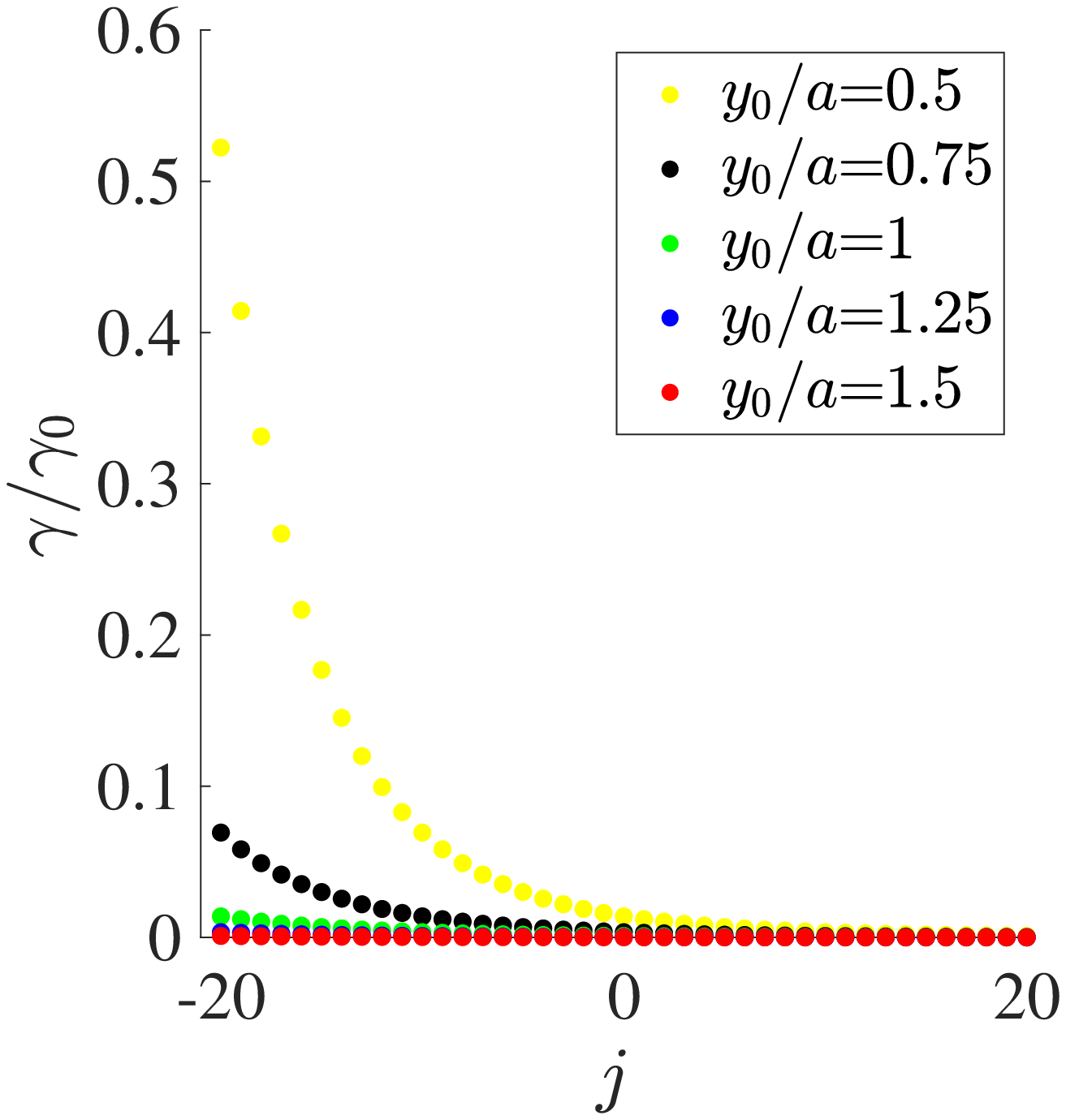}}
\subfigure[]{
\includegraphics[width=0.3\textwidth]{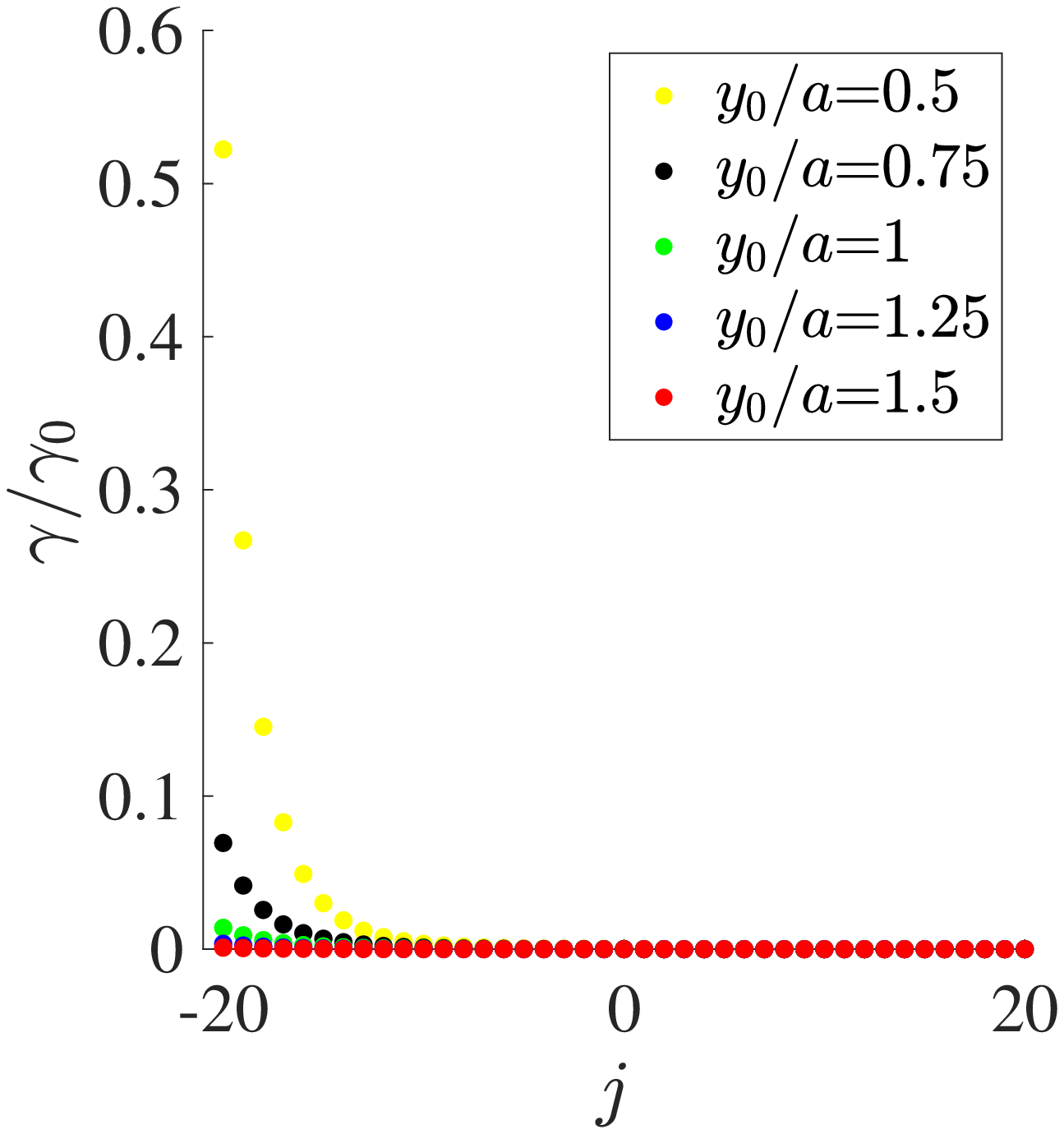}}
\subfigure[]{
\includegraphics[width=0.3\textwidth]{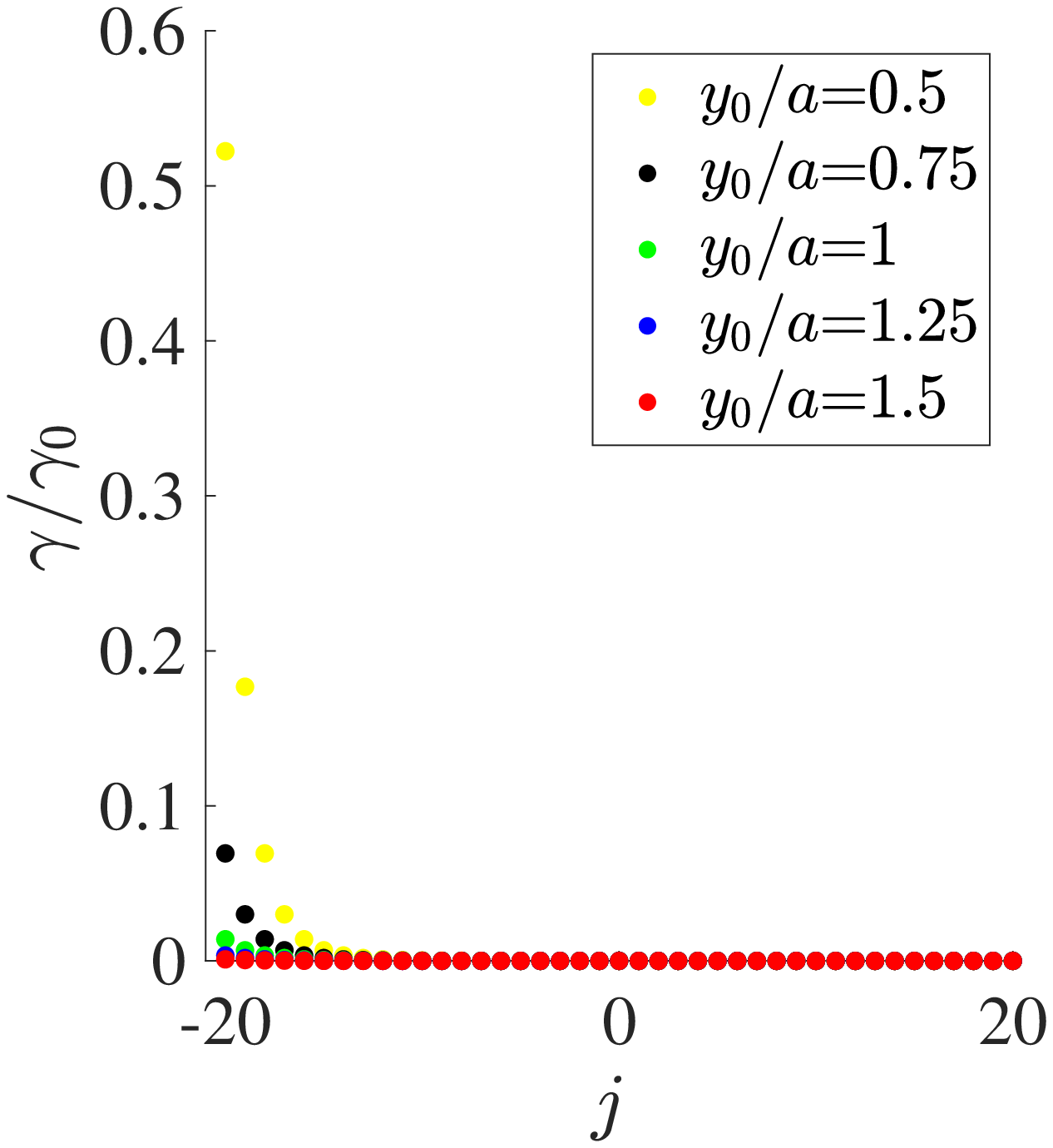}}\\
    \small{Fig. E2: The numerical results of the spontaneous decay rate of guided modes for a single waveguide at the position of each atom. Each decay is normalized by spontaneous decay rate in free-space. (a)--(c) The decay rate with varying $H/a=1,2,3,4$ and 5, and constant (a) $y_0/a=0.5$, (b) $y_0/a=1$ and (c) $y_0/a=2$, as labelled in yellow, cyan, green, blue and red dots, respectively. (d)--(f) The decay rate with varying $y_0/a=0.5,0.75,1,1.25$ and 1.5, and constant (d) $H/a=1$, (e) $H/a=3$ and (f) $H/a=5$, as labelled in yellow, cyan, green, blue and red dots, respectively.}
\end{figure}

In our system, $H$ and $y_0$ reflect how quickly the single- and cross-atom couplings vary and how strong they are. We plot the the numerical results of the spontaneous decay rate determined by the distance $r-a$ from the atom to the top waveguide surface in Fig. E2. When $y_0$ is a constant, larger $H$ indicates the decrease rate of coupling strengths $\gamma$ is faster, as exhibited in Fig. E2(a)--(c). Similarly, when $H$ is constant, larger $y_0$ means lower couplings, as exhibited in Fig. E2(d)--(f). One can notice that $y_0$ determines where the value of $\gamma$ starts at $j=-20$. $H/a$ and $y_0/a$ together determine the varying rate of $\gamma$ on atomic positions, while here we find the varying rate of $H/a$ matters more than $y_0/a$'s, as exhibited in Fig. E2(d)--(f). Here we choose $d/\lambda=10.65, N=41$ in simulations.

We also calculate the FWHM with varying $y_0/a$ and $H/a$, as illustrated in Fig. E3. According to Fig. E3(a), as $y_0/a$ increases, the FWHM roughly stays constant until $y_0/a$ reaches 2. As to Fig. E3(b), as $H/a$ increases, FWHM$_{\text{min}}$ decreases which means the concentration increases, and roughly stays constant after $H/a=4.6$, whereas $\left \langle \text{FWHM} \right \rangle$ increase at first and then decrease after $H/a$ reaches 4.6.

\setcounter{subfigure}{0}
\begin{figure}[!htp]
\centering
\subfigure[]{
\includegraphics[width=0.48\textwidth]{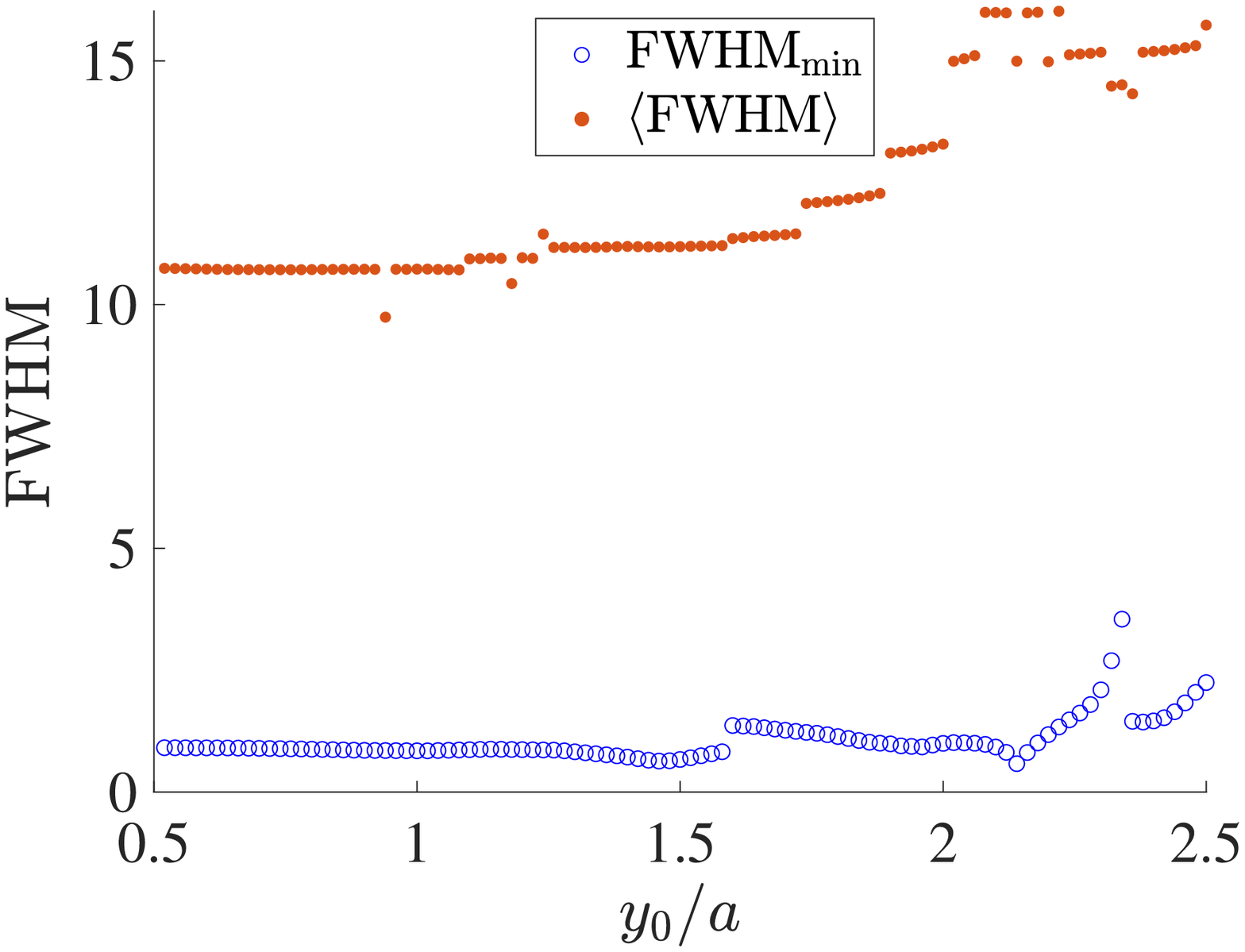}}
\subfigure[]{
\includegraphics[width=0.48\textwidth]{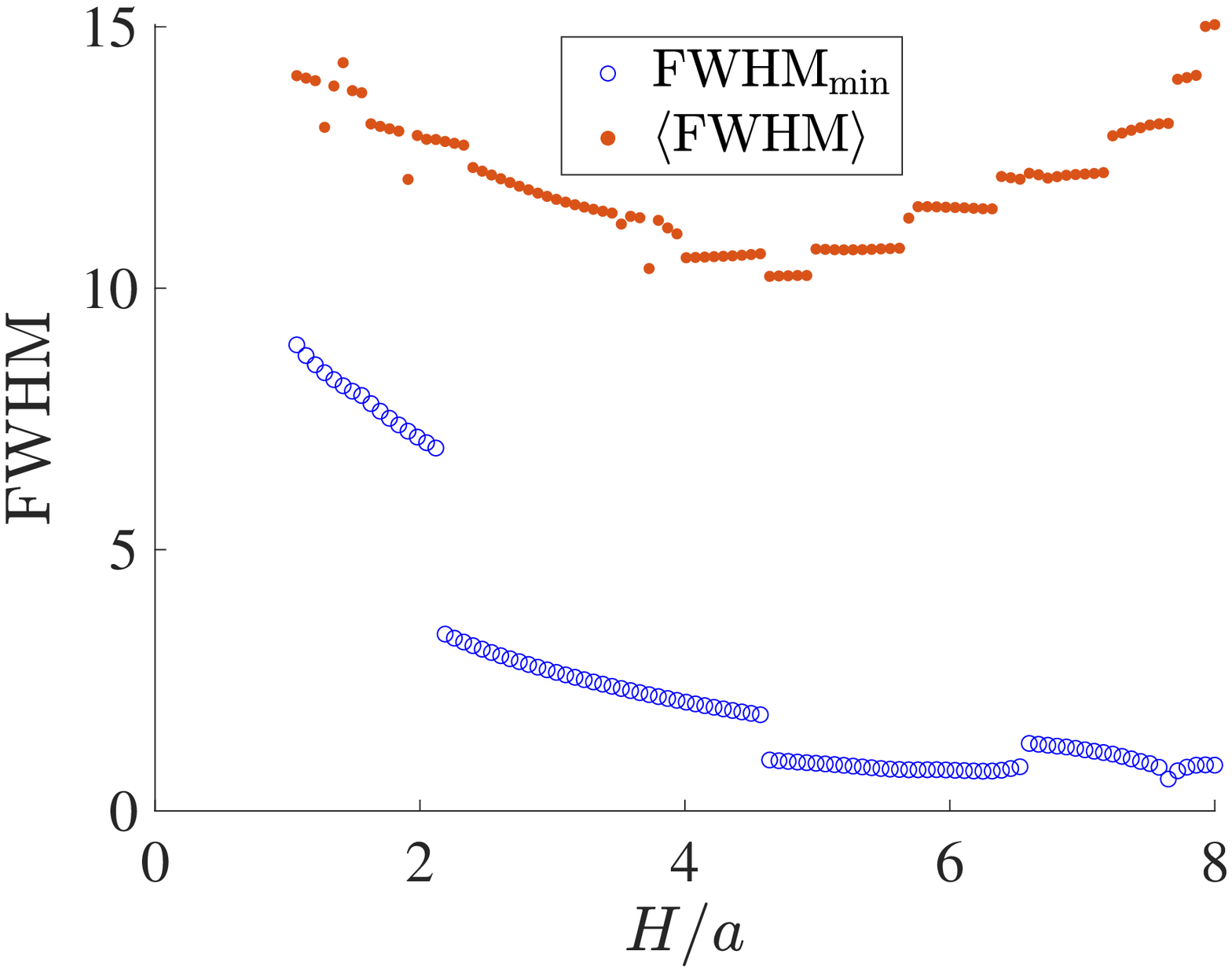}}\\
\small{Fig. E3: (a) FWHM$_{\text{min}}$ and $\left \langle \text{FWHM} \right \rangle$ versus $y_0/a$ when $H/a=5$. (b) FWHM$_{\text{min}}$ and $\left \langle \text{FWHM} \right \rangle$ versus $H/a$ when $y_0=0.5$}
\end{figure}

Moreover, we make a heatmap on FWHM$_{\text{min}}$, as exhibited in Fig. E4. The cool colors in the Fig. E4 indicate better concentrations than the warm colors. One can explicitly see the transition in Fig. E4. For example, when $H/a\lesssim 2$, it shows that eigenstates are not concentrated, while for $H/a > 2$, it indicates that the eigenstates exhibit smaller FWHM$_{min}$, meaning the concentration. Therefore, we find that the decrease of the vertical width of the atomic array in Fig. E1, $H$, brings the transition from concentrated states to non-concentrated states.

\begin{figure}[!htp]
    \centering
    \includegraphics[width=0.6\textwidth]{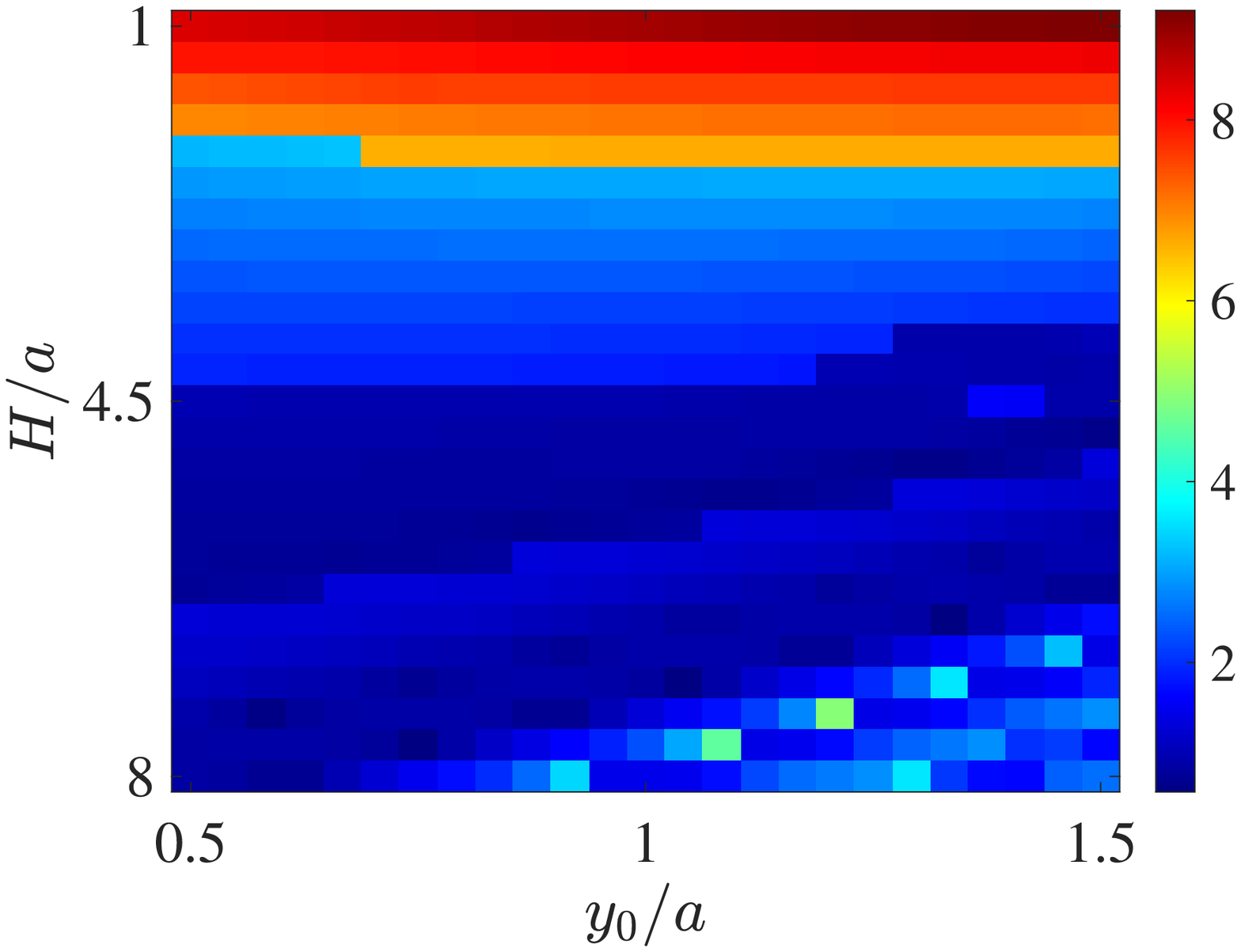}\\
    \small{Fig. E4: The heatmap of FWHM$_{\text{min}}$. The colors indicate the amplitude of FWHM$_{\text{min}}$.}
\end{figure}

We further plot $\left \langle \tau \right \rangle$, as exhibited in Fig. E5. As one can see in Fig. E5, $\left \langle \tau \right \rangle$ gets longer as $y_0/a$ increases, and $\left \langle \tau \right \rangle$ gets longer as well when $H/a$ increases.

\setcounter{subfigure}{0}
\begin{figure}[!htp]
\centering
\subfigure[]{
\includegraphics[width=0.48\textwidth]{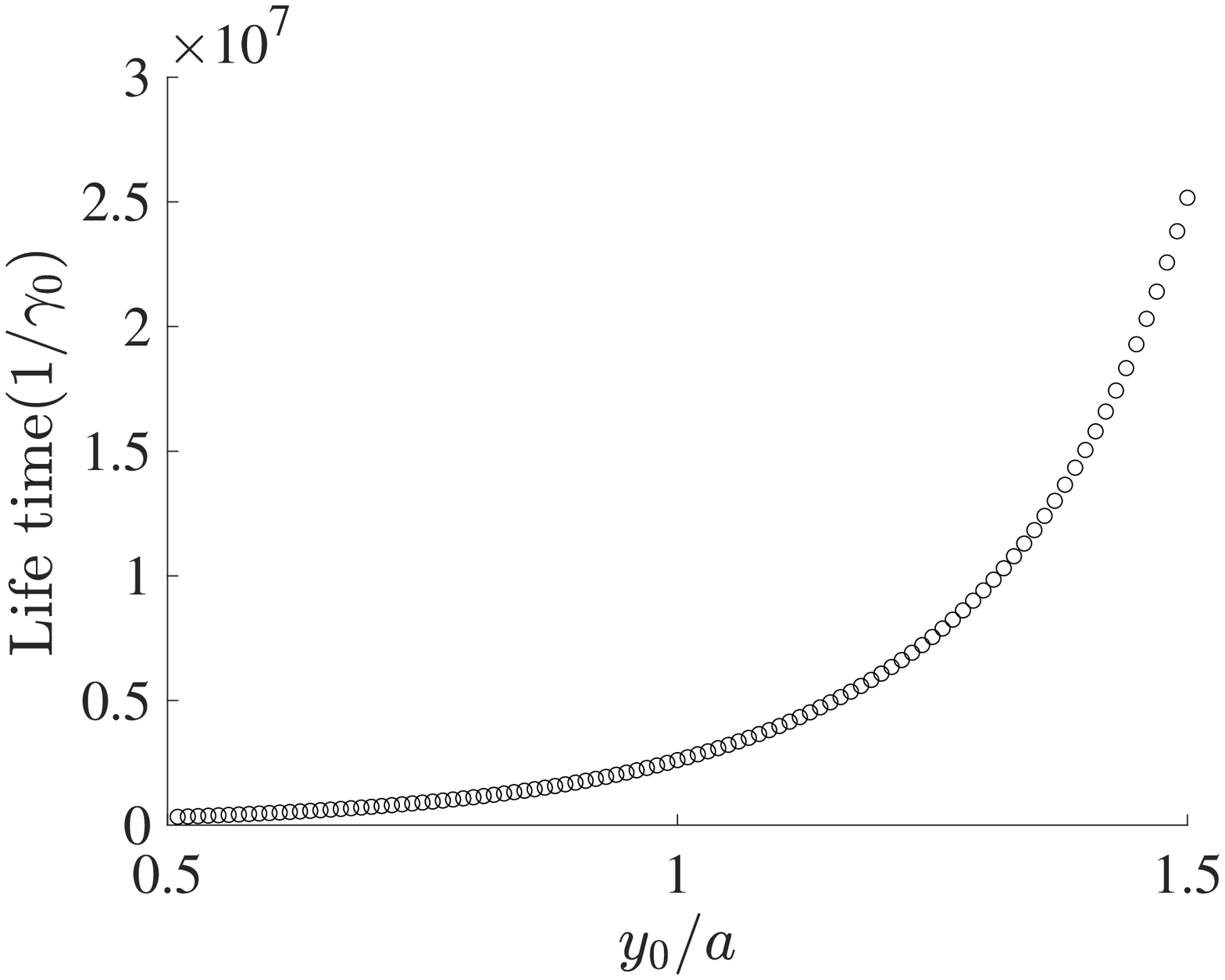}}
\subfigure[]{
\includegraphics[width=0.48\textwidth]{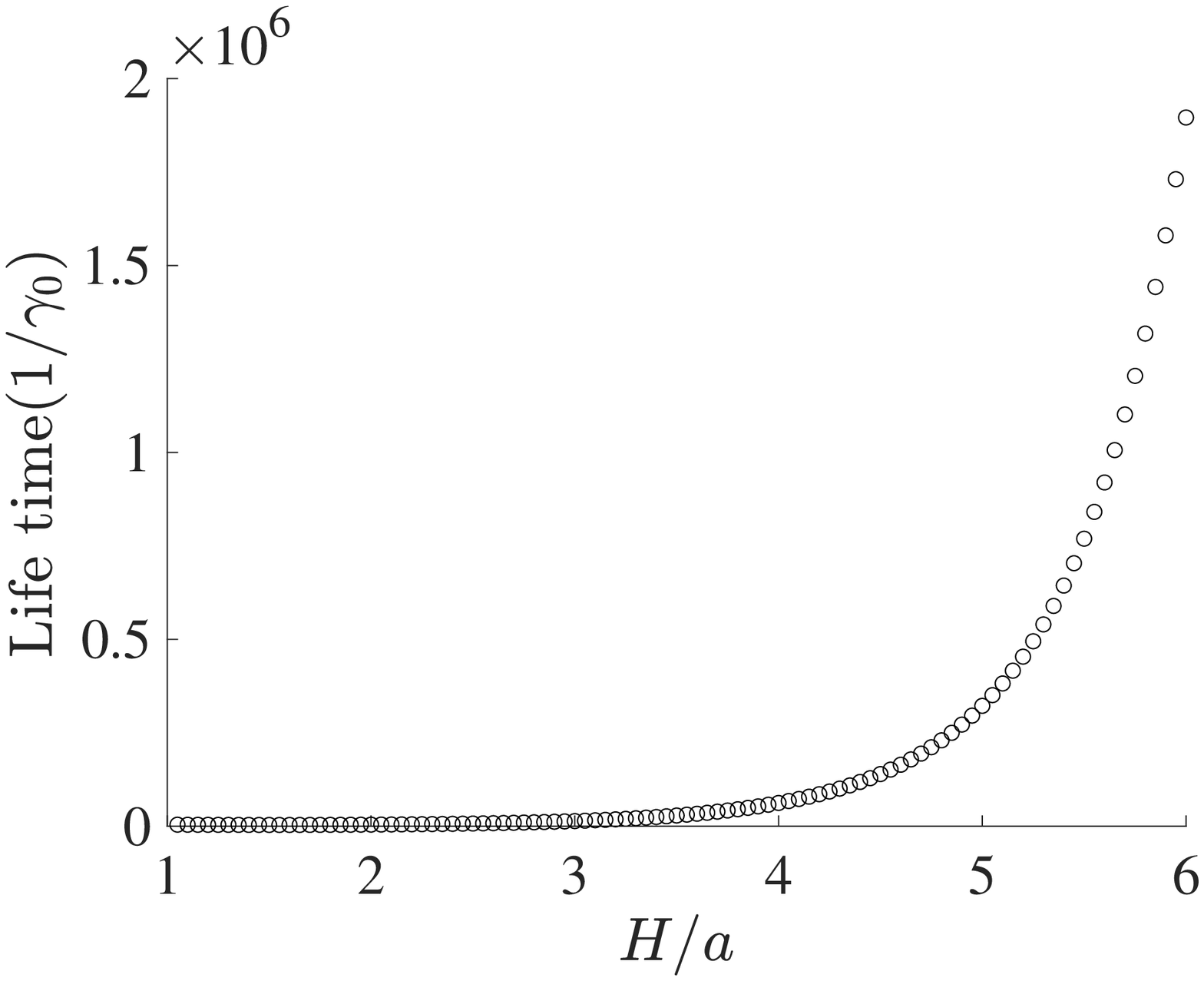}}\\
\small{Fig. E5: (a) The life time $\left \langle \tau \right \rangle$ versus $y_0/a$, when $H/a=5$. (b) The life time $\left \langle \tau \right \rangle$ versus $H/a$, when $y_0/a=0.5$}
\end{figure}

Here we select parameters ($y_0/a=0.5,H/a=1$), ($y_0/a=0.5,H/a=5$), ($y_0/a=2.2,H/a=1$), and ($y_0/a=2.2,H/a=5$) to show the corresponding localization, which are demonstrated in Fig. E6. From Fig. E6(a) and (c) , one can clearly see that when $H/a=1$, these eigenstates exhibit properties as bulk states. Though the minimal loss states, i.e., mode $m=1$, is still concentrated, FWHM$_{\text{min}}$ is very big so the concentration feature disappear. In Fig. E6(b) and (d), one can see that the models exhibit suggested phenomena including well concentrated states. One can clearly see from Fig. E2(a)--(c), no matter what $y_0/a$ is, the shape of the $\gamma$ curve for the same $H/a$ is roughly the same, and this is why Fig. E6(a) and (c) behave similar in terms of the localization, while the life time of these states are different. The similar arguments also work for Fig. E6(b) and (d). 

\setcounter{subfigure}{0}
\begin{figure}[!htp]
\centering
\subfigure[]{
\includegraphics[width=0.4\textwidth]{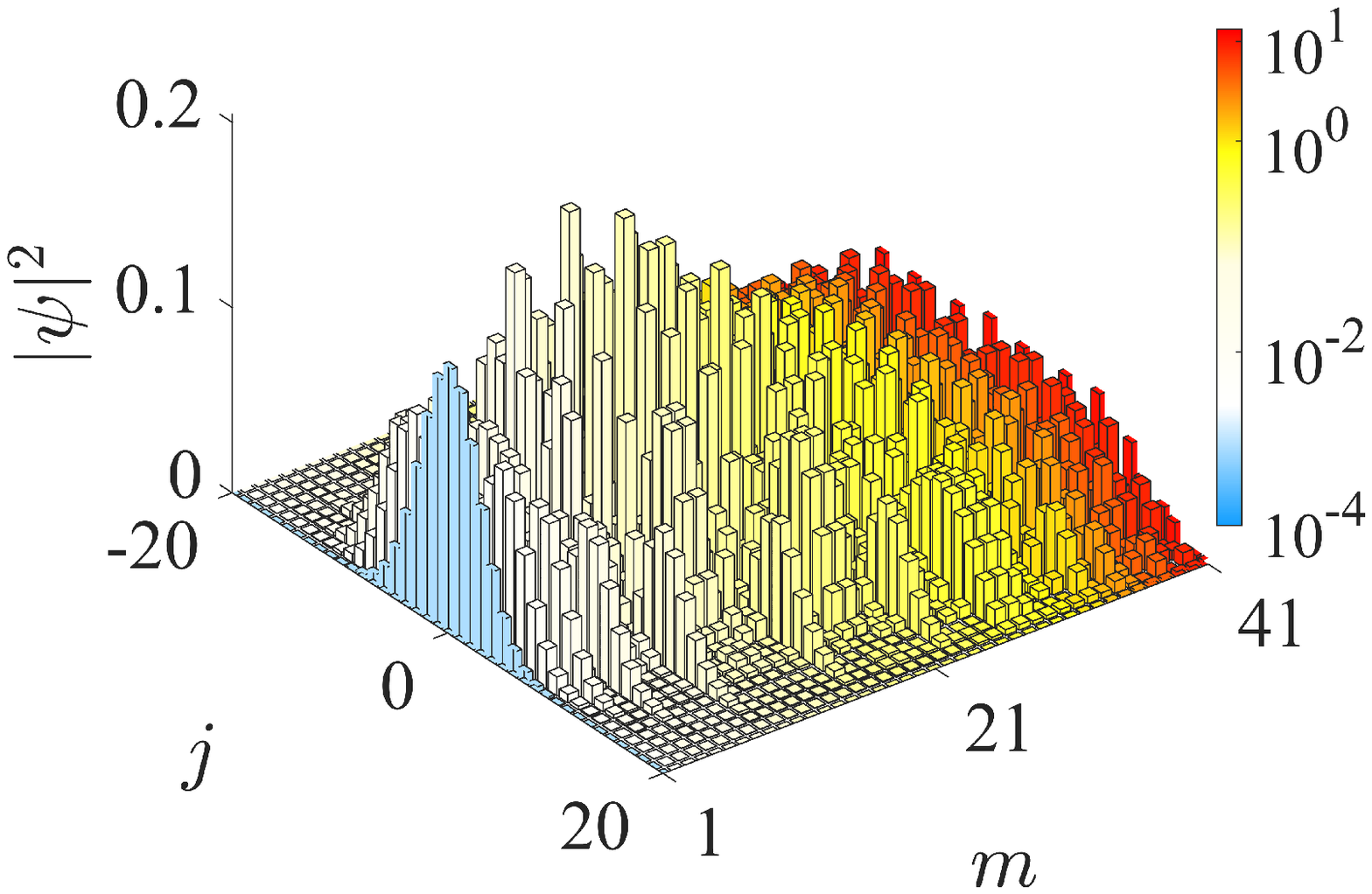}}
\subfigure[]{
\includegraphics[width=0.4\textwidth]{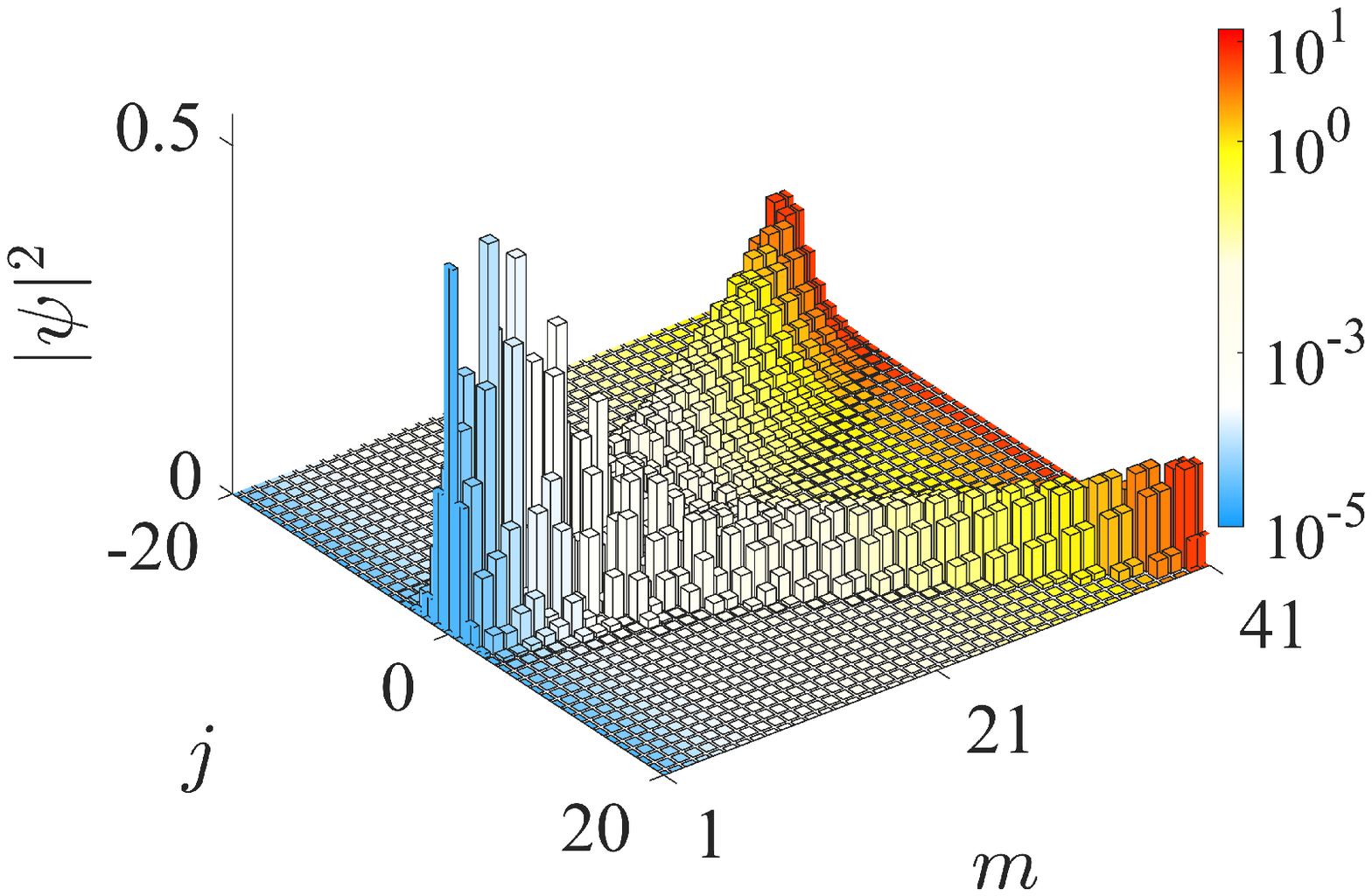}}
\subfigure[]{
\includegraphics[width=0.4\textwidth]{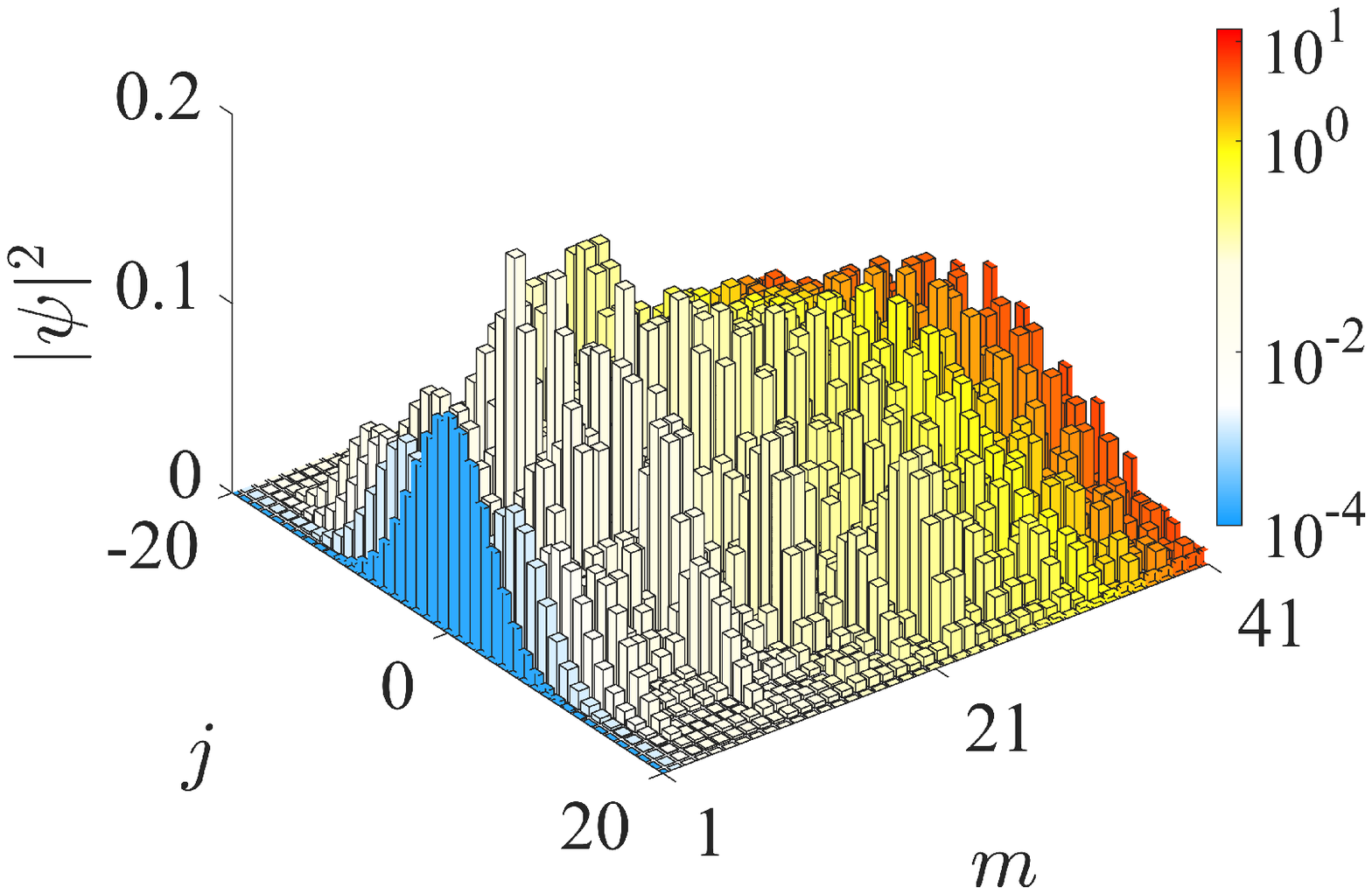}}
\subfigure[]{
\includegraphics[width=0.4\textwidth]{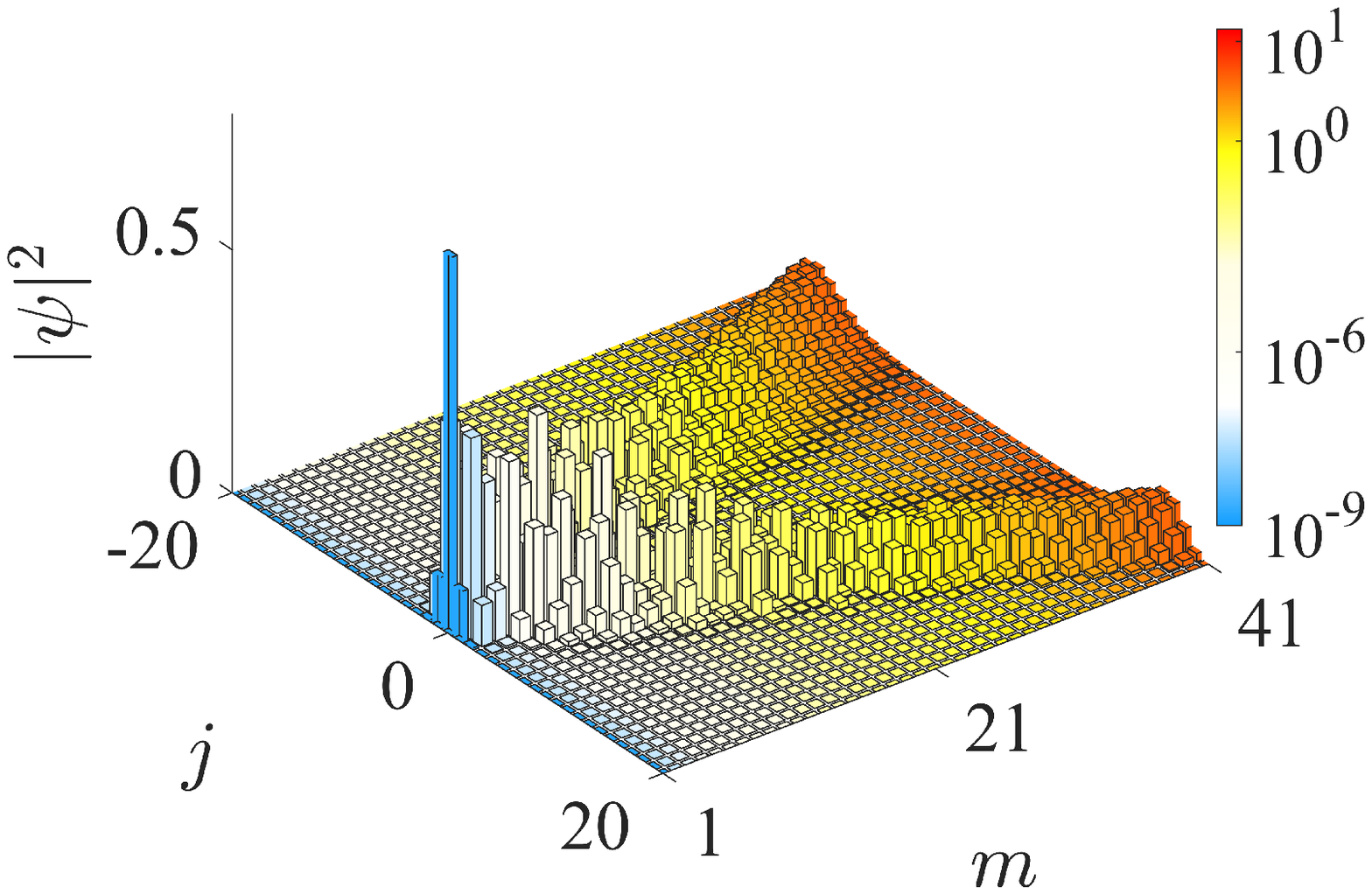}}\\
\small{Fig. E6: Intensity distributions versus atom site $j$ for all eigen-states with $m=1,2,\ldots,41$ at (a) $y_0/a=0.5, H/a=1$, (b) $y_0/a=0.5, H/a=5$, (c) $y_0/a=2.2, H/a=1$, and (d) $y_0/a=2.2, H/a=5$. Colors for each eigen-state indicate the collective decay rates.}
\end{figure}

To better illustrate the effect caused by $H/a$, we further plot Fig. E7. Here we take $y_0/a=0.5$ as a constant since $y_0/a$ does not change the tendencies of couplings but only changes the amplitudes. As shown in Fig. E7, the localization gradually changes from bulk states to the suggested phenomena including well-concentrated states as $H/a$ increases.

\setcounter{subfigure}{0}
\begin{figure}[H]
\centering
\subfigure[]{
\includegraphics[width=0.4\textwidth]{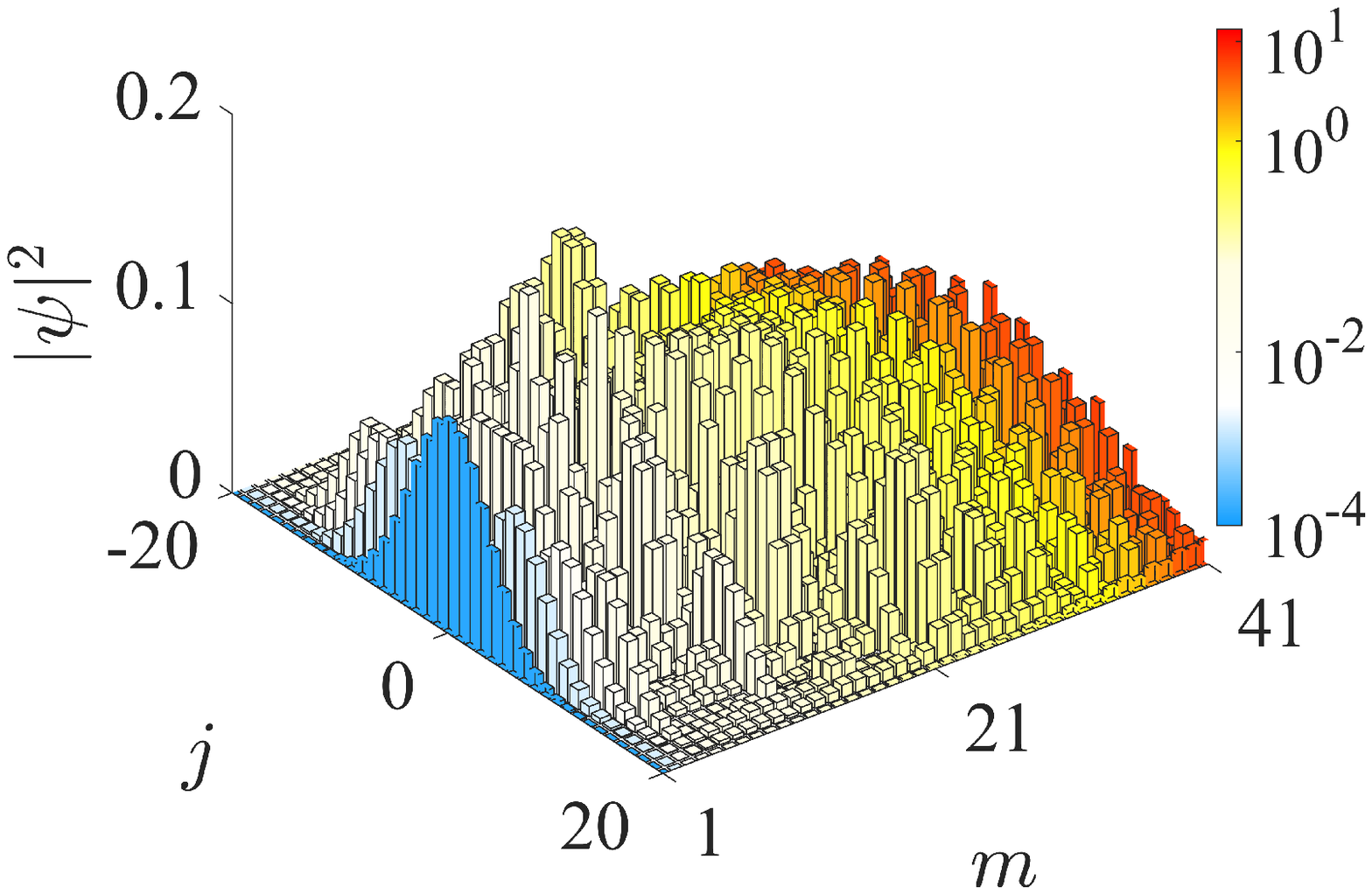}}
\subfigure[]{
\includegraphics[width=0.4\textwidth]{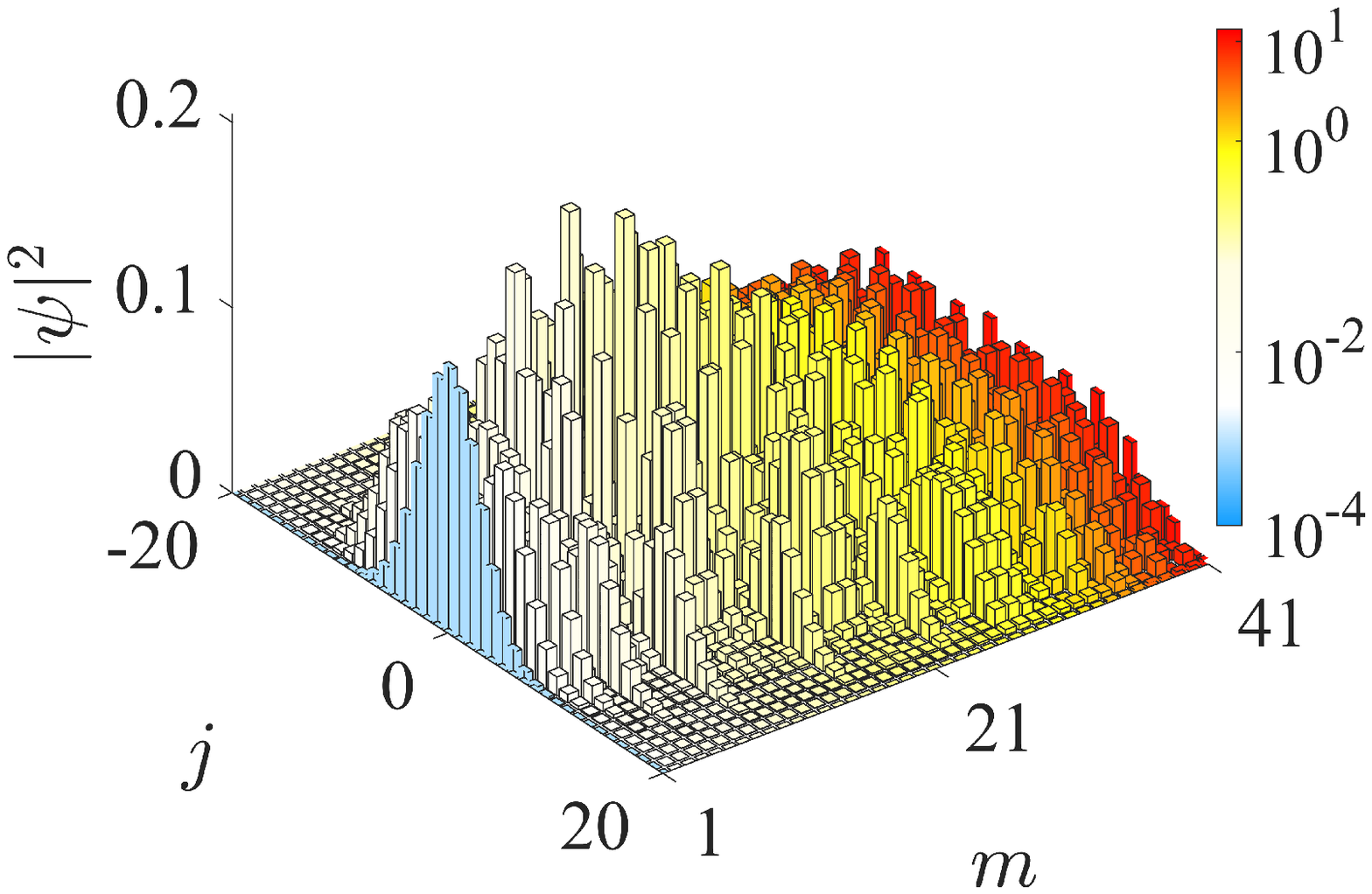}}
\subfigure[]{
\includegraphics[width=0.4\textwidth]{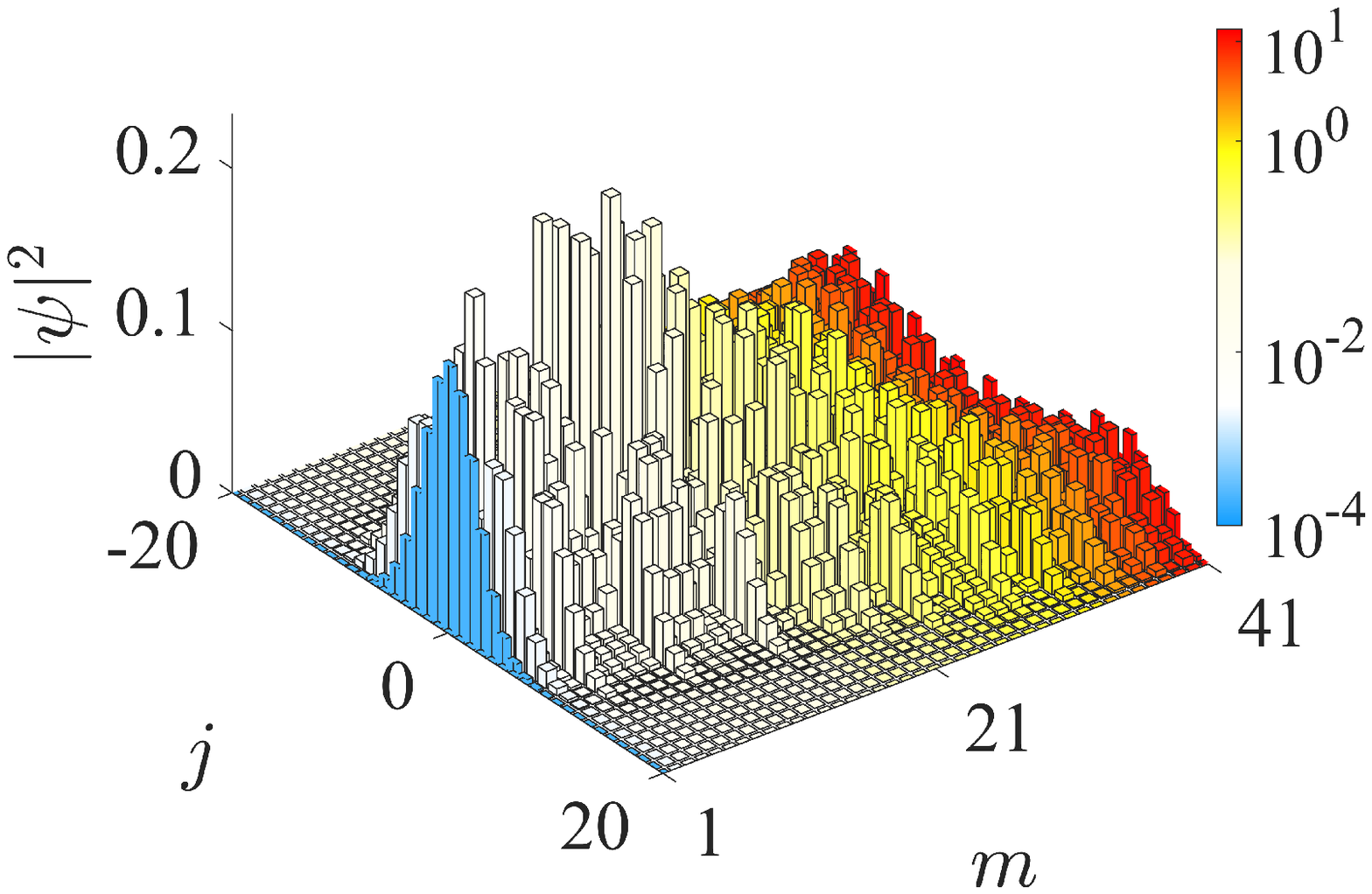}}
\subfigure[]{
\includegraphics[width=0.4\textwidth]{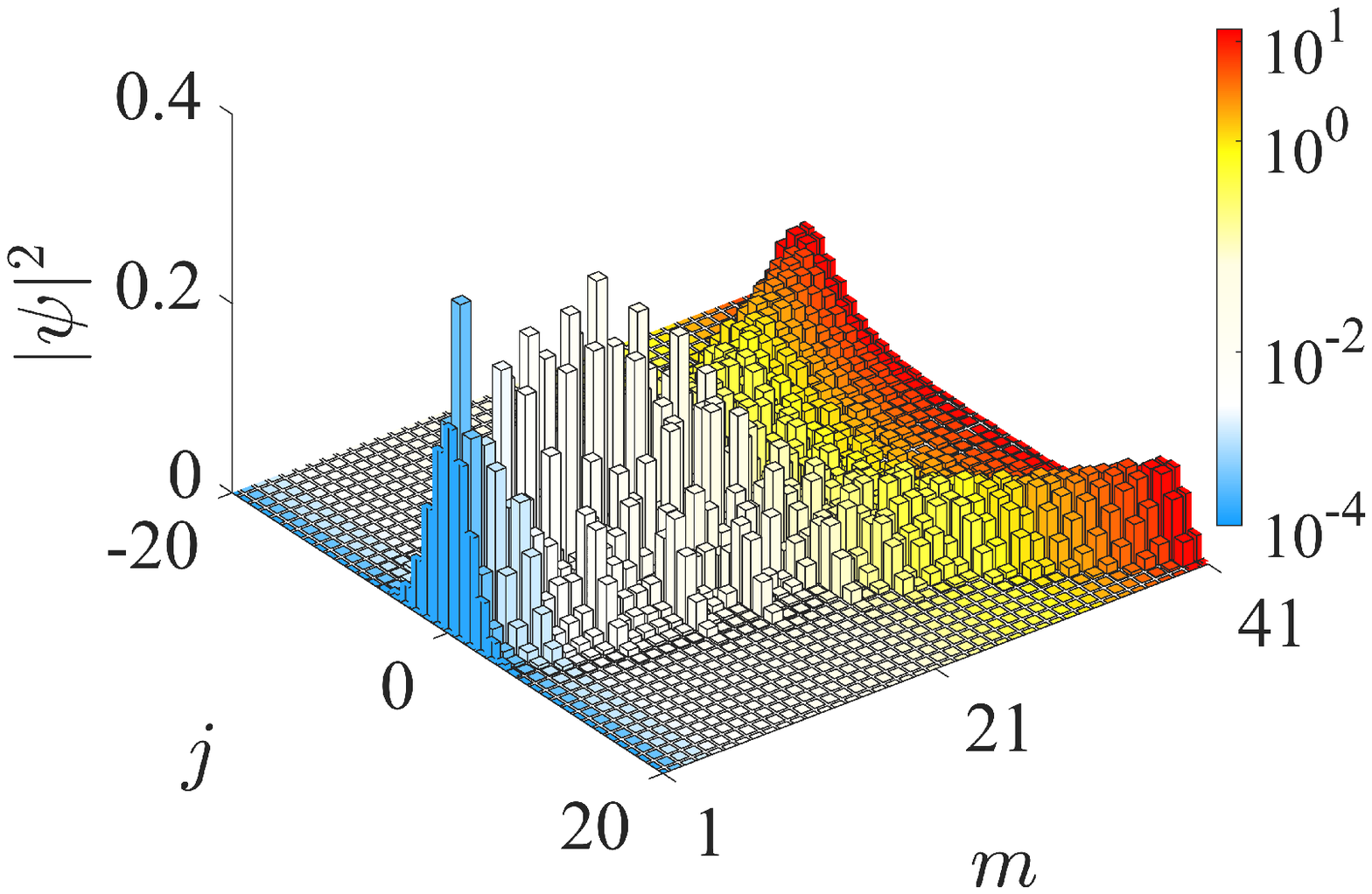}}\\
\small{Fig. E7: Intensity distributions versus atom site $j$ for all eigen-states with $m=1,2,\ldots,41$ at $y_0/a=0.5$ and (a)  $H/a=0.5$, (b) $H/a=1$, (c) $H/a=1.5$, and (d) $H/a=3$. Colors for each eigen-state indicate the collective decay rates.}
\end{figure}

From what has been discussed above, we come to the following conclusion: there is no phase transition in our system as $H/a$ and $y_0/a$ change. The occurrence of the suggested phenomena including well concentrated states mainly depends on the decrease rate of $\gamma$, which is closely related to $H/a$. Specifically, if $\gamma$ decreases too slowly, the eigenstates behave as bulk states; on the other hand, the phenomena of the concentrated states the phenomena will emerge. $y_0/a$ mainly tunes the life time of the eigenstates.


\subsection*{E2. Effects of parameter $d$}

We set $d>10\lambda = 8520$nm such that $d>>\lambda=852$nm to reduce the impact from the environment as much as possible. Here we choose $H/a=3, y_0/a=0.5, N=41$ in calculations.

\setcounter{subfigure}{0}
\begin{figure}[H]
\centering
\subfigure[]{
\includegraphics[width=0.48\textwidth]{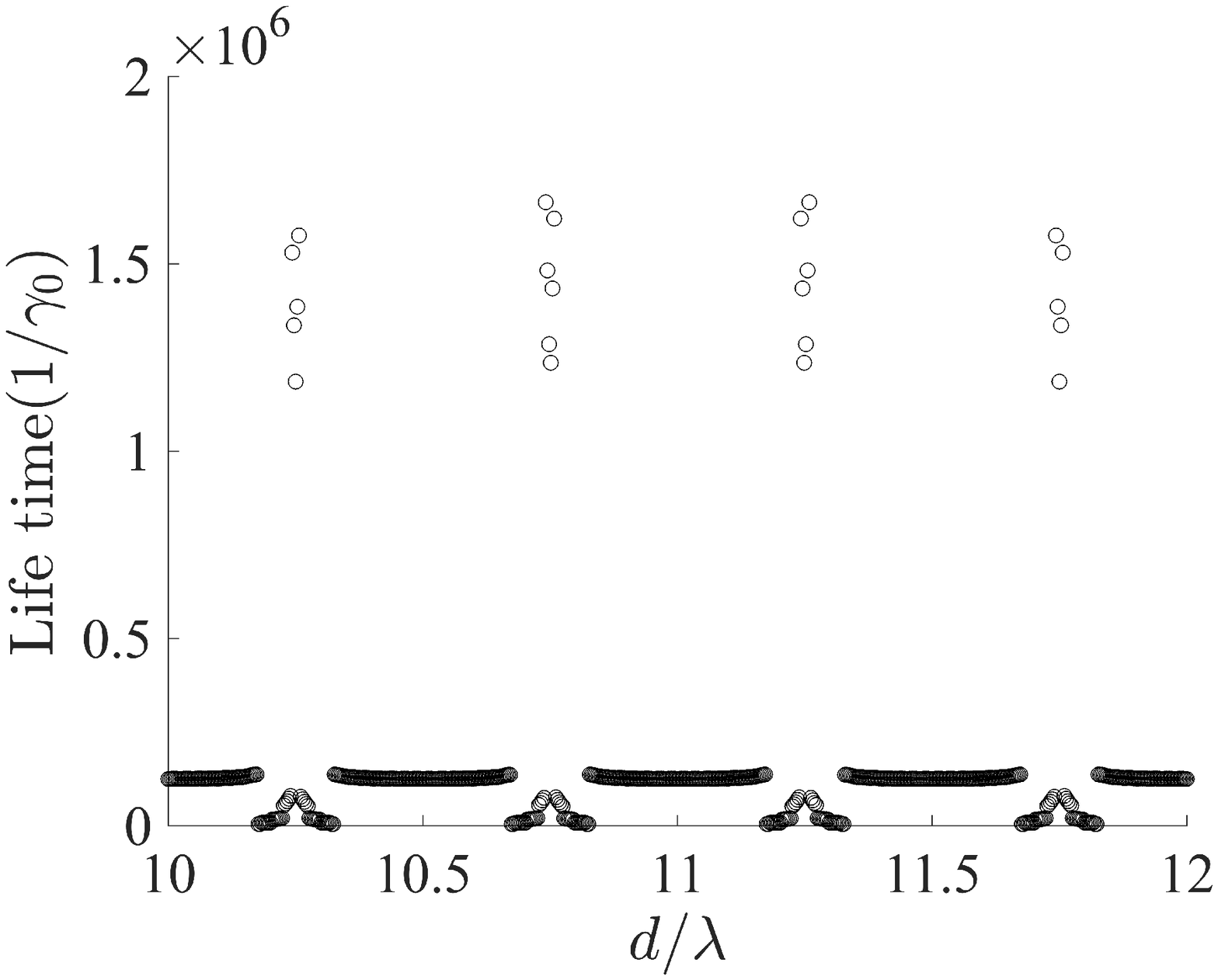}}
\subfigure[]{
\includegraphics[width=0.48\textwidth]{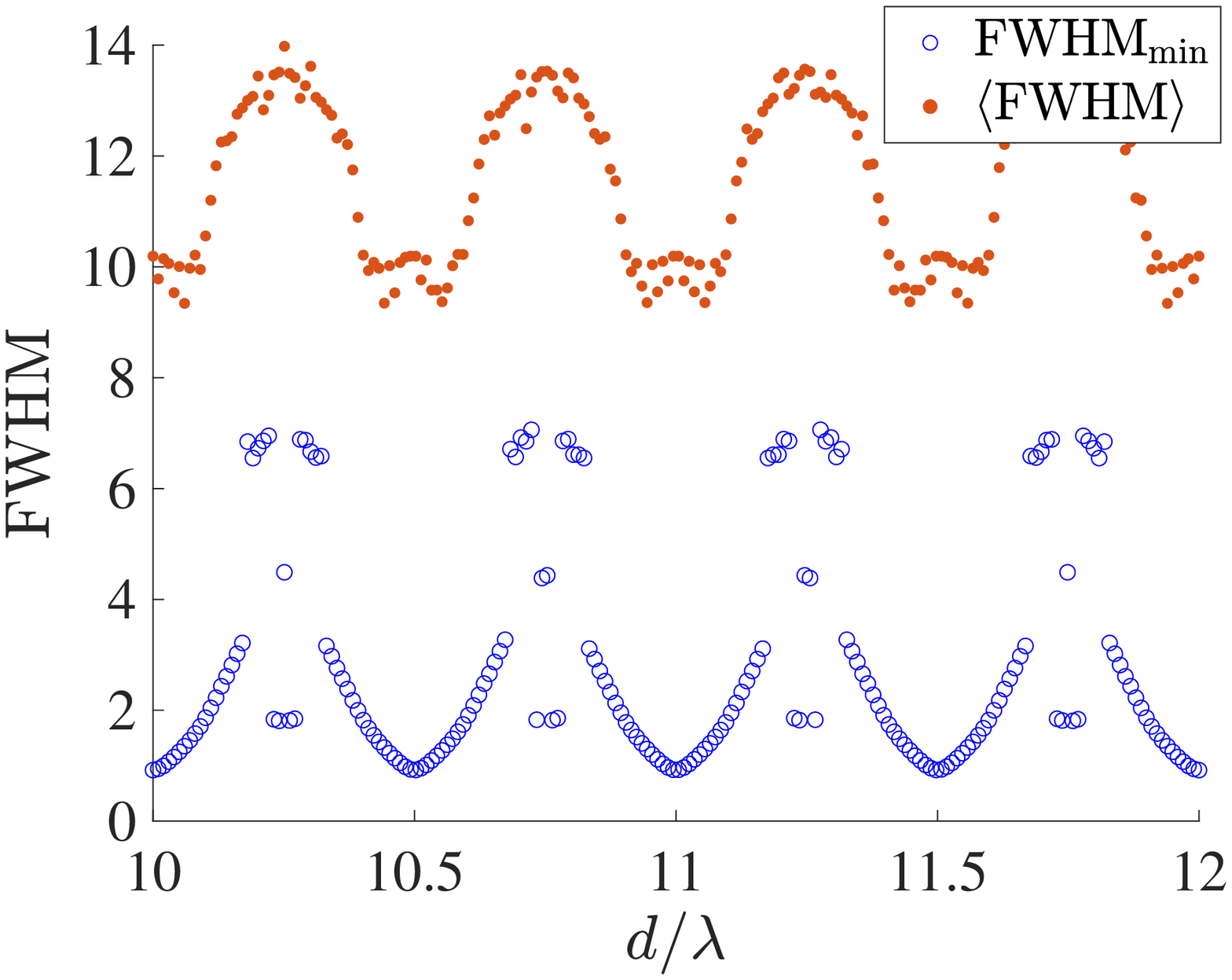}}\\
    \small{Fig. E8: (a) $\left\langle \tau \right \rangle$ versus $d/\lambda$. (b) FWHM$_{\text{min}}$ and $\left\langle \text{FWHM} \right \rangle$ versus $d/\lambda$, as indicated in blue dots and red dots, respectively.}
\end{figure}

We plot $\left\langle\tau\right\rangle$ shown in Fig. E8(a), and also make plots on FWHM$_{\text{min}}$ and $\left\langle \text{FWHM} \right \rangle$ exhibited in Fig. E8(b). We find the periodicity roughly $0.5d/\lambda$. In our proposed model, the Hamiltonian is written as 

\begin{equation}
\hat{H}_{\mathrm{eff}}=-\frac{i}{2}\sum_{j}(\gamma_{Lj}+\gamma_{Rj})\hat{\sigma}_{j}^{\dagger}\hat{\sigma}_{j}-i\sum_{j>l}\sqrt{\gamma_{Ll}\gamma_{Lj}}\hat{\sigma}_{l}^{\dagger}\hat{\sigma}_{j} e^{ik(x_{j}-x_{l})}-i\sum_{j>l}\sqrt{\gamma_{Rl}\gamma_{Rj}}\hat{\sigma}_{j}^{\dagger}\hat{\sigma}_{l} e^{ik(x_{j}-x_{l})},
    \label{R3}
\end{equation}
where $k(x_{j}-x_{l})=2\pi(j-l) d/\lambda$, so $d/\lambda$ is the phase factor of the cross-atom couplings.


According to Fig. E8(b), $\left\langle \text{FWHM} \right \rangle$ and FWHM$_{\text{min}}$ is in the range of $[9,14]$ and $[1,7]$, respectively, which indicates a good concentration when varying $d$. We further choose two spacing $d/\lambda=10.25$, $d/\lambda=10.72$ and $d/\lambda=10.73$, which correspond the widest $\left\langle \text{FWHM} \right \rangle$ = 13.98 and FWHM$_{\text{min}}$=7.06, and the narrowest FWHM$_{\text{min}}$=1.83, respectively, as plotted in Fig. E9. In Fig. E9, we find the main feature of the concentrated subradiant states and extended superradiant states persists, but the details of the distributions localization inevitably changes due to the varying phase of the interactions. In Fig. E9(b) and (c), the corresponding atoms spacing, which are $d/\lambda=10.72$ and $d/\lambda=10.73$, respectively, are very close. While both of their FWHM$_{\text{min}}$ and localizations are so different and hence indicate a transition at $d/\lambda= 10.72\sim 10.73$. Additionally, we can find this kind of transition periodically.

\setcounter{subfigure}{0}
\begin{figure}[H]
\subfigure[]{
\includegraphics[width=0.32\textwidth]{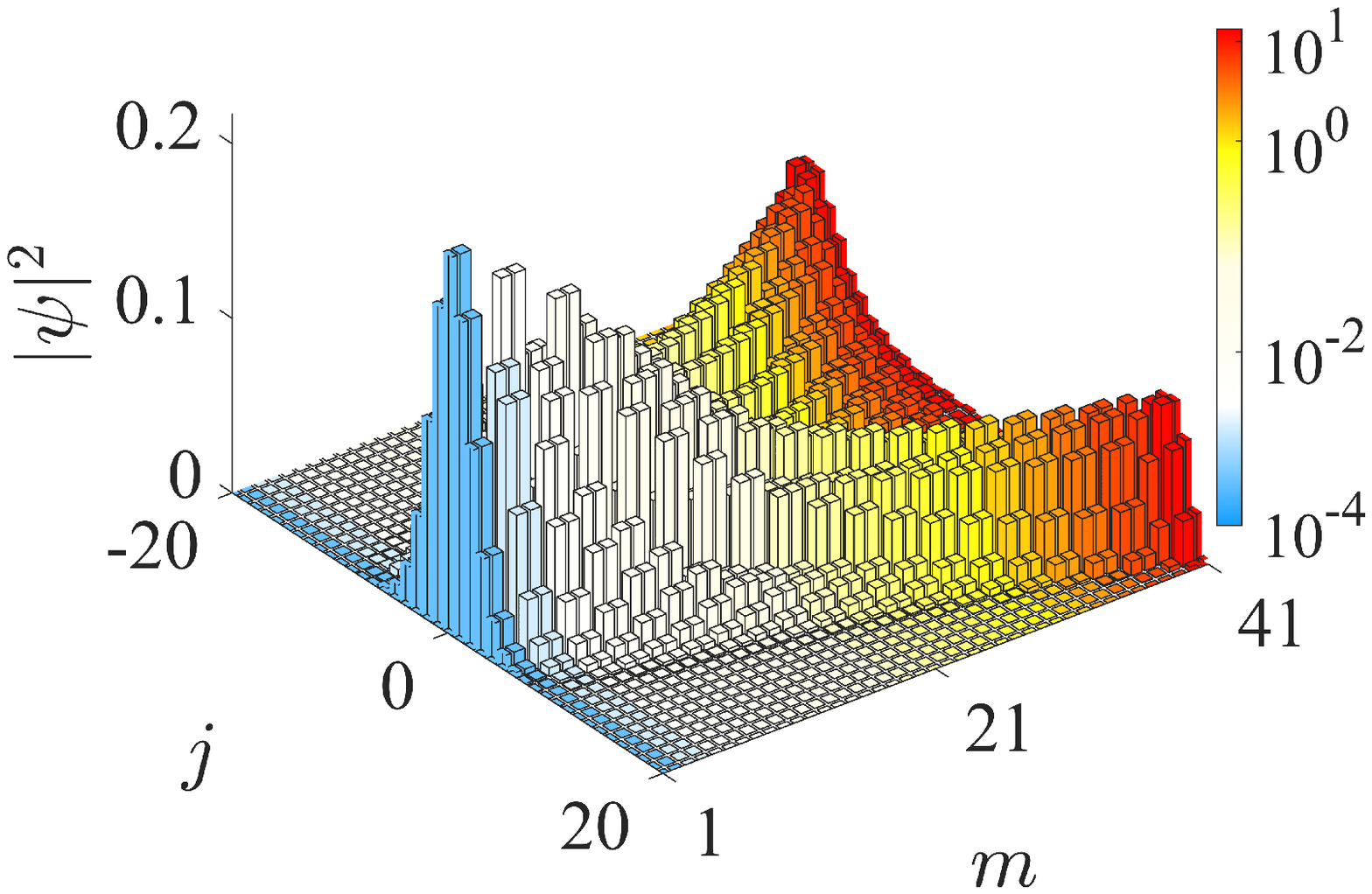}}
\subfigure[]{
\includegraphics[width=0.32\textwidth]{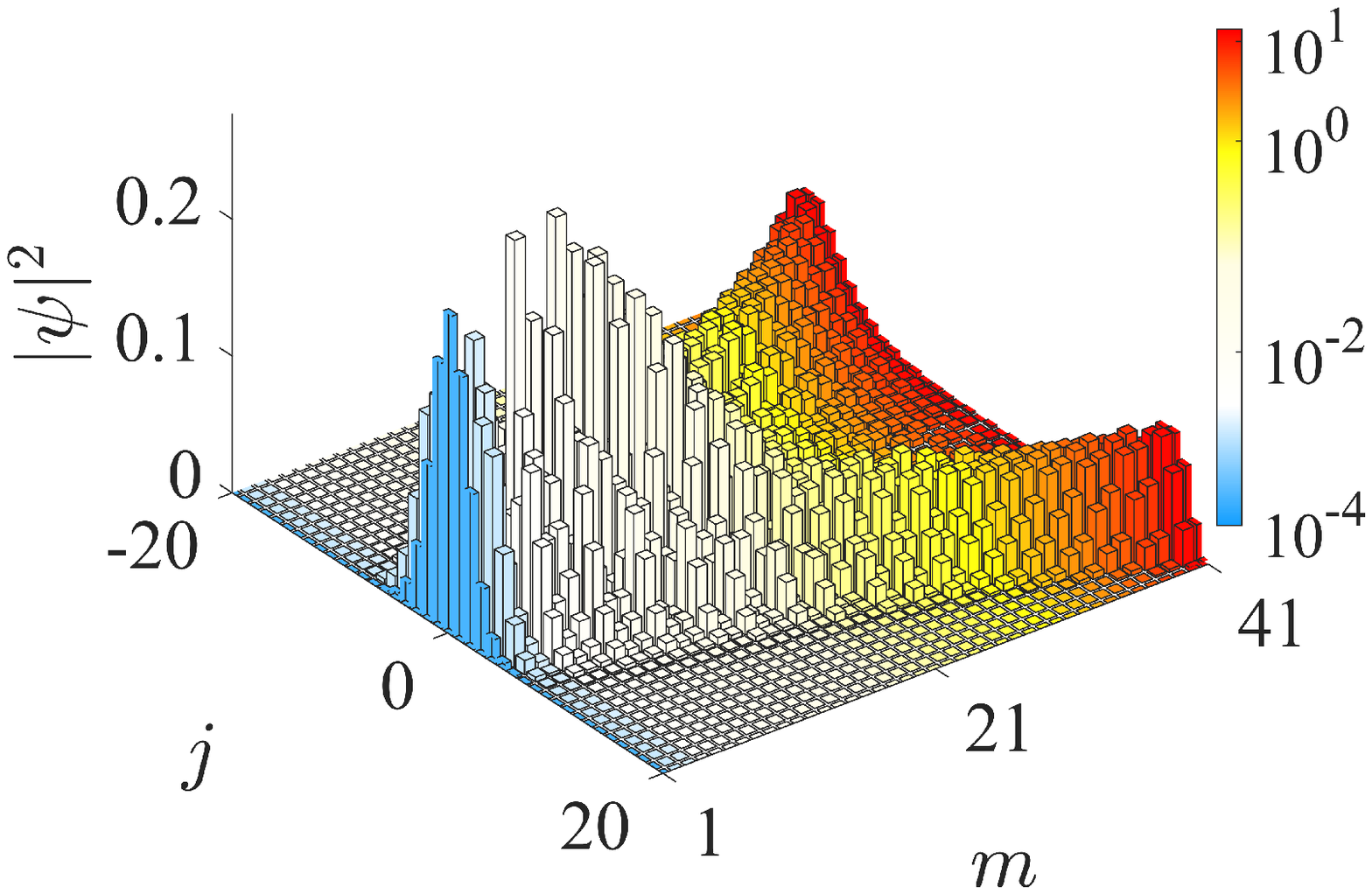}}
\subfigure[]{
\includegraphics[width=0.32\textwidth]{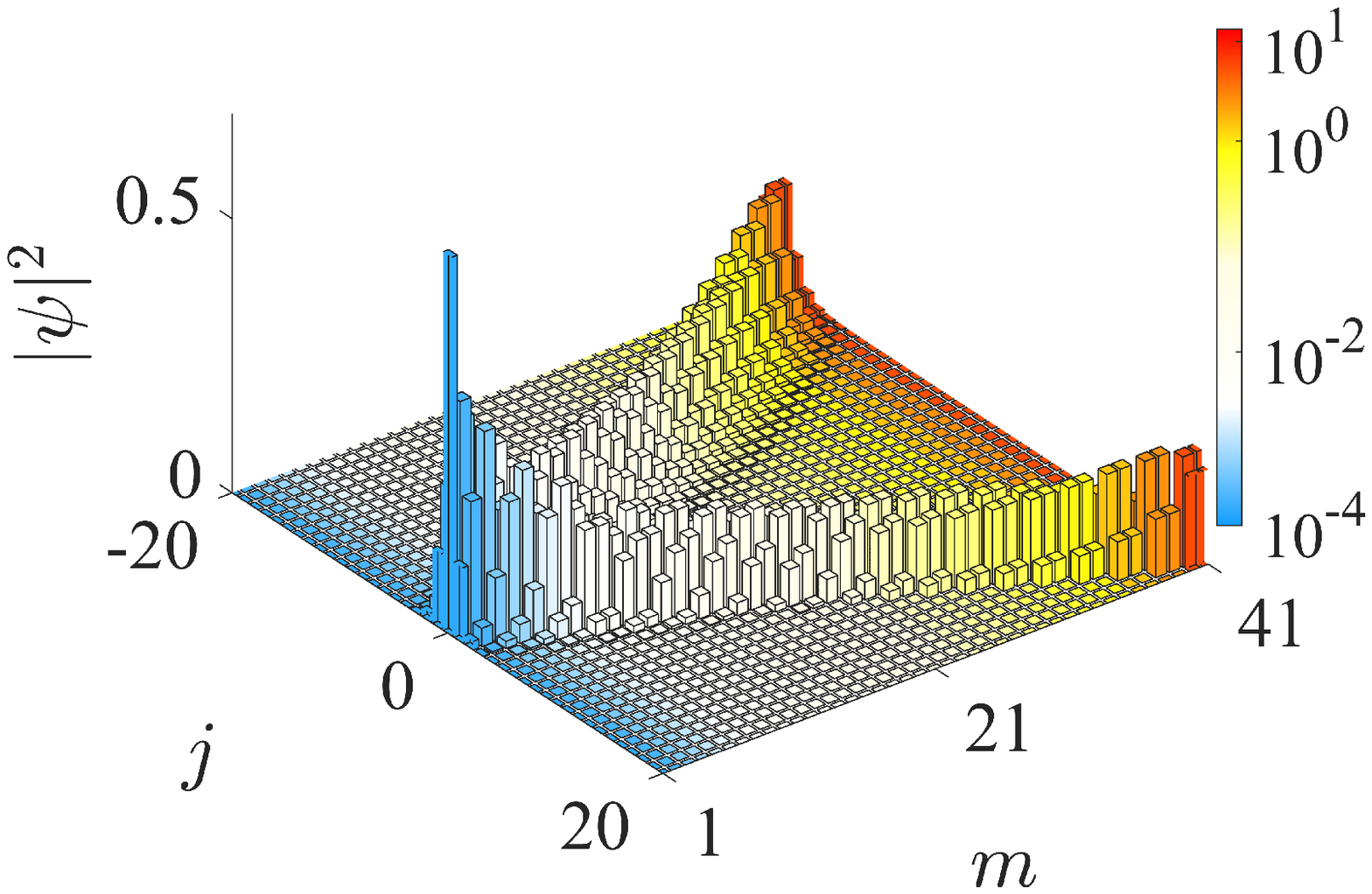}}\\
    \small{Fig. E9: Intensity distributions versus atom site $j$ for all eigen-states with $m=1,2,\ldots,41$ at (a) $d/\lambda=10.25$, $\left\langle \text{FWHM} \right \rangle$= 13.98, (b) $d/\lambda=10.72$, FWHM$_{\text{min}}$ = 7.06, and (c) $d/\lambda=10.73$, FWHM$_{\text{min}}$ = 1.83. Colors for each eigen-state indicate the collective decay rates.}
\end{figure}


\subsection*{E3. Effects of parameter $N$}\label{para}

\setcounter{subfigure}{0}
\begin{figure}[H]
    \centering
    \subfigure[]{
    \includegraphics[width=0.48\textwidth]{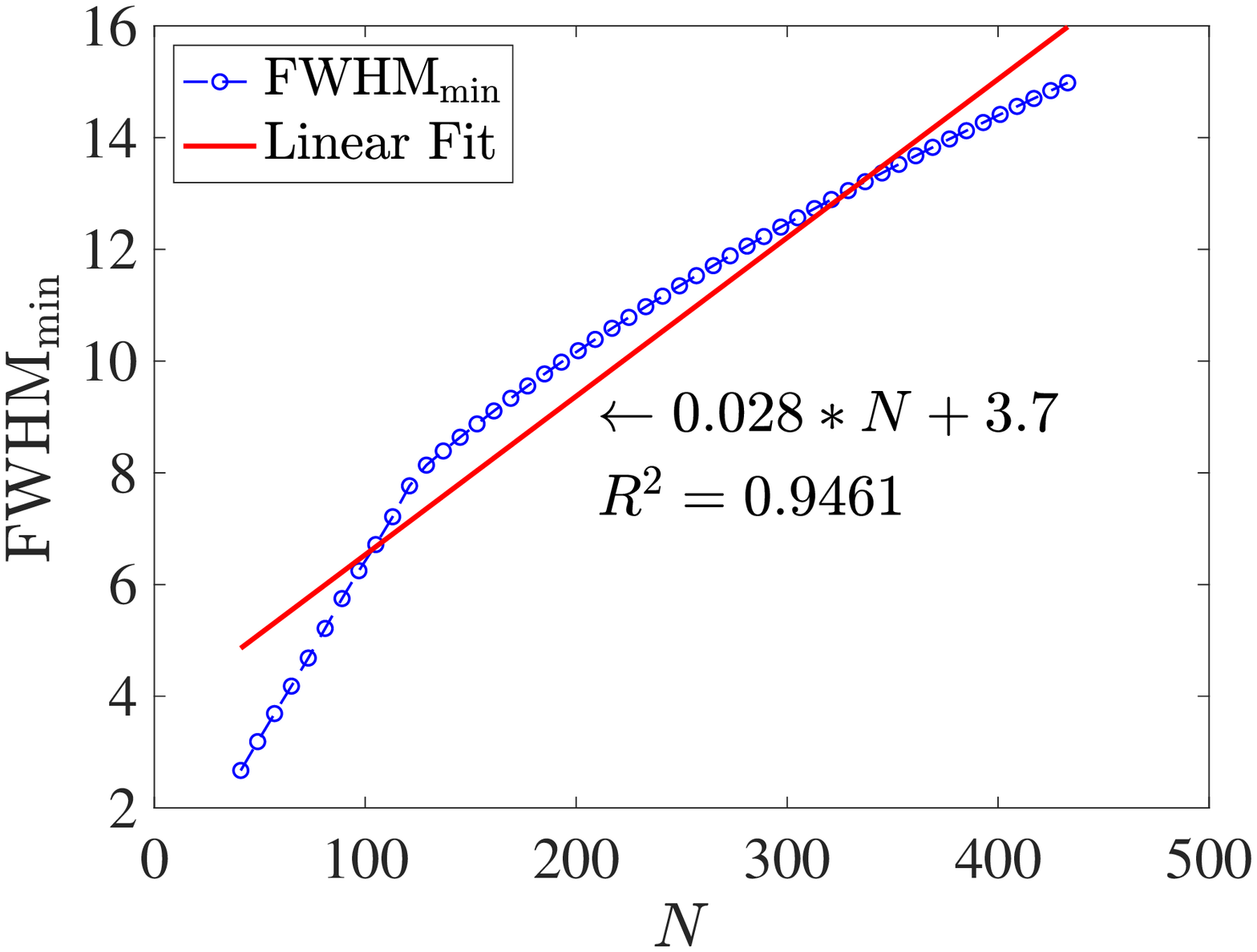}}
    \subfigure[]{
    \includegraphics[width=0.48\textwidth]{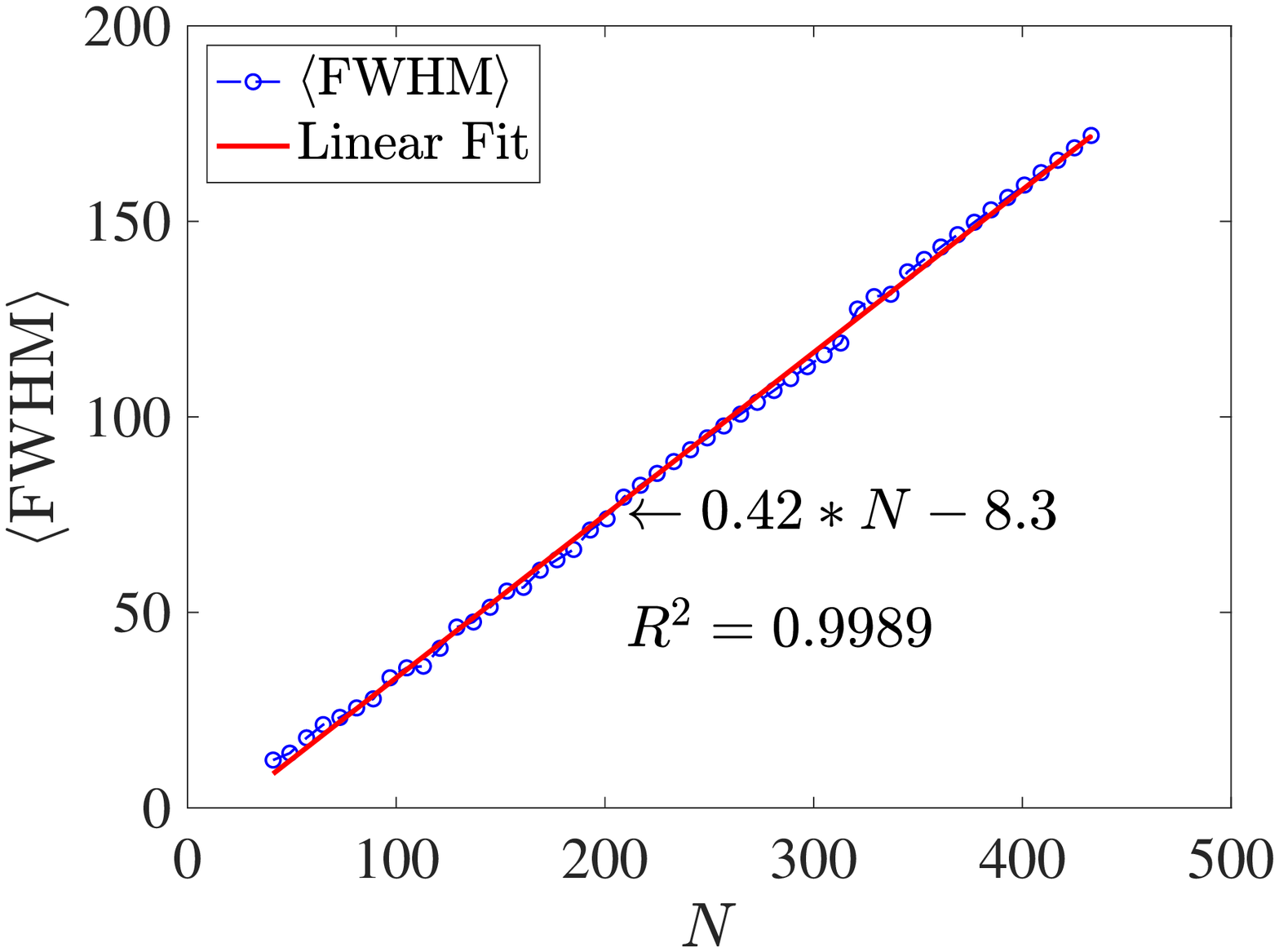}}\\
    \small{Fig. E10: (a) FWHM$_{\text{min}}$ and the corresponding linearly fitted curves, as labelled in blue dots-line and red line, respectively. The R-square of the linear fitting is 0.9461. (b) $\left\langle \text{FWHM} \right \rangle$ and the corresponding linearly fitted curves, as labelled in blue dots-line and red line, respectively. The R-square of the linear fitting is 0.9989.}
\end{figure}

Here we choose $H/a=3, y_0/a=0.5$ and $d/\lambda=10.65$ and perform simulations with different $N$. $\left\langle \text{FWHM} \right \rangle$ and FWHM$_{\text{min}}$ and their corresponding linearly fitted curves are plotted in Fig. E10. One can see the goodness of linear fitting, i.e., $R^2=0.9461$ and 0.9989 $\rightarrow 1$, indicating the change of $N$ does not change the main features of our system. In particular, we further plot localizations of eigenstates for four different $N$, including $N=41, 101, 151$ and 201, as illustrated in Fig. E11.

\setcounter{subfigure}{0}
\begin{figure}[H]
\centering
\subfigure[]{
\includegraphics[width=0.4\textwidth]{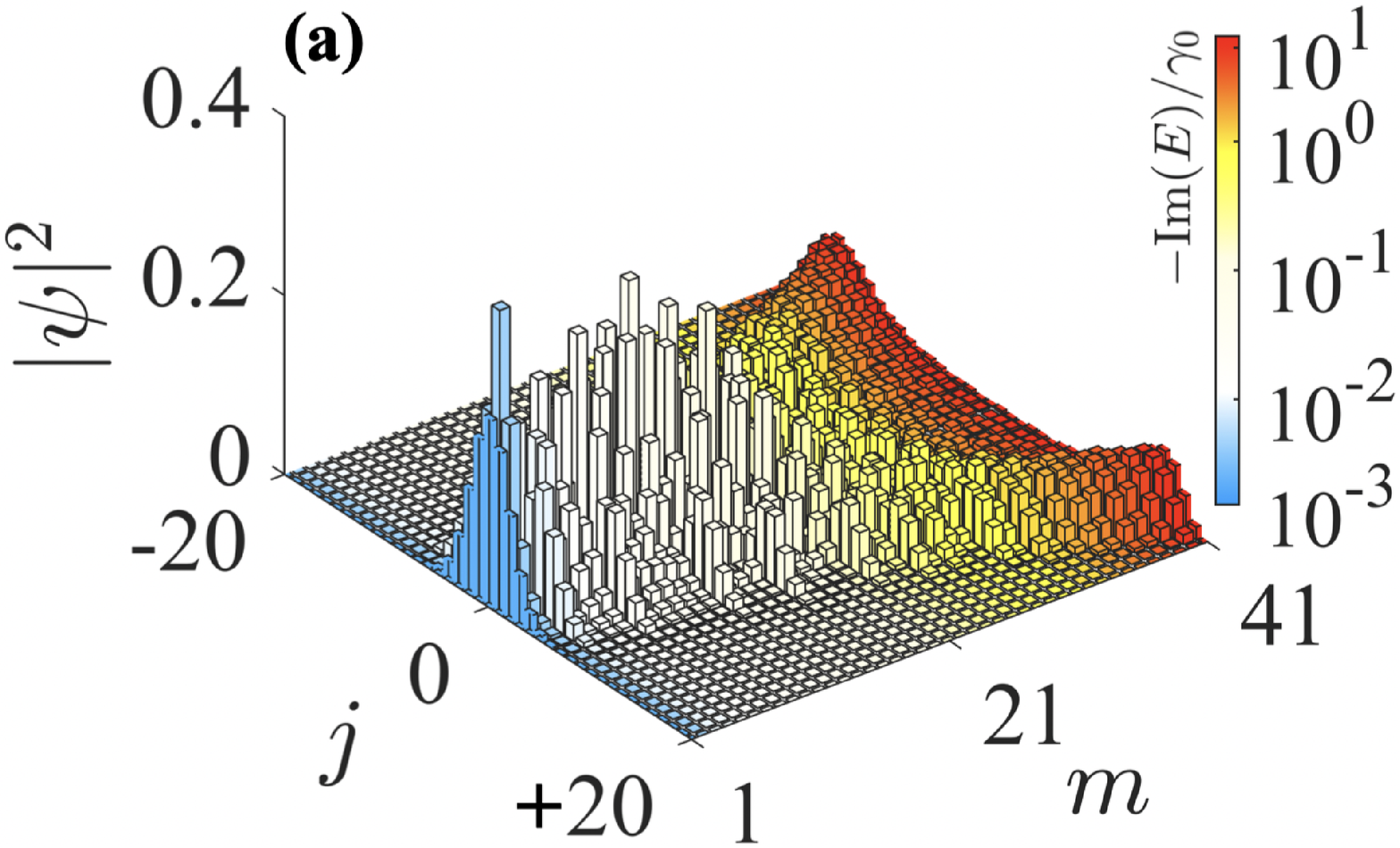}}
\subfigure[]{
\includegraphics[width=0.4\textwidth]{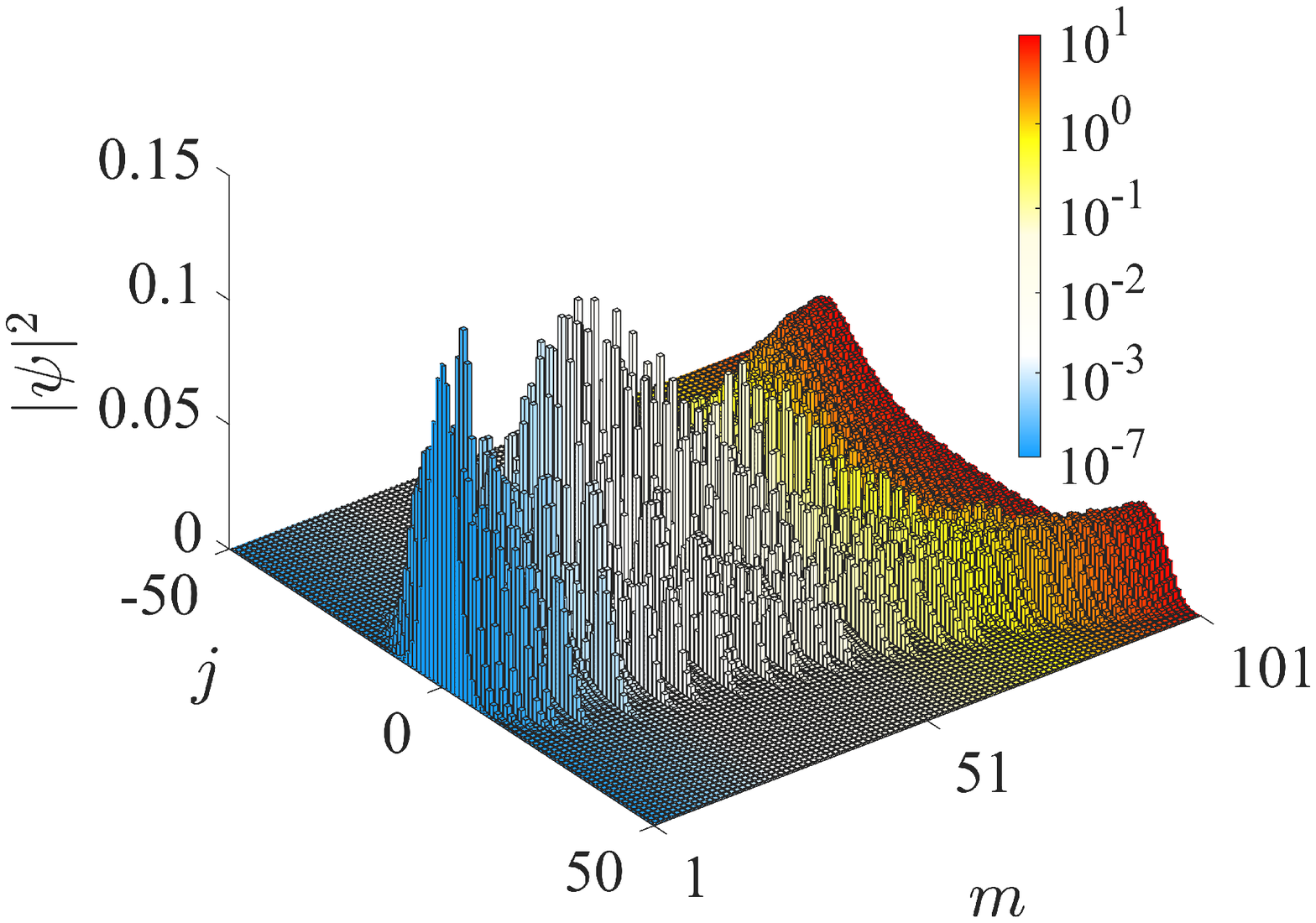}}
\subfigure[]{
\includegraphics[width=0.4\textwidth]{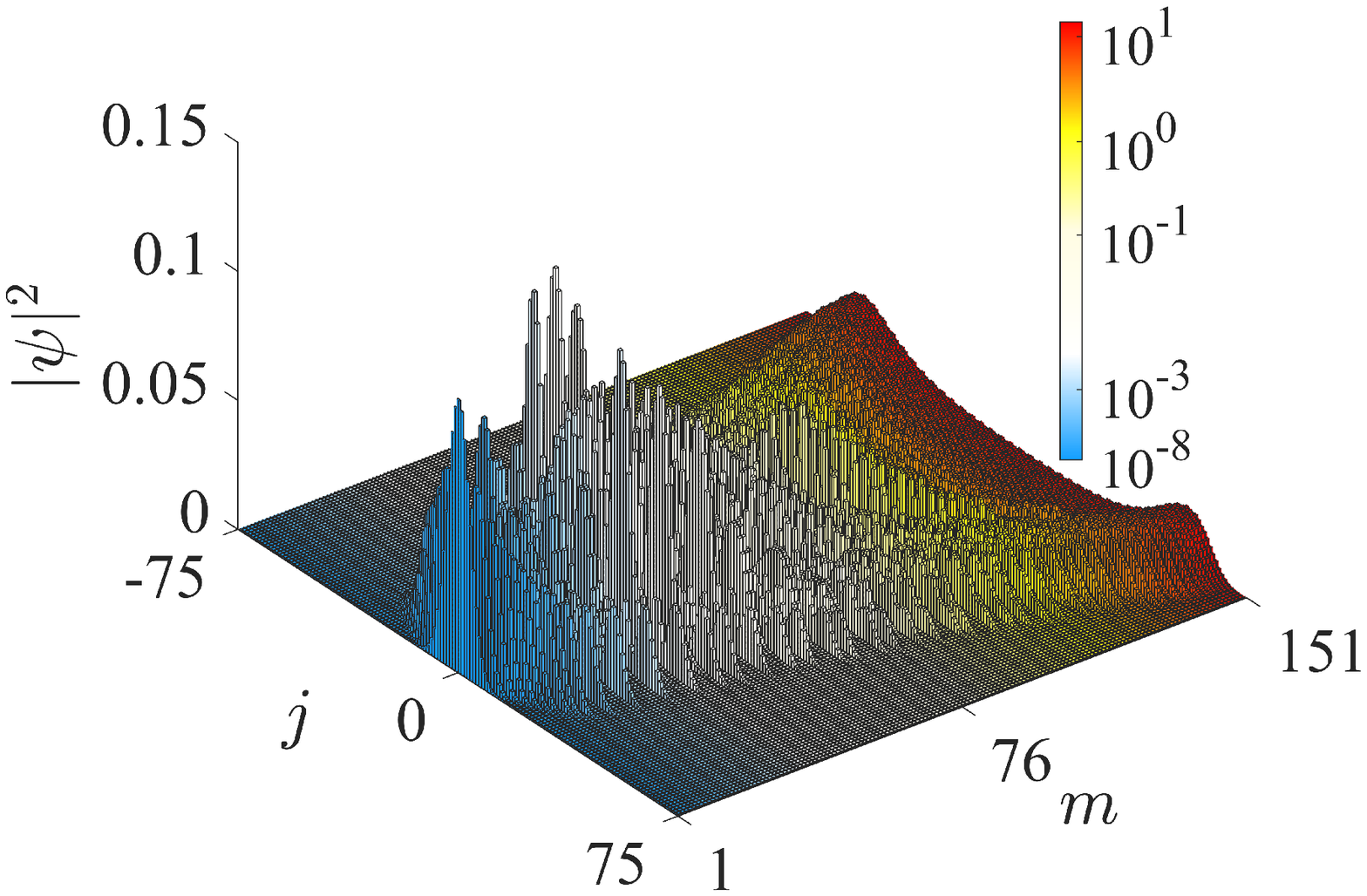}}
\subfigure[]{
\includegraphics[width=0.4\textwidth]{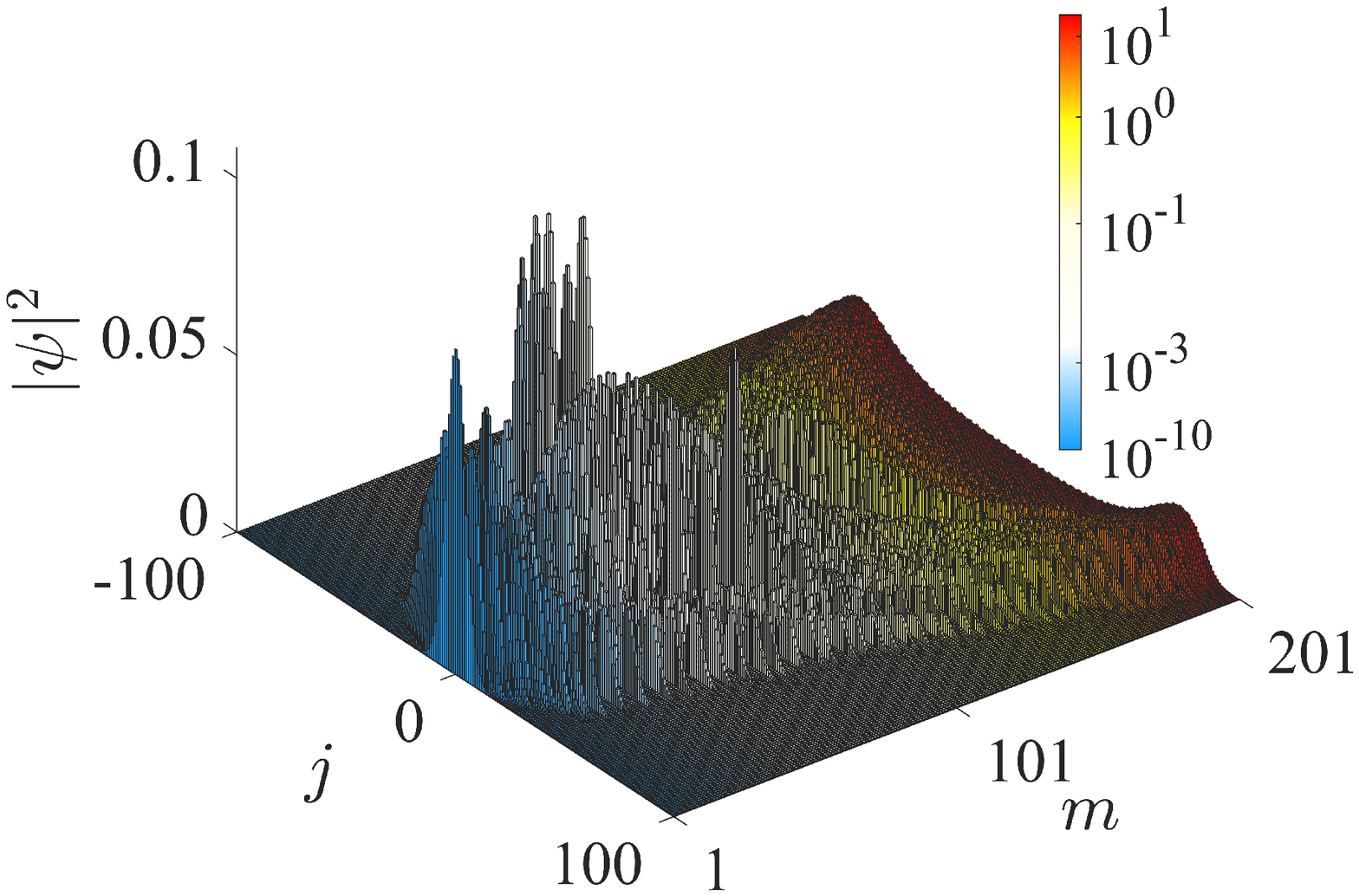}}\\
\small{Fig. E11: Intensity distributions versus atom site $j$ for all eigen-states at (a) $N=41$, (b) $N=101$, (c) $N=151$ and (d) $N=201$. Colors for each eigen-state indicate the collective decay rates.}
\end{figure}

Additionally, in our system, we have $(N-1)\theta d=H$ in geometry. When $N$ is too large, the tilted angle $\theta$ will be extremely small, which is potentially beyond the experimental limit. Therefore, we choose $N=41$ to show the main feature of the system in our paper.

In summary, we study the model over a broad range of parameters. We find that the found existence of concentrated states with subradiant features persists for the scale of the system $N\rightarrow\infty$, which provides further evidence of connecting to NHSE \cite{yokomizo2021non}. Moreover, the occurrence of the found interesting phenomena mainly depends on the varying rate of $\gamma_{R(L)j}$ on atomic positions, which is a result of the geometry of the atomic array. In other words, if $\gamma_{R(L)j}$ varies largely (such as the yellow dot lines in Fig. E2(e) and (f)), the distributions of eigenstates show the phenomena of concentrates subradiant states and extended superradiant states (see Fig. E6(b) and (d), and Fig. E7(d)). However, when the varying rate of $\gamma_{R(L)j}$ decreases (such as the yellow dot lines in Fig. E2(a)--(d)), a transition occurs and the system exhibits bulk states (see Fig. E6(a) and (c), and Fig. E7(a) and (b)). Besides, the faster the decrease is, the better the concentration of the system will perform.


\begin{thebibliography}{104}%
\makeatletter
\providecommand \@ifxundefined [1]{%
 \@ifx{#1\undefined}
}%
\providecommand \@ifnum [1]{%
 \ifnum #1\expandafter \@firstoftwo
 \else \expandafter \@secondoftwo
 \fi
}%
\providecommand \@ifx [1]{%
 \ifx #1\expandafter \@firstoftwo
 \else \expandafter \@secondoftwo
 \fi
}%
\providecommand \natexlab [1]{#1}%
\providecommand \enquote  [1]{``#1''}%
\providecommand \bibnamefont  [1]{#1}%
\providecommand \bibfnamefont [1]{#1}%
\providecommand \citenamefont [1]{#1}%
\providecommand \href@noop [0]{\@secondoftwo}%
\providecommand \href [0]{\begingroup \@sanitize@url \@href}%
\providecommand \@href[1]{\@@startlink{#1}\@@href}%
\providecommand \@@href[1]{\endgroup#1\@@endlink}%
\providecommand \@sanitize@url [0]{\catcode `\\12\catcode `\$12\catcode
  `\&12\catcode `\#12\catcode `\^12\catcode `\_12\catcode `\%12\relax}%
\providecommand \@@startlink[1]{}%
\providecommand \@@endlink[0]{}%
\providecommand \url  [0]{\begingroup\@sanitize@url \@url }%
\providecommand \@url [1]{\endgroup\@href {#1}{\urlprefix }}%
\providecommand \urlprefix  [0]{URL }%
\providecommand \Eprint [0]{\href }%
\providecommand \doibase [0]{https://doi.org/}%
\providecommand \selectlanguage [0]{\@gobble}%
\providecommand \bibinfo  [0]{\@secondoftwo}%
\providecommand \bibfield  [0]{\@secondoftwo}%
\providecommand \translation [1]{[#1]}%
\providecommand \BibitemOpen [0]{}%
\providecommand \bibitemStop [0]{}%
\providecommand \bibitemNoStop [0]{.\EOS\space}%
\providecommand \EOS [0]{\spacefactor3000\relax}%
\providecommand \BibitemShut  [1]{\csname bibitem#1\endcsname}%
\let\auto@bib@innerbib\@empty
\bibitem [{\citenamefont {Gorini}\ \emph {et~al.}(1976)\citenamefont {Gorini},
  \citenamefont {Kossakowski},\ and\ \citenamefont
  {Sudarshan}}]{gorini1976completely}%
  \BibitemOpen
  \bibfield  {author} {\bibinfo {author} {\bibfnamefont {V.}~\bibnamefont
  {Gorini}}, \bibinfo {author} {\bibfnamefont {A.}~\bibnamefont
  {Kossakowski}},\ and\ \bibinfo {author} {\bibfnamefont {E.~C.~G.}\
  \bibnamefont {Sudarshan}},\ }\href@noop {} {\bibfield  {journal} {\bibinfo
  {journal} {Journal of Mathematical Physics}\ }\textbf {\bibinfo {volume}
  {17}},\ \bibinfo {pages} {821} (\bibinfo {year} {1976})}\BibitemShut
  {NoStop}%
\bibitem [{\citenamefont {Lindblad}(1976)}]{lindblad1976generators}%
  \BibitemOpen
  \bibfield  {author} {\bibinfo {author} {\bibfnamefont {G.}~\bibnamefont
  {Lindblad}},\ }\href@noop {} {\bibfield  {journal} {\bibinfo  {journal}
  {Communications in Mathematical Physics}\ }\textbf {\bibinfo {volume} {48}},\
  \bibinfo {pages} {119} (\bibinfo {year} {1976})}\BibitemShut {NoStop}%
\bibitem [{\citenamefont {Carusotto}\ and\ \citenamefont
  {Ciuti}(2013)}]{carusotto2013quantum}%
  \BibitemOpen
  \bibfield  {author} {\bibinfo {author} {\bibfnamefont {I.}~\bibnamefont
  {Carusotto}}\ and\ \bibinfo {author} {\bibfnamefont {C.}~\bibnamefont
  {Ciuti}},\ }\href@noop {} {\bibfield  {journal} {\bibinfo  {journal} {Reviews
  of Modern Physics}\ }\textbf {\bibinfo {volume} {85}},\ \bibinfo {pages}
  {299} (\bibinfo {year} {2013})}\BibitemShut {NoStop}%
\bibitem [{\citenamefont {Marchetti}\ \emph {et~al.}(2013)\citenamefont
  {Marchetti}, \citenamefont {Joanny}, \citenamefont {Ramaswamy}, \citenamefont
  {Liverpool}, \citenamefont {Prost}, \citenamefont {Rao},\ and\ \citenamefont
  {Simha}}]{marchetti2013hydrodynamics}%
  \BibitemOpen
  \bibfield  {author} {\bibinfo {author} {\bibfnamefont {M.~C.}\ \bibnamefont
  {Marchetti}}, \bibinfo {author} {\bibfnamefont {J.-F.}\ \bibnamefont
  {Joanny}}, \bibinfo {author} {\bibfnamefont {S.}~\bibnamefont {Ramaswamy}},
  \bibinfo {author} {\bibfnamefont {T.~B.}\ \bibnamefont {Liverpool}}, \bibinfo
  {author} {\bibfnamefont {J.}~\bibnamefont {Prost}}, \bibinfo {author}
  {\bibfnamefont {M.}~\bibnamefont {Rao}},\ and\ \bibinfo {author}
  {\bibfnamefont {R.~A.}\ \bibnamefont {Simha}},\ }\href@noop {} {\bibfield
  {journal} {\bibinfo  {journal} {Reviews of Modern Physics}\ }\textbf
  {\bibinfo {volume} {85}},\ \bibinfo {pages} {1143} (\bibinfo {year}
  {2013})}\BibitemShut {NoStop}%
\bibitem [{\citenamefont {Feng}\ \emph {et~al.}(2011)\citenamefont {Feng},
  \citenamefont {Ayache}, \citenamefont {Huang}, \citenamefont {Xu},
  \citenamefont {Lu}, \citenamefont {Chen}, \citenamefont {Fainman},\ and\
  \citenamefont {Scherer}}]{feng2011nonreciprocal}%
  \BibitemOpen
  \bibfield  {author} {\bibinfo {author} {\bibfnamefont {L.}~\bibnamefont
  {Feng}}, \bibinfo {author} {\bibfnamefont {M.}~\bibnamefont {Ayache}},
  \bibinfo {author} {\bibfnamefont {J.}~\bibnamefont {Huang}}, \bibinfo
  {author} {\bibfnamefont {Y.-L.}\ \bibnamefont {Xu}}, \bibinfo {author}
  {\bibfnamefont {M.-H.}\ \bibnamefont {Lu}}, \bibinfo {author} {\bibfnamefont
  {Y.-F.}\ \bibnamefont {Chen}}, \bibinfo {author} {\bibfnamefont
  {Y.}~\bibnamefont {Fainman}},\ and\ \bibinfo {author} {\bibfnamefont
  {A.}~\bibnamefont {Scherer}},\ }\href@noop {} {\bibfield  {journal} {\bibinfo
   {journal} {Science}\ }\textbf {\bibinfo {volume} {333}},\ \bibinfo {pages}
  {729} (\bibinfo {year} {2011})}\BibitemShut {NoStop}%
\bibitem [{\citenamefont {Popa}\ and\ \citenamefont
  {Cummer}(2014)}]{popa2014non}%
  \BibitemOpen
  \bibfield  {author} {\bibinfo {author} {\bibfnamefont {B.-I.}\ \bibnamefont
  {Popa}}\ and\ \bibinfo {author} {\bibfnamefont {S.~A.}\ \bibnamefont
  {Cummer}},\ }\href@noop {} {\bibfield  {journal} {\bibinfo  {journal} {Nature
  Communications}\ }\textbf {\bibinfo {volume} {5}},\ \bibinfo {pages} {1}
  (\bibinfo {year} {2014})}\BibitemShut {NoStop}%
\bibitem [{\citenamefont {Longhi}\ \emph {et~al.}(2015)\citenamefont {Longhi},
  \citenamefont {Gatti},\ and\ \citenamefont {Della~Valle}}]{longhi2015non}%
  \BibitemOpen
  \bibfield  {author} {\bibinfo {author} {\bibfnamefont {S.}~\bibnamefont
  {Longhi}}, \bibinfo {author} {\bibfnamefont {D.}~\bibnamefont {Gatti}},\ and\
  \bibinfo {author} {\bibfnamefont {G.}~\bibnamefont {Della~Valle}},\
  }\href@noop {} {\bibfield  {journal} {\bibinfo  {journal} {Physical Review
  B}\ }\textbf {\bibinfo {volume} {92}},\ \bibinfo {pages} {094204} (\bibinfo
  {year} {2015})}\BibitemShut {NoStop}%
\bibitem [{\citenamefont {Brandenbourger}\ \emph {et~al.}(2019)\citenamefont
  {Brandenbourger}, \citenamefont {Locsin}, \citenamefont {Lerner},\ and\
  \citenamefont {Coulais}}]{brandenbourger2019non}%
  \BibitemOpen
  \bibfield  {author} {\bibinfo {author} {\bibfnamefont {M.}~\bibnamefont
  {Brandenbourger}}, \bibinfo {author} {\bibfnamefont {X.}~\bibnamefont
  {Locsin}}, \bibinfo {author} {\bibfnamefont {E.}~\bibnamefont {Lerner}},\
  and\ \bibinfo {author} {\bibfnamefont {C.}~\bibnamefont {Coulais}},\
  }\href@noop {} {\bibfield  {journal} {\bibinfo  {journal} {Nature
  Communications}\ }\textbf {\bibinfo {volume} {10}},\ \bibinfo {pages} {1}
  (\bibinfo {year} {2019})}\BibitemShut {NoStop}%
\bibitem [{\citenamefont {Ghatak}\ \emph {et~al.}(2020)\citenamefont {Ghatak},
  \citenamefont {Brandenbourger}, \citenamefont {Van~Wezel},\ and\
  \citenamefont {Coulais}}]{ghatak2020observation}%
  \BibitemOpen
  \bibfield  {author} {\bibinfo {author} {\bibfnamefont {A.}~\bibnamefont
  {Ghatak}}, \bibinfo {author} {\bibfnamefont {M.}~\bibnamefont
  {Brandenbourger}}, \bibinfo {author} {\bibfnamefont {J.}~\bibnamefont
  {Van~Wezel}},\ and\ \bibinfo {author} {\bibfnamefont {C.}~\bibnamefont
  {Coulais}},\ }\href@noop {} {\bibfield  {journal} {\bibinfo  {journal}
  {Proceedings of the National Academy of Sciences}\ }\textbf {\bibinfo
  {volume} {117}},\ \bibinfo {pages} {29561} (\bibinfo {year}
  {2020})}\BibitemShut {NoStop}%
\bibitem [{\citenamefont {Hofmann}\ \emph {et~al.}(2020)\citenamefont
  {Hofmann}, \citenamefont {Helbig}, \citenamefont {Schindler}, \citenamefont
  {Salgo}, \citenamefont {Brzezi{\'n}ska}, \citenamefont {Greiter},
  \citenamefont {Kiessling}, \citenamefont {Wolf}, \citenamefont {Vollhardt},
  \citenamefont {Kaba{\v{s}}i} \emph {et~al.}}]{hofmann2020reciprocal}%
  \BibitemOpen
  \bibfield  {author} {\bibinfo {author} {\bibfnamefont {T.}~\bibnamefont
  {Hofmann}}, \bibinfo {author} {\bibfnamefont {T.}~\bibnamefont {Helbig}},
  \bibinfo {author} {\bibfnamefont {F.}~\bibnamefont {Schindler}}, \bibinfo
  {author} {\bibfnamefont {N.}~\bibnamefont {Salgo}}, \bibinfo {author}
  {\bibfnamefont {M.}~\bibnamefont {Brzezi{\'n}ska}}, \bibinfo {author}
  {\bibfnamefont {M.}~\bibnamefont {Greiter}}, \bibinfo {author} {\bibfnamefont
  {T.}~\bibnamefont {Kiessling}}, \bibinfo {author} {\bibfnamefont
  {D.}~\bibnamefont {Wolf}}, \bibinfo {author} {\bibfnamefont {A.}~\bibnamefont
  {Vollhardt}}, \bibinfo {author} {\bibfnamefont {A.}~\bibnamefont
  {Kaba{\v{s}}i}}, \emph {et~al.},\ }\href@noop {} {\bibfield  {journal}
  {\bibinfo  {journal} {Physical Review Research}\ }\textbf {\bibinfo {volume}
  {2}},\ \bibinfo {pages} {023265} (\bibinfo {year} {2020})}\BibitemShut
  {NoStop}%
\bibitem [{\citenamefont {Li}\ \emph {et~al.}(2020{\natexlab{a}})\citenamefont
  {Li}, \citenamefont {Lee}, \citenamefont {Mu},\ and\ \citenamefont
  {Gong}}]{li2020critical}%
  \BibitemOpen
  \bibfield  {author} {\bibinfo {author} {\bibfnamefont {L.}~\bibnamefont
  {Li}}, \bibinfo {author} {\bibfnamefont {C.~H.}\ \bibnamefont {Lee}},
  \bibinfo {author} {\bibfnamefont {S.}~\bibnamefont {Mu}},\ and\ \bibinfo
  {author} {\bibfnamefont {J.}~\bibnamefont {Gong}},\ }\href@noop {} {\bibfield
   {journal} {\bibinfo  {journal} {Nature Communications}\ }\textbf {\bibinfo
  {volume} {11}},\ \bibinfo {pages} {1} (\bibinfo {year}
  {2020}{\natexlab{a}})}\BibitemShut {NoStop}%
\bibitem [{\citenamefont {Zou}\ \emph {et~al.}(2021)\citenamefont {Zou},
  \citenamefont {Chen}, \citenamefont {He}, \citenamefont {Bao}, \citenamefont
  {Lee}, \citenamefont {Sun},\ and\ \citenamefont
  {Zhang}}]{zou2021observation}%
  \BibitemOpen
  \bibfield  {author} {\bibinfo {author} {\bibfnamefont {D.}~\bibnamefont
  {Zou}}, \bibinfo {author} {\bibfnamefont {T.}~\bibnamefont {Chen}}, \bibinfo
  {author} {\bibfnamefont {W.}~\bibnamefont {He}}, \bibinfo {author}
  {\bibfnamefont {J.}~\bibnamefont {Bao}}, \bibinfo {author} {\bibfnamefont
  {C.~H.}\ \bibnamefont {Lee}}, \bibinfo {author} {\bibfnamefont
  {H.}~\bibnamefont {Sun}},\ and\ \bibinfo {author} {\bibfnamefont
  {X.}~\bibnamefont {Zhang}},\ }\href@noop {} {\bibfield  {journal} {\bibinfo
  {journal} {Nature Communications}\ }\textbf {\bibinfo {volume} {12}},\
  \bibinfo {pages} {1} (\bibinfo {year} {2021})}\BibitemShut {NoStop}%
\bibitem [{\citenamefont {Ma}\ and\ \citenamefont
  {Sheng}(2016)}]{ma2016acoustic}%
  \BibitemOpen
  \bibfield  {author} {\bibinfo {author} {\bibfnamefont {G.}~\bibnamefont
  {Ma}}\ and\ \bibinfo {author} {\bibfnamefont {P.}~\bibnamefont {Sheng}},\
  }\href@noop {} {\bibfield  {journal} {\bibinfo  {journal} {Science Advances}\
  }\textbf {\bibinfo {volume} {2}},\ \bibinfo {pages} {e1501595} (\bibinfo
  {year} {2016})}\BibitemShut {NoStop}%
\bibitem [{\citenamefont {Cummer}\ \emph {et~al.}(2016)\citenamefont {Cummer},
  \citenamefont {Christensen},\ and\ \citenamefont
  {Al{\`u}}}]{cummer2016controlling}%
  \BibitemOpen
  \bibfield  {author} {\bibinfo {author} {\bibfnamefont {S.~A.}\ \bibnamefont
  {Cummer}}, \bibinfo {author} {\bibfnamefont {J.}~\bibnamefont
  {Christensen}},\ and\ \bibinfo {author} {\bibfnamefont {A.}~\bibnamefont
  {Al{\`u}}},\ }\href@noop {} {\bibfield  {journal} {\bibinfo  {journal}
  {Nature Reviews Materials}\ }\textbf {\bibinfo {volume} {1}},\ \bibinfo
  {pages} {1} (\bibinfo {year} {2016})}\BibitemShut {NoStop}%
\bibitem [{\citenamefont {Zhang}\ \emph
  {et~al.}(2021{\natexlab{a}})\citenamefont {Zhang}, \citenamefont {Yang},
  \citenamefont {Ge}, \citenamefont {Guan}, \citenamefont {Chen}, \citenamefont
  {Yan}, \citenamefont {Chen}, \citenamefont {Xi}, \citenamefont {Li},
  \citenamefont {Jia} \emph {et~al.}}]{zhang2021acoustic}%
  \BibitemOpen
  \bibfield  {author} {\bibinfo {author} {\bibfnamefont {L.}~\bibnamefont
  {Zhang}}, \bibinfo {author} {\bibfnamefont {Y.}~\bibnamefont {Yang}},
  \bibinfo {author} {\bibfnamefont {Y.}~\bibnamefont {Ge}}, \bibinfo {author}
  {\bibfnamefont {Y.-J.}\ \bibnamefont {Guan}}, \bibinfo {author}
  {\bibfnamefont {Q.}~\bibnamefont {Chen}}, \bibinfo {author} {\bibfnamefont
  {Q.}~\bibnamefont {Yan}}, \bibinfo {author} {\bibfnamefont {F.}~\bibnamefont
  {Chen}}, \bibinfo {author} {\bibfnamefont {R.}~\bibnamefont {Xi}}, \bibinfo
  {author} {\bibfnamefont {Y.}~\bibnamefont {Li}}, \bibinfo {author}
  {\bibfnamefont {D.}~\bibnamefont {Jia}}, \emph {et~al.},\ }\href@noop {}
  {\bibfield  {journal} {\bibinfo  {journal} {Nature Communications}\ }\textbf
  {\bibinfo {volume} {12}},\ \bibinfo {pages} {1} (\bibinfo {year}
  {2021}{\natexlab{a}})}\BibitemShut {NoStop}%
\bibitem [{\citenamefont {Nelson}\ and\ \citenamefont
  {Shnerb}(1998)}]{nelson1998non}%
  \BibitemOpen
  \bibfield  {author} {\bibinfo {author} {\bibfnamefont {D.~R.}\ \bibnamefont
  {Nelson}}\ and\ \bibinfo {author} {\bibfnamefont {N.~M.}\ \bibnamefont
  {Shnerb}},\ }\href@noop {} {\bibfield  {journal} {\bibinfo  {journal}
  {Physical Review E}\ }\textbf {\bibinfo {volume} {58}},\ \bibinfo {pages}
  {1383} (\bibinfo {year} {1998})}\BibitemShut {NoStop}%
\bibitem [{\citenamefont {Amir}\ \emph {et~al.}(2016)\citenamefont {Amir},
  \citenamefont {Hatano},\ and\ \citenamefont {Nelson}}]{amir2016non}%
  \BibitemOpen
  \bibfield  {author} {\bibinfo {author} {\bibfnamefont {A.}~\bibnamefont
  {Amir}}, \bibinfo {author} {\bibfnamefont {N.}~\bibnamefont {Hatano}},\ and\
  \bibinfo {author} {\bibfnamefont {D.~R.}\ \bibnamefont {Nelson}},\
  }\href@noop {} {\bibfield  {journal} {\bibinfo  {journal} {Physical Review
  E}\ }\textbf {\bibinfo {volume} {93}},\ \bibinfo {pages} {042310} (\bibinfo
  {year} {2016})}\BibitemShut {NoStop}%
\bibitem [{\citenamefont {Cao}\ and\ \citenamefont
  {Wiersig}(2015)}]{cao2015dielectric}%
  \BibitemOpen
  \bibfield  {author} {\bibinfo {author} {\bibfnamefont {H.}~\bibnamefont
  {Cao}}\ and\ \bibinfo {author} {\bibfnamefont {J.}~\bibnamefont {Wiersig}},\
  }\href@noop {} {\bibfield  {journal} {\bibinfo  {journal} {Reviews of Modern
  Physics}\ }\textbf {\bibinfo {volume} {87}},\ \bibinfo {pages} {61} (\bibinfo
  {year} {2015})}\BibitemShut {NoStop}%
\bibitem [{\citenamefont {Budich}\ \emph {et~al.}(2019)\citenamefont {Budich},
  \citenamefont {Carlstr{\"o}m}, \citenamefont {Kunst},\ and\ \citenamefont
  {Bergholtz}}]{budich2019symmetry}%
  \BibitemOpen
  \bibfield  {author} {\bibinfo {author} {\bibfnamefont {J.~C.}\ \bibnamefont
  {Budich}}, \bibinfo {author} {\bibfnamefont {J.}~\bibnamefont
  {Carlstr{\"o}m}}, \bibinfo {author} {\bibfnamefont {F.~K.}\ \bibnamefont
  {Kunst}},\ and\ \bibinfo {author} {\bibfnamefont {E.~J.}\ \bibnamefont
  {Bergholtz}},\ }\href@noop {} {\bibfield  {journal} {\bibinfo  {journal}
  {Physical Review B}\ }\textbf {\bibinfo {volume} {99}},\ \bibinfo {pages}
  {041406} (\bibinfo {year} {2019})}\BibitemShut {NoStop}%
\bibitem [{\citenamefont {Lee}\ and\ \citenamefont
  {Thomale}(2019)}]{lee2019anatomy}%
  \BibitemOpen
  \bibfield  {author} {\bibinfo {author} {\bibfnamefont {C.~H.}\ \bibnamefont
  {Lee}}\ and\ \bibinfo {author} {\bibfnamefont {R.}~\bibnamefont {Thomale}},\
  }\href@noop {} {\bibfield  {journal} {\bibinfo  {journal} {Physical Review
  B}\ }\textbf {\bibinfo {volume} {99}},\ \bibinfo {pages} {201103} (\bibinfo
  {year} {2019})}\BibitemShut {NoStop}%
\bibitem [{\citenamefont {Edvardsson}\ \emph {et~al.}(2019)\citenamefont
  {Edvardsson}, \citenamefont {Kunst},\ and\ \citenamefont
  {Bergholtz}}]{edvardsson2019non}%
  \BibitemOpen
  \bibfield  {author} {\bibinfo {author} {\bibfnamefont {E.}~\bibnamefont
  {Edvardsson}}, \bibinfo {author} {\bibfnamefont {F.~K.}\ \bibnamefont
  {Kunst}},\ and\ \bibinfo {author} {\bibfnamefont {E.~J.}\ \bibnamefont
  {Bergholtz}},\ }\href@noop {} {\bibfield  {journal} {\bibinfo  {journal}
  {Physical Review B}\ }\textbf {\bibinfo {volume} {99}},\ \bibinfo {pages}
  {081302} (\bibinfo {year} {2019})}\BibitemShut {NoStop}%
\bibitem [{\citenamefont {Ashida}\ \emph {et~al.}(2020)\citenamefont {Ashida},
  \citenamefont {Gong},\ and\ \citenamefont {Ueda}}]{ashida2020non}%
  \BibitemOpen
  \bibfield  {author} {\bibinfo {author} {\bibfnamefont {Y.}~\bibnamefont
  {Ashida}}, \bibinfo {author} {\bibfnamefont {Z.}~\bibnamefont {Gong}},\ and\
  \bibinfo {author} {\bibfnamefont {M.}~\bibnamefont {Ueda}},\ }\href@noop {}
  {\bibfield  {journal} {\bibinfo  {journal} {Advances in Physics}\ }\textbf
  {\bibinfo {volume} {69}},\ \bibinfo {pages} {249} (\bibinfo {year}
  {2020})}\BibitemShut {NoStop}%
\bibitem [{\citenamefont {Borgnia}\ \emph {et~al.}(2020)\citenamefont
  {Borgnia}, \citenamefont {Kruchkov},\ and\ \citenamefont
  {Slager}}]{borgnia2020non}%
  \BibitemOpen
  \bibfield  {author} {\bibinfo {author} {\bibfnamefont {D.~S.}\ \bibnamefont
  {Borgnia}}, \bibinfo {author} {\bibfnamefont {A.~J.}\ \bibnamefont
  {Kruchkov}},\ and\ \bibinfo {author} {\bibfnamefont {R.-J.}\ \bibnamefont
  {Slager}},\ }\href@noop {} {\bibfield  {journal} {\bibinfo  {journal}
  {Physical Review Letters}\ }\textbf {\bibinfo {volume} {124}},\ \bibinfo
  {pages} {056802} (\bibinfo {year} {2020})}\BibitemShut {NoStop}%
\bibitem [{\citenamefont {Bergholtz}\ \emph {et~al.}(2021)\citenamefont
  {Bergholtz}, \citenamefont {Budich},\ and\ \citenamefont
  {Kunst}}]{bergholtz2021exceptional}%
  \BibitemOpen
  \bibfield  {author} {\bibinfo {author} {\bibfnamefont {E.~J.}\ \bibnamefont
  {Bergholtz}}, \bibinfo {author} {\bibfnamefont {J.~C.}\ \bibnamefont
  {Budich}},\ and\ \bibinfo {author} {\bibfnamefont {F.~K.}\ \bibnamefont
  {Kunst}},\ }\href@noop {} {\bibfield  {journal} {\bibinfo  {journal} {Reviews
  of Modern Physics}\ }\textbf {\bibinfo {volume} {93}},\ \bibinfo {pages}
  {015005} (\bibinfo {year} {2021})}\BibitemShut {NoStop}%
\bibitem [{\citenamefont {Lee}(2016)}]{lee2016anomalous}%
  \BibitemOpen
  \bibfield  {author} {\bibinfo {author} {\bibfnamefont {T.~E.}\ \bibnamefont
  {Lee}},\ }\href@noop {} {\bibfield  {journal} {\bibinfo  {journal} {Physical
  Review Letters}\ }\textbf {\bibinfo {volume} {116}},\ \bibinfo {pages}
  {133903} (\bibinfo {year} {2016})}\BibitemShut {NoStop}%
\bibitem [{\citenamefont {Alvarez}\ \emph {et~al.}(2018)\citenamefont
  {Alvarez}, \citenamefont {Vargas},\ and\ \citenamefont
  {Torres}}]{alvarez2018non}%
  \BibitemOpen
  \bibfield  {author} {\bibinfo {author} {\bibfnamefont {V.~M.}\ \bibnamefont
  {Alvarez}}, \bibinfo {author} {\bibfnamefont {J.~B.}\ \bibnamefont
  {Vargas}},\ and\ \bibinfo {author} {\bibfnamefont {L.~F.}\ \bibnamefont
  {Torres}},\ }\href@noop {} {\bibfield  {journal} {\bibinfo  {journal}
  {Physical Review B}\ }\textbf {\bibinfo {volume} {97}},\ \bibinfo {pages}
  {121401} (\bibinfo {year} {2018})}\BibitemShut {NoStop}%
\bibitem [{\citenamefont {Lee}\ \emph {et~al.}(2019)\citenamefont {Lee},
  \citenamefont {Li},\ and\ \citenamefont {Gong}}]{lee2019hybrid}%
  \BibitemOpen
  \bibfield  {author} {\bibinfo {author} {\bibfnamefont {C.~H.}\ \bibnamefont
  {Lee}}, \bibinfo {author} {\bibfnamefont {L.}~\bibnamefont {Li}},\ and\
  \bibinfo {author} {\bibfnamefont {J.}~\bibnamefont {Gong}},\ }\href@noop {}
  {\bibfield  {journal} {\bibinfo  {journal} {Physical Review Letters}\
  }\textbf {\bibinfo {volume} {123}},\ \bibinfo {pages} {016805} (\bibinfo
  {year} {2019})}\BibitemShut {NoStop}%
\bibitem [{\citenamefont {Song}\ \emph {et~al.}(2019)\citenamefont {Song},
  \citenamefont {Yao},\ and\ \citenamefont {Wang}}]{song2019non}%
  \BibitemOpen
  \bibfield  {author} {\bibinfo {author} {\bibfnamefont {F.}~\bibnamefont
  {Song}}, \bibinfo {author} {\bibfnamefont {S.}~\bibnamefont {Yao}},\ and\
  \bibinfo {author} {\bibfnamefont {Z.}~\bibnamefont {Wang}},\ }\href@noop {}
  {\bibfield  {journal} {\bibinfo  {journal} {Physical Review Letters}\
  }\textbf {\bibinfo {volume} {123}},\ \bibinfo {pages} {170401} (\bibinfo
  {year} {2019})}\BibitemShut {NoStop}%
\bibitem [{\citenamefont {Okuma}\ \emph {et~al.}(2020)\citenamefont {Okuma},
  \citenamefont {Kawabata}, \citenamefont {Shiozaki},\ and\ \citenamefont
  {Sato}}]{okuma2020topological}%
  \BibitemOpen
  \bibfield  {author} {\bibinfo {author} {\bibfnamefont {N.}~\bibnamefont
  {Okuma}}, \bibinfo {author} {\bibfnamefont {K.}~\bibnamefont {Kawabata}},
  \bibinfo {author} {\bibfnamefont {K.}~\bibnamefont {Shiozaki}},\ and\
  \bibinfo {author} {\bibfnamefont {M.}~\bibnamefont {Sato}},\ }\href@noop {}
  {\bibfield  {journal} {\bibinfo  {journal} {Physical Review Letters}\
  }\textbf {\bibinfo {volume} {124}},\ \bibinfo {pages} {086801} (\bibinfo
  {year} {2020})}\BibitemShut {NoStop}%
\bibitem [{\citenamefont {Weidemann}\ \emph {et~al.}(2020)\citenamefont
  {Weidemann}, \citenamefont {Kremer}, \citenamefont {Helbig}, \citenamefont
  {Hofmann}, \citenamefont {Stegmaier}, \citenamefont {Greiter}, \citenamefont
  {Thomale},\ and\ \citenamefont {Szameit}}]{weidemann2020topological}%
  \BibitemOpen
  \bibfield  {author} {\bibinfo {author} {\bibfnamefont {S.}~\bibnamefont
  {Weidemann}}, \bibinfo {author} {\bibfnamefont {M.}~\bibnamefont {Kremer}},
  \bibinfo {author} {\bibfnamefont {T.}~\bibnamefont {Helbig}}, \bibinfo
  {author} {\bibfnamefont {T.}~\bibnamefont {Hofmann}}, \bibinfo {author}
  {\bibfnamefont {A.}~\bibnamefont {Stegmaier}}, \bibinfo {author}
  {\bibfnamefont {M.}~\bibnamefont {Greiter}}, \bibinfo {author} {\bibfnamefont
  {R.}~\bibnamefont {Thomale}},\ and\ \bibinfo {author} {\bibfnamefont
  {A.}~\bibnamefont {Szameit}},\ }\href@noop {} {\bibfield  {journal} {\bibinfo
   {journal} {Science}\ }\textbf {\bibinfo {volume} {368}},\ \bibinfo {pages}
  {311} (\bibinfo {year} {2020})}\BibitemShut {NoStop}%
\bibitem [{\citenamefont {Kunst}\ \emph {et~al.}(2018)\citenamefont {Kunst},
  \citenamefont {Edvardsson}, \citenamefont {Budich},\ and\ \citenamefont
  {Bergholtz}}]{kunst2018biorthogonal}%
  \BibitemOpen
  \bibfield  {author} {\bibinfo {author} {\bibfnamefont {F.~K.}\ \bibnamefont
  {Kunst}}, \bibinfo {author} {\bibfnamefont {E.}~\bibnamefont {Edvardsson}},
  \bibinfo {author} {\bibfnamefont {J.~C.}\ \bibnamefont {Budich}},\ and\
  \bibinfo {author} {\bibfnamefont {E.~J.}\ \bibnamefont {Bergholtz}},\
  }\href@noop {} {\bibfield  {journal} {\bibinfo  {journal} {Physical Review
  Letters}\ }\textbf {\bibinfo {volume} {121}},\ \bibinfo {pages} {026808}
  (\bibinfo {year} {2018})}\BibitemShut {NoStop}%
\bibitem [{\citenamefont {Yao}\ and\ \citenamefont {Wang}(2018)}]{yao2018edge}%
  \BibitemOpen
  \bibfield  {author} {\bibinfo {author} {\bibfnamefont {S.}~\bibnamefont
  {Yao}}\ and\ \bibinfo {author} {\bibfnamefont {Z.}~\bibnamefont {Wang}},\
  }\href@noop {} {\bibfield  {journal} {\bibinfo  {journal} {Physical Review
  Letters}\ }\textbf {\bibinfo {volume} {121}},\ \bibinfo {pages} {086803}
  (\bibinfo {year} {2018})}\BibitemShut {NoStop}%
\bibitem [{\citenamefont {Zhu}\ \emph {et~al.}(2020)\citenamefont {Zhu},
  \citenamefont {Wang}, \citenamefont {Gupta}, \citenamefont {Zhang},
  \citenamefont {Xie}, \citenamefont {Lu},\ and\ \citenamefont
  {Chen}}]{zhu2020photonic}%
  \BibitemOpen
  \bibfield  {author} {\bibinfo {author} {\bibfnamefont {X.}~\bibnamefont
  {Zhu}}, \bibinfo {author} {\bibfnamefont {H.}~\bibnamefont {Wang}}, \bibinfo
  {author} {\bibfnamefont {S.~K.}\ \bibnamefont {Gupta}}, \bibinfo {author}
  {\bibfnamefont {H.}~\bibnamefont {Zhang}}, \bibinfo {author} {\bibfnamefont
  {B.}~\bibnamefont {Xie}}, \bibinfo {author} {\bibfnamefont {M.}~\bibnamefont
  {Lu}},\ and\ \bibinfo {author} {\bibfnamefont {Y.}~\bibnamefont {Chen}},\
  }\href@noop {} {\bibfield  {journal} {\bibinfo  {journal} {Physical Review
  Research}\ }\textbf {\bibinfo {volume} {2}},\ \bibinfo {pages} {013280}
  (\bibinfo {year} {2020})}\BibitemShut {NoStop}%
\bibitem [{\citenamefont {Xiao}\ \emph
  {et~al.}(2020{\natexlab{a}})\citenamefont {Xiao}, \citenamefont {Deng},
  \citenamefont {Wang}, \citenamefont {Zhu}, \citenamefont {Wang},
  \citenamefont {Yi},\ and\ \citenamefont {Xue}}]{xiao2020non}%
  \BibitemOpen
  \bibfield  {author} {\bibinfo {author} {\bibfnamefont {L.}~\bibnamefont
  {Xiao}}, \bibinfo {author} {\bibfnamefont {T.}~\bibnamefont {Deng}}, \bibinfo
  {author} {\bibfnamefont {K.}~\bibnamefont {Wang}}, \bibinfo {author}
  {\bibfnamefont {G.}~\bibnamefont {Zhu}}, \bibinfo {author} {\bibfnamefont
  {Z.}~\bibnamefont {Wang}}, \bibinfo {author} {\bibfnamefont {W.}~\bibnamefont
  {Yi}},\ and\ \bibinfo {author} {\bibfnamefont {P.}~\bibnamefont {Xue}},\
  }\href@noop {} {\bibfield  {journal} {\bibinfo  {journal} {Nature Physics}\
  }\textbf {\bibinfo {volume} {16}},\ \bibinfo {pages} {761} (\bibinfo {year}
  {2020}{\natexlab{a}})}\BibitemShut {NoStop}%
\bibitem [{\citenamefont {Helbig}\ \emph {et~al.}(2020)\citenamefont {Helbig},
  \citenamefont {Hofmann}, \citenamefont {Imhof}, \citenamefont {Abdelghany},
  \citenamefont {Kiessling}, \citenamefont {Molenkamp}, \citenamefont {Lee},
  \citenamefont {Szameit}, \citenamefont {Greiter},\ and\ \citenamefont
  {Thomale}}]{helbig2020generalized}%
  \BibitemOpen
  \bibfield  {author} {\bibinfo {author} {\bibfnamefont {T.}~\bibnamefont
  {Helbig}}, \bibinfo {author} {\bibfnamefont {T.}~\bibnamefont {Hofmann}},
  \bibinfo {author} {\bibfnamefont {S.}~\bibnamefont {Imhof}}, \bibinfo
  {author} {\bibfnamefont {M.}~\bibnamefont {Abdelghany}}, \bibinfo {author}
  {\bibfnamefont {T.}~\bibnamefont {Kiessling}}, \bibinfo {author}
  {\bibfnamefont {L.}~\bibnamefont {Molenkamp}}, \bibinfo {author}
  {\bibfnamefont {C.}~\bibnamefont {Lee}}, \bibinfo {author} {\bibfnamefont
  {A.}~\bibnamefont {Szameit}}, \bibinfo {author} {\bibfnamefont
  {M.}~\bibnamefont {Greiter}},\ and\ \bibinfo {author} {\bibfnamefont
  {R.}~\bibnamefont {Thomale}},\ }\href@noop {} {\bibfield  {journal} {\bibinfo
   {journal} {Nature Physics}\ }\textbf {\bibinfo {volume} {16}},\ \bibinfo
  {pages} {747} (\bibinfo {year} {2020})}\BibitemShut {NoStop}%
\bibitem [{\citenamefont {Zhang}\ \emph
  {et~al.}(2021{\natexlab{b}})\citenamefont {Zhang}, \citenamefont {Tian},
  \citenamefont {Jiang}, \citenamefont {Lu},\ and\ \citenamefont
  {Chen}}]{zhang2021observation}%
  \BibitemOpen
  \bibfield  {author} {\bibinfo {author} {\bibfnamefont {X.}~\bibnamefont
  {Zhang}}, \bibinfo {author} {\bibfnamefont {Y.}~\bibnamefont {Tian}},
  \bibinfo {author} {\bibfnamefont {J.-H.}\ \bibnamefont {Jiang}}, \bibinfo
  {author} {\bibfnamefont {M.-H.}\ \bibnamefont {Lu}},\ and\ \bibinfo {author}
  {\bibfnamefont {Y.-F.}\ \bibnamefont {Chen}},\ }\href@noop {} {\bibfield
  {journal} {\bibinfo  {journal} {Nature Communications}\ }\textbf {\bibinfo
  {volume} {12}},\ \bibinfo {pages} {1} (\bibinfo {year}
  {2021}{\natexlab{b}})}\BibitemShut {NoStop}%
\bibitem [{\citenamefont {Torres}(2019)}]{torres2019perspective}%
  \BibitemOpen
  \bibfield  {author} {\bibinfo {author} {\bibfnamefont {L.~E.~F.}\
  \bibnamefont {Torres}},\ }\href@noop {} {\bibfield  {journal} {\bibinfo
  {journal} {Journal of Physics: Materials}\ }\textbf {\bibinfo {volume} {3}},\
  \bibinfo {pages} {014002} (\bibinfo {year} {2019})}\BibitemShut {NoStop}%
\bibitem [{\citenamefont {Haroche}\ and\ \citenamefont
  {Raimond}(2006)}]{haroche2006exploring}%
  \BibitemOpen
  \bibfield  {author} {\bibinfo {author} {\bibfnamefont {S.}~\bibnamefont
  {Haroche}}\ and\ \bibinfo {author} {\bibfnamefont {J.-M.}\ \bibnamefont
  {Raimond}},\ }\href@noop {} {\emph {\bibinfo {title} {Exploring the quantum:
  atoms, cavities, and photons}}}\ (\bibinfo  {publisher} {Oxford university
  press},\ \bibinfo {year} {2006})\BibitemShut {NoStop}%
\bibitem [{\citenamefont {Le~Kien}\ \emph
  {et~al.}(2005{\natexlab{a}})\citenamefont {Le~Kien}, \citenamefont {Gupta},
  \citenamefont {Balykin},\ and\ \citenamefont {Hakuta}}]{le2005spontaneous}%
  \BibitemOpen
  \bibfield  {author} {\bibinfo {author} {\bibfnamefont {F.}~\bibnamefont
  {Le~Kien}}, \bibinfo {author} {\bibfnamefont {S.~D.}\ \bibnamefont {Gupta}},
  \bibinfo {author} {\bibfnamefont {V.}~\bibnamefont {Balykin}},\ and\ \bibinfo
  {author} {\bibfnamefont {K.}~\bibnamefont {Hakuta}},\ }\href@noop {}
  {\bibfield  {journal} {\bibinfo  {journal} {Physical Review A}\ }\textbf
  {\bibinfo {volume} {72}},\ \bibinfo {pages} {032509} (\bibinfo {year}
  {2005}{\natexlab{a}})}\BibitemShut {NoStop}%
\bibitem [{\citenamefont {Shen}\ and\ \citenamefont
  {Fan}(2005{\natexlab{a}})}]{shen2005coherent}%
  \BibitemOpen
  \bibfield  {author} {\bibinfo {author} {\bibfnamefont {J.-T.}\ \bibnamefont
  {Shen}}\ and\ \bibinfo {author} {\bibfnamefont {S.}~\bibnamefont {Fan}},\
  }\href@noop {} {\bibfield  {journal} {\bibinfo  {journal} {Physical Review
  Letters}\ }\textbf {\bibinfo {volume} {95}},\ \bibinfo {pages} {213001}
  (\bibinfo {year} {2005}{\natexlab{a}})}\BibitemShut {NoStop}%
\bibitem [{\citenamefont {Shen}\ and\ \citenamefont
  {Fan}(2005{\natexlab{b}})}]{shen2005coherentspontaneous}%
  \BibitemOpen
  \bibfield  {author} {\bibinfo {author} {\bibfnamefont {J.-t.}\ \bibnamefont
  {Shen}}\ and\ \bibinfo {author} {\bibfnamefont {S.}~\bibnamefont {Fan}},\
  }\href@noop {} {\bibfield  {journal} {\bibinfo  {journal} {Optics Letters}\
  }\textbf {\bibinfo {volume} {30}},\ \bibinfo {pages} {2001} (\bibinfo {year}
  {2005}{\natexlab{b}})}\BibitemShut {NoStop}%
\bibitem [{\citenamefont {Zheng}\ \emph {et~al.}(2010)\citenamefont {Zheng},
  \citenamefont {Gauthier},\ and\ \citenamefont
  {Baranger}}]{zheng2010waveguide}%
  \BibitemOpen
  \bibfield  {author} {\bibinfo {author} {\bibfnamefont {H.}~\bibnamefont
  {Zheng}}, \bibinfo {author} {\bibfnamefont {D.~J.}\ \bibnamefont
  {Gauthier}},\ and\ \bibinfo {author} {\bibfnamefont {H.~U.}\ \bibnamefont
  {Baranger}},\ }\href@noop {} {\bibfield  {journal} {\bibinfo  {journal}
  {Physical Review A}\ }\textbf {\bibinfo {volume} {82}},\ \bibinfo {pages}
  {063816} (\bibinfo {year} {2010})}\BibitemShut {NoStop}%
\bibitem [{\citenamefont {Yuan}\ \emph {et~al.}(2015)\citenamefont {Yuan},
  \citenamefont {Xu},\ and\ \citenamefont {Fan}}]{yuan2015achieving}%
  \BibitemOpen
  \bibfield  {author} {\bibinfo {author} {\bibfnamefont {L.}~\bibnamefont
  {Yuan}}, \bibinfo {author} {\bibfnamefont {S.}~\bibnamefont {Xu}},\ and\
  \bibinfo {author} {\bibfnamefont {S.}~\bibnamefont {Fan}},\ }\href@noop {}
  {\bibfield  {journal} {\bibinfo  {journal} {Optics Letters}\ }\textbf
  {\bibinfo {volume} {40}},\ \bibinfo {pages} {5140} (\bibinfo {year}
  {2015})}\BibitemShut {NoStop}%
\bibitem [{\citenamefont {Kockum}\ \emph {et~al.}(2018)\citenamefont {Kockum},
  \citenamefont {Johansson},\ and\ \citenamefont
  {Nori}}]{kockum2018decoherence}%
  \BibitemOpen
  \bibfield  {author} {\bibinfo {author} {\bibfnamefont {A.~F.}\ \bibnamefont
  {Kockum}}, \bibinfo {author} {\bibfnamefont {G.}~\bibnamefont {Johansson}},\
  and\ \bibinfo {author} {\bibfnamefont {F.}~\bibnamefont {Nori}},\ }\href@noop
  {} {\bibfield  {journal} {\bibinfo  {journal} {Physical Review Letters}\
  }\textbf {\bibinfo {volume} {120}},\ \bibinfo {pages} {140404} (\bibinfo
  {year} {2018})}\BibitemShut {NoStop}%
\bibitem [{\citenamefont {Xiao}\ \emph
  {et~al.}(2020{\natexlab{b}})\citenamefont {Xiao}, \citenamefont {Wang},
  \citenamefont {Yuan},\ and\ \citenamefont {Chen}}]{xiao2020frequency}%
  \BibitemOpen
  \bibfield  {author} {\bibinfo {author} {\bibfnamefont {H.}~\bibnamefont
  {Xiao}}, \bibinfo {author} {\bibfnamefont {L.}~\bibnamefont {Wang}}, \bibinfo
  {author} {\bibfnamefont {L.}~\bibnamefont {Yuan}},\ and\ \bibinfo {author}
  {\bibfnamefont {X.}~\bibnamefont {Chen}},\ }\href@noop {} {\bibfield
  {journal} {\bibinfo  {journal} {ACS Photonics}\ }\textbf {\bibinfo {volume}
  {7}},\ \bibinfo {pages} {2010} (\bibinfo {year}
  {2020}{\natexlab{b}})}\BibitemShut {NoStop}%
\bibitem [{\citenamefont {Le~Kien}\ \emph
  {et~al.}(2005{\natexlab{b}})\citenamefont {Le~Kien}, \citenamefont {Gupta},
  \citenamefont {Nayak},\ and\ \citenamefont {Hakuta}}]{le2005nanofiber}%
  \BibitemOpen
  \bibfield  {author} {\bibinfo {author} {\bibfnamefont {F.}~\bibnamefont
  {Le~Kien}}, \bibinfo {author} {\bibfnamefont {S.~D.}\ \bibnamefont {Gupta}},
  \bibinfo {author} {\bibfnamefont {K.}~\bibnamefont {Nayak}},\ and\ \bibinfo
  {author} {\bibfnamefont {K.}~\bibnamefont {Hakuta}},\ }\href@noop {}
  {\bibfield  {journal} {\bibinfo  {journal} {Physical Review A}\ }\textbf
  {\bibinfo {volume} {72}},\ \bibinfo {pages} {063815} (\bibinfo {year}
  {2005}{\natexlab{b}})}\BibitemShut {NoStop}%
\bibitem [{\citenamefont {Lalumiere}\ \emph {et~al.}(2013)\citenamefont
  {Lalumiere}, \citenamefont {Sanders}, \citenamefont {van Loo}, \citenamefont
  {Fedorov}, \citenamefont {Wallraff},\ and\ \citenamefont
  {Blais}}]{lalumiere2013input}%
  \BibitemOpen
  \bibfield  {author} {\bibinfo {author} {\bibfnamefont {K.}~\bibnamefont
  {Lalumiere}}, \bibinfo {author} {\bibfnamefont {B.~C.}\ \bibnamefont
  {Sanders}}, \bibinfo {author} {\bibfnamefont {A.~F.}\ \bibnamefont {van
  Loo}}, \bibinfo {author} {\bibfnamefont {A.}~\bibnamefont {Fedorov}},
  \bibinfo {author} {\bibfnamefont {A.}~\bibnamefont {Wallraff}},\ and\
  \bibinfo {author} {\bibfnamefont {A.}~\bibnamefont {Blais}},\ }\href@noop {}
  {\bibfield  {journal} {\bibinfo  {journal} {Physical Review A}\ }\textbf
  {\bibinfo {volume} {88}},\ \bibinfo {pages} {043806} (\bibinfo {year}
  {2013})}\BibitemShut {NoStop}%
\bibitem [{\citenamefont {Shahmoon}\ and\ \citenamefont
  {Kurizki}(2013)}]{shahmoon2013nonradiative}%
  \BibitemOpen
  \bibfield  {author} {\bibinfo {author} {\bibfnamefont {E.}~\bibnamefont
  {Shahmoon}}\ and\ \bibinfo {author} {\bibfnamefont {G.}~\bibnamefont
  {Kurizki}},\ }\href@noop {} {\bibfield  {journal} {\bibinfo  {journal}
  {Physical Review A}\ }\textbf {\bibinfo {volume} {87}},\ \bibinfo {pages}
  {033831} (\bibinfo {year} {2013})}\BibitemShut {NoStop}%
\bibitem [{\citenamefont {Masson}\ and\ \citenamefont
  {Asenjo-Garcia}(2020)}]{masson2020atomic}%
  \BibitemOpen
  \bibfield  {author} {\bibinfo {author} {\bibfnamefont {S.~J.}\ \bibnamefont
  {Masson}}\ and\ \bibinfo {author} {\bibfnamefont {A.}~\bibnamefont
  {Asenjo-Garcia}},\ }\href@noop {} {\bibfield  {journal} {\bibinfo  {journal}
  {Physical Review Research}\ }\textbf {\bibinfo {volume} {2}},\ \bibinfo
  {pages} {043213} (\bibinfo {year} {2020})}\BibitemShut {NoStop}%
\bibitem [{\citenamefont {Corzo}\ \emph {et~al.}(2019)\citenamefont {Corzo},
  \citenamefont {Raskop}, \citenamefont {Chandra}, \citenamefont {Sheremet},
  \citenamefont {Gouraud},\ and\ \citenamefont {Laurat}}]{corzo2019waveguide}%
  \BibitemOpen
  \bibfield  {author} {\bibinfo {author} {\bibfnamefont {N.~V.}\ \bibnamefont
  {Corzo}}, \bibinfo {author} {\bibfnamefont {J.}~\bibnamefont {Raskop}},
  \bibinfo {author} {\bibfnamefont {A.}~\bibnamefont {Chandra}}, \bibinfo
  {author} {\bibfnamefont {A.~S.}\ \bibnamefont {Sheremet}}, \bibinfo {author}
  {\bibfnamefont {B.}~\bibnamefont {Gouraud}},\ and\ \bibinfo {author}
  {\bibfnamefont {J.}~\bibnamefont {Laurat}},\ }\href@noop {} {\bibfield
  {journal} {\bibinfo  {journal} {Nature}\ }\textbf {\bibinfo {volume} {566}},\
  \bibinfo {pages} {359} (\bibinfo {year} {2019})}\BibitemShut {NoStop}%
\bibitem [{\citenamefont {Zhang}\ and\ \citenamefont
  {M{\o}lmer}(2019)}]{zhang2019theory}%
  \BibitemOpen
  \bibfield  {author} {\bibinfo {author} {\bibfnamefont {Y.-X.}\ \bibnamefont
  {Zhang}}\ and\ \bibinfo {author} {\bibfnamefont {K.}~\bibnamefont
  {M{\o}lmer}},\ }\href@noop {} {\bibfield  {journal} {\bibinfo  {journal}
  {Physical Review Letters}\ }\textbf {\bibinfo {volume} {122}},\ \bibinfo
  {pages} {203605} (\bibinfo {year} {2019})}\BibitemShut {NoStop}%
\bibitem [{\citenamefont {Albrecht}\ \emph {et~al.}(2019)\citenamefont
  {Albrecht}, \citenamefont {Henriet}, \citenamefont {Asenjo-Garcia},
  \citenamefont {Dieterle}, \citenamefont {Painter},\ and\ \citenamefont
  {Chang}}]{albrecht2019subradiant}%
  \BibitemOpen
  \bibfield  {author} {\bibinfo {author} {\bibfnamefont {A.}~\bibnamefont
  {Albrecht}}, \bibinfo {author} {\bibfnamefont {L.}~\bibnamefont {Henriet}},
  \bibinfo {author} {\bibfnamefont {A.}~\bibnamefont {Asenjo-Garcia}}, \bibinfo
  {author} {\bibfnamefont {P.~B.}\ \bibnamefont {Dieterle}}, \bibinfo {author}
  {\bibfnamefont {O.}~\bibnamefont {Painter}},\ and\ \bibinfo {author}
  {\bibfnamefont {D.~E.}\ \bibnamefont {Chang}},\ }\href@noop {} {\bibfield
  {journal} {\bibinfo  {journal} {New Journal of Physics}\ }\textbf {\bibinfo
  {volume} {21}},\ \bibinfo {pages} {025003} (\bibinfo {year}
  {2019})}\BibitemShut {NoStop}%
\bibitem [{\citenamefont {Goban}\ \emph {et~al.}(2015)\citenamefont {Goban},
  \citenamefont {Hung}, \citenamefont {Hood}, \citenamefont {Yu}, \citenamefont
  {Muniz}, \citenamefont {Painter},\ and\ \citenamefont
  {Kimble}}]{goban2015superradiance}%
  \BibitemOpen
  \bibfield  {author} {\bibinfo {author} {\bibfnamefont {A.}~\bibnamefont
  {Goban}}, \bibinfo {author} {\bibfnamefont {C.-L.}\ \bibnamefont {Hung}},
  \bibinfo {author} {\bibfnamefont {J.}~\bibnamefont {Hood}}, \bibinfo {author}
  {\bibfnamefont {S.-P.}\ \bibnamefont {Yu}}, \bibinfo {author} {\bibfnamefont
  {J.}~\bibnamefont {Muniz}}, \bibinfo {author} {\bibfnamefont
  {O.}~\bibnamefont {Painter}},\ and\ \bibinfo {author} {\bibfnamefont
  {H.}~\bibnamefont {Kimble}},\ }\href@noop {} {\bibfield  {journal} {\bibinfo
  {journal} {Physical Review Letters}\ }\textbf {\bibinfo {volume} {115}},\
  \bibinfo {pages} {063601} (\bibinfo {year} {2015})}\BibitemShut {NoStop}%
\bibitem [{\citenamefont {Solano}\ \emph {et~al.}(2017)\citenamefont {Solano},
  \citenamefont {Barberis-Blostein}, \citenamefont {Fatemi}, \citenamefont
  {Orozco},\ and\ \citenamefont {Rolston}}]{solano2017super}%
  \BibitemOpen
  \bibfield  {author} {\bibinfo {author} {\bibfnamefont {P.}~\bibnamefont
  {Solano}}, \bibinfo {author} {\bibfnamefont {P.}~\bibnamefont
  {Barberis-Blostein}}, \bibinfo {author} {\bibfnamefont {F.~K.}\ \bibnamefont
  {Fatemi}}, \bibinfo {author} {\bibfnamefont {L.~A.}\ \bibnamefont {Orozco}},\
  and\ \bibinfo {author} {\bibfnamefont {S.~L.}\ \bibnamefont {Rolston}},\
  }\href@noop {} {\bibfield  {journal} {\bibinfo  {journal} {Nature
  Communications}\ }\textbf {\bibinfo {volume} {8}},\ \bibinfo {pages} {1}
  (\bibinfo {year} {2017})}\BibitemShut {NoStop}%
\bibitem [{\citenamefont {Nie}\ \emph {et~al.}(2021)\citenamefont {Nie},
  \citenamefont {Shi}, \citenamefont {Nori},\ and\ \citenamefont
  {Liu}}]{nie2021topology}%
  \BibitemOpen
  \bibfield  {author} {\bibinfo {author} {\bibfnamefont {W.}~\bibnamefont
  {Nie}}, \bibinfo {author} {\bibfnamefont {T.}~\bibnamefont {Shi}}, \bibinfo
  {author} {\bibfnamefont {F.}~\bibnamefont {Nori}},\ and\ \bibinfo {author}
  {\bibfnamefont {Y.-x.}\ \bibnamefont {Liu}},\ }\href@noop {} {\bibfield
  {journal} {\bibinfo  {journal} {Physical Review Applied}\ }\textbf {\bibinfo
  {volume} {15}},\ \bibinfo {pages} {044041} (\bibinfo {year}
  {2021})}\BibitemShut {NoStop}%
\bibitem [{\citenamefont {Shahmoon}\ \emph {et~al.}(2016)\citenamefont
  {Shahmoon}, \citenamefont {Gri{\v{s}}ins}, \citenamefont {Stimming},
  \citenamefont {Mazets},\ and\ \citenamefont {Kurizki}}]{shahmoon2016highly}%
  \BibitemOpen
  \bibfield  {author} {\bibinfo {author} {\bibfnamefont {E.}~\bibnamefont
  {Shahmoon}}, \bibinfo {author} {\bibfnamefont {P.}~\bibnamefont
  {Gri{\v{s}}ins}}, \bibinfo {author} {\bibfnamefont {H.~P.}\ \bibnamefont
  {Stimming}}, \bibinfo {author} {\bibfnamefont {I.}~\bibnamefont {Mazets}},\
  and\ \bibinfo {author} {\bibfnamefont {G.}~\bibnamefont {Kurizki}},\
  }\href@noop {} {\bibfield  {journal} {\bibinfo  {journal} {Optica}\ }\textbf
  {\bibinfo {volume} {3}},\ \bibinfo {pages} {725} (\bibinfo {year}
  {2016})}\BibitemShut {NoStop}%
\bibitem [{\citenamefont {Roy}(2011)}]{roy2011two}%
  \BibitemOpen
  \bibfield  {author} {\bibinfo {author} {\bibfnamefont {D.}~\bibnamefont
  {Roy}},\ }\href@noop {} {\bibfield  {journal} {\bibinfo  {journal} {Physical
  Review Letters}\ }\textbf {\bibinfo {volume} {106}},\ \bibinfo {pages}
  {053601} (\bibinfo {year} {2011})}\BibitemShut {NoStop}%
\bibitem [{\citenamefont {Song}\ \emph {et~al.}(2017)\citenamefont {Song},
  \citenamefont {Munro}, \citenamefont {Nie}, \citenamefont {Deng},
  \citenamefont {Yang},\ and\ \citenamefont {Kwek}}]{song2017photon}%
  \BibitemOpen
  \bibfield  {author} {\bibinfo {author} {\bibfnamefont {G.-Z.}\ \bibnamefont
  {Song}}, \bibinfo {author} {\bibfnamefont {E.}~\bibnamefont {Munro}},
  \bibinfo {author} {\bibfnamefont {W.}~\bibnamefont {Nie}}, \bibinfo {author}
  {\bibfnamefont {F.-G.}\ \bibnamefont {Deng}}, \bibinfo {author}
  {\bibfnamefont {G.-J.}\ \bibnamefont {Yang}},\ and\ \bibinfo {author}
  {\bibfnamefont {L.-C.}\ \bibnamefont {Kwek}},\ }\href@noop {} {\bibfield
  {journal} {\bibinfo  {journal} {Physical Review A}\ }\textbf {\bibinfo
  {volume} {96}},\ \bibinfo {pages} {043872} (\bibinfo {year}
  {2017})}\BibitemShut {NoStop}%
\bibitem [{\citenamefont {Lodahl}\ \emph {et~al.}(2017)\citenamefont {Lodahl},
  \citenamefont {Mahmoodian}, \citenamefont {Stobbe}, \citenamefont
  {Rauschenbeutel}, \citenamefont {Schneeweiss}, \citenamefont {Volz},
  \citenamefont {Pichler},\ and\ \citenamefont {Zoller}}]{lodahl2017chiral}%
  \BibitemOpen
  \bibfield  {author} {\bibinfo {author} {\bibfnamefont {P.}~\bibnamefont
  {Lodahl}}, \bibinfo {author} {\bibfnamefont {S.}~\bibnamefont {Mahmoodian}},
  \bibinfo {author} {\bibfnamefont {S.}~\bibnamefont {Stobbe}}, \bibinfo
  {author} {\bibfnamefont {A.}~\bibnamefont {Rauschenbeutel}}, \bibinfo
  {author} {\bibfnamefont {P.}~\bibnamefont {Schneeweiss}}, \bibinfo {author}
  {\bibfnamefont {J.}~\bibnamefont {Volz}}, \bibinfo {author} {\bibfnamefont
  {H.}~\bibnamefont {Pichler}},\ and\ \bibinfo {author} {\bibfnamefont
  {P.}~\bibnamefont {Zoller}},\ }\href@noop {} {\bibfield  {journal} {\bibinfo
  {journal} {Nature}\ }\textbf {\bibinfo {volume} {541}},\ \bibinfo {pages}
  {473} (\bibinfo {year} {2017})}\BibitemShut {NoStop}%
\bibitem [{\citenamefont {Rodr{\'\i}guez-Fortu{\~n}o}\ \emph
  {et~al.}(2013)\citenamefont {Rodr{\'\i}guez-Fortu{\~n}o}, \citenamefont
  {Marino}, \citenamefont {Ginzburg}, \citenamefont {O’Connor}, \citenamefont
  {Mart{\'\i}nez}, \citenamefont {Wurtz},\ and\ \citenamefont
  {Zayats}}]{rodriguez2013near}%
  \BibitemOpen
  \bibfield  {author} {\bibinfo {author} {\bibfnamefont {F.~J.}\ \bibnamefont
  {Rodr{\'\i}guez-Fortu{\~n}o}}, \bibinfo {author} {\bibfnamefont
  {G.}~\bibnamefont {Marino}}, \bibinfo {author} {\bibfnamefont
  {P.}~\bibnamefont {Ginzburg}}, \bibinfo {author} {\bibfnamefont
  {D.}~\bibnamefont {O’Connor}}, \bibinfo {author} {\bibfnamefont
  {A.}~\bibnamefont {Mart{\'\i}nez}}, \bibinfo {author} {\bibfnamefont {G.~A.}\
  \bibnamefont {Wurtz}},\ and\ \bibinfo {author} {\bibfnamefont {A.~V.}\
  \bibnamefont {Zayats}},\ }\href@noop {} {\bibfield  {journal} {\bibinfo
  {journal} {Science}\ }\textbf {\bibinfo {volume} {340}},\ \bibinfo {pages}
  {328} (\bibinfo {year} {2013})}\BibitemShut {NoStop}%
\bibitem [{\citenamefont {Mitsch}\ \emph {et~al.}(2014)\citenamefont {Mitsch},
  \citenamefont {Sayrin}, \citenamefont {Albrecht}, \citenamefont
  {Schneeweiss},\ and\ \citenamefont {Rauschenbeutel}}]{mitsch2014quantum}%
  \BibitemOpen
  \bibfield  {author} {\bibinfo {author} {\bibfnamefont {R.}~\bibnamefont
  {Mitsch}}, \bibinfo {author} {\bibfnamefont {C.}~\bibnamefont {Sayrin}},
  \bibinfo {author} {\bibfnamefont {B.}~\bibnamefont {Albrecht}}, \bibinfo
  {author} {\bibfnamefont {P.}~\bibnamefont {Schneeweiss}},\ and\ \bibinfo
  {author} {\bibfnamefont {A.}~\bibnamefont {Rauschenbeutel}},\ }\href@noop {}
  {\bibfield  {journal} {\bibinfo  {journal} {Nature Communications}\ }\textbf
  {\bibinfo {volume} {5}},\ \bibinfo {pages} {1} (\bibinfo {year}
  {2014})}\BibitemShut {NoStop}%
\bibitem [{\citenamefont {Petersen}\ \emph {et~al.}(2014)\citenamefont
  {Petersen}, \citenamefont {Volz},\ and\ \citenamefont
  {Rauschenbeutel}}]{petersen2014chiral}%
  \BibitemOpen
  \bibfield  {author} {\bibinfo {author} {\bibfnamefont {J.}~\bibnamefont
  {Petersen}}, \bibinfo {author} {\bibfnamefont {J.}~\bibnamefont {Volz}},\
  and\ \bibinfo {author} {\bibfnamefont {A.}~\bibnamefont {Rauschenbeutel}},\
  }\href@noop {} {\bibfield  {journal} {\bibinfo  {journal} {Science}\ }\textbf
  {\bibinfo {volume} {346}},\ \bibinfo {pages} {67} (\bibinfo {year}
  {2014})}\BibitemShut {NoStop}%
\bibitem [{\citenamefont {S{\"o}llner}\ \emph {et~al.}(2015)\citenamefont
  {S{\"o}llner}, \citenamefont {Mahmoodian}, \citenamefont {Hansen},
  \citenamefont {Midolo}, \citenamefont {Javadi}, \citenamefont
  {Kir{\v{s}}ansk{\.e}}, \citenamefont {Pregnolato}, \citenamefont {El-Ella},
  \citenamefont {Lee}, \citenamefont {Song} \emph
  {et~al.}}]{sollner2015deterministic}%
  \BibitemOpen
  \bibfield  {author} {\bibinfo {author} {\bibfnamefont {I.}~\bibnamefont
  {S{\"o}llner}}, \bibinfo {author} {\bibfnamefont {S.}~\bibnamefont
  {Mahmoodian}}, \bibinfo {author} {\bibfnamefont {S.~L.}\ \bibnamefont
  {Hansen}}, \bibinfo {author} {\bibfnamefont {L.}~\bibnamefont {Midolo}},
  \bibinfo {author} {\bibfnamefont {A.}~\bibnamefont {Javadi}}, \bibinfo
  {author} {\bibfnamefont {G.}~\bibnamefont {Kir{\v{s}}ansk{\.e}}}, \bibinfo
  {author} {\bibfnamefont {T.}~\bibnamefont {Pregnolato}}, \bibinfo {author}
  {\bibfnamefont {H.}~\bibnamefont {El-Ella}}, \bibinfo {author} {\bibfnamefont
  {E.~H.}\ \bibnamefont {Lee}}, \bibinfo {author} {\bibfnamefont {J.~D.}\
  \bibnamefont {Song}}, \emph {et~al.},\ }\href@noop {} {\bibfield  {journal}
  {\bibinfo  {journal} {Nature Nanotechnology}\ }\textbf {\bibinfo {volume}
  {10}},\ \bibinfo {pages} {775} (\bibinfo {year} {2015})}\BibitemShut
  {NoStop}%
\bibitem [{\citenamefont {Coles}\ \emph {et~al.}(2016)\citenamefont {Coles},
  \citenamefont {Price}, \citenamefont {Dixon}, \citenamefont {Royall},
  \citenamefont {Clarke}, \citenamefont {Kok}, \citenamefont {Skolnick},
  \citenamefont {Fox},\ and\ \citenamefont {Makhonin}}]{coles2016chirality}%
  \BibitemOpen
  \bibfield  {author} {\bibinfo {author} {\bibfnamefont {R.}~\bibnamefont
  {Coles}}, \bibinfo {author} {\bibfnamefont {D.}~\bibnamefont {Price}},
  \bibinfo {author} {\bibfnamefont {J.}~\bibnamefont {Dixon}}, \bibinfo
  {author} {\bibfnamefont {B.}~\bibnamefont {Royall}}, \bibinfo {author}
  {\bibfnamefont {E.}~\bibnamefont {Clarke}}, \bibinfo {author} {\bibfnamefont
  {P.}~\bibnamefont {Kok}}, \bibinfo {author} {\bibfnamefont {M.}~\bibnamefont
  {Skolnick}}, \bibinfo {author} {\bibfnamefont {A.}~\bibnamefont {Fox}},\ and\
  \bibinfo {author} {\bibfnamefont {M.}~\bibnamefont {Makhonin}},\ }\href@noop
  {} {\bibfield  {journal} {\bibinfo  {journal} {Nature Communications}\
  }\textbf {\bibinfo {volume} {7}},\ \bibinfo {pages} {1} (\bibinfo {year}
  {2016})}\BibitemShut {NoStop}%
\bibitem [{\citenamefont {Javadi}\ \emph {et~al.}(2018)\citenamefont {Javadi},
  \citenamefont {Ding}, \citenamefont {Appel}, \citenamefont {Mahmoodian},
  \citenamefont {L{\"o}bl}, \citenamefont {S{\"o}llner}, \citenamefont
  {Schott}, \citenamefont {Papon}, \citenamefont {Pregnolato}, \citenamefont
  {Stobbe} \emph {et~al.}}]{javadi2018spin}%
  \BibitemOpen
  \bibfield  {author} {\bibinfo {author} {\bibfnamefont {A.}~\bibnamefont
  {Javadi}}, \bibinfo {author} {\bibfnamefont {D.}~\bibnamefont {Ding}},
  \bibinfo {author} {\bibfnamefont {M.~H.}\ \bibnamefont {Appel}}, \bibinfo
  {author} {\bibfnamefont {S.}~\bibnamefont {Mahmoodian}}, \bibinfo {author}
  {\bibfnamefont {M.~C.}\ \bibnamefont {L{\"o}bl}}, \bibinfo {author}
  {\bibfnamefont {I.}~\bibnamefont {S{\"o}llner}}, \bibinfo {author}
  {\bibfnamefont {R.}~\bibnamefont {Schott}}, \bibinfo {author} {\bibfnamefont
  {C.}~\bibnamefont {Papon}}, \bibinfo {author} {\bibfnamefont
  {T.}~\bibnamefont {Pregnolato}}, \bibinfo {author} {\bibfnamefont
  {S.}~\bibnamefont {Stobbe}}, \emph {et~al.},\ }\href@noop {} {\bibfield
  {journal} {\bibinfo  {journal} {Nature Nanotechnology}\ }\textbf {\bibinfo
  {volume} {13}},\ \bibinfo {pages} {398} (\bibinfo {year} {2018})}\BibitemShut
  {NoStop}%
\bibitem [{\citenamefont {Mirza}\ and\ \citenamefont
  {Schotland}(2016)}]{Mirza2016}%
  \BibitemOpen
  \bibfield  {author} {\bibinfo {author} {\bibfnamefont {I.~M.}\ \bibnamefont
  {Mirza}}\ and\ \bibinfo {author} {\bibfnamefont {J.~C.}\ \bibnamefont
  {Schotland}},\ }\href@noop {} {\bibfield  {journal} {\bibinfo  {journal}
  {Physical Review A}\ }\textbf {\bibinfo {volume} {94}},\ \bibinfo {pages}
  {012309} (\bibinfo {year} {2016})}\BibitemShut {NoStop}%
\bibitem [{\citenamefont {Mahmoodian}\ \emph {et~al.}(2020)\citenamefont
  {Mahmoodian}, \citenamefont {Calaj{\'o}}, \citenamefont {Chang},
  \citenamefont {Hammerer},\ and\ \citenamefont
  {S{\o}rensen}}]{mahmoodian2020dynamics}%
  \BibitemOpen
  \bibfield  {author} {\bibinfo {author} {\bibfnamefont {S.}~\bibnamefont
  {Mahmoodian}}, \bibinfo {author} {\bibfnamefont {G.}~\bibnamefont
  {Calaj{\'o}}}, \bibinfo {author} {\bibfnamefont {D.~E.}\ \bibnamefont
  {Chang}}, \bibinfo {author} {\bibfnamefont {K.}~\bibnamefont {Hammerer}},\
  and\ \bibinfo {author} {\bibfnamefont {A.~S.}\ \bibnamefont {S{\o}rensen}},\
  }\href@noop {} {\bibfield  {journal} {\bibinfo  {journal} {Physical Review
  X}\ }\textbf {\bibinfo {volume} {10}},\ \bibinfo {pages} {031011} (\bibinfo
  {year} {2020})}\BibitemShut {NoStop}%
\bibitem [{\citenamefont {Wang}\ \emph {et~al.}(2020)\citenamefont {Wang},
  \citenamefont {Yuan}, \citenamefont {Chen},\ and\ \citenamefont
  {Fan}}]{wang2020single}%
  \BibitemOpen
  \bibfield  {author} {\bibinfo {author} {\bibfnamefont {L.}~\bibnamefont
  {Wang}}, \bibinfo {author} {\bibfnamefont {L.}~\bibnamefont {Yuan}}, \bibinfo
  {author} {\bibfnamefont {X.}~\bibnamefont {Chen}},\ and\ \bibinfo {author}
  {\bibfnamefont {S.}~\bibnamefont {Fan}},\ }\href@noop {} {\bibfield
  {journal} {\bibinfo  {journal} {Physical Review Applied}\ }\textbf {\bibinfo
  {volume} {14}},\ \bibinfo {pages} {014063} (\bibinfo {year}
  {2020})}\BibitemShut {NoStop}%
\bibitem [{\citenamefont {Le~Kien}\ and\ \citenamefont
  {Rauschenbeutel}(2017)}]{LeKien2017}%
  \BibitemOpen
  \bibfield  {author} {\bibinfo {author} {\bibfnamefont {F.}~\bibnamefont
  {Le~Kien}}\ and\ \bibinfo {author} {\bibfnamefont {A.}~\bibnamefont
  {Rauschenbeutel}},\ }\href@noop {} {\bibfield  {journal} {\bibinfo  {journal}
  {Physical Review A}\ }\textbf {\bibinfo {volume} {95}},\ \bibinfo {pages}
  {023838} (\bibinfo {year} {2017})}\BibitemShut {NoStop}%
\bibitem [{\citenamefont {Le~Kien}\ \emph {et~al.}(2006)\citenamefont
  {Le~Kien}, \citenamefont {Balykin},\ and\ \citenamefont
  {Hakuta}}]{le2006scattering}%
  \BibitemOpen
  \bibfield  {author} {\bibinfo {author} {\bibfnamefont {F.}~\bibnamefont
  {Le~Kien}}, \bibinfo {author} {\bibfnamefont {V.}~\bibnamefont {Balykin}},\
  and\ \bibinfo {author} {\bibfnamefont {K.}~\bibnamefont {Hakuta}},\
  }\href@noop {} {\bibfield  {journal} {\bibinfo  {journal} {Physical Review
  A}\ }\textbf {\bibinfo {volume} {73}},\ \bibinfo {pages} {013819} (\bibinfo
  {year} {2006})}\BibitemShut {NoStop}%
\bibitem [{\citenamefont {Scheel}\ \emph {et~al.}(2015)\citenamefont {Scheel},
  \citenamefont {Buhmann}, \citenamefont {Clausen},\ and\ \citenamefont
  {Schneeweiss}}]{Scheel2015}%
  \BibitemOpen
  \bibfield  {author} {\bibinfo {author} {\bibfnamefont {S.}~\bibnamefont
  {Scheel}}, \bibinfo {author} {\bibfnamefont {S.~Y.}\ \bibnamefont {Buhmann}},
  \bibinfo {author} {\bibfnamefont {C.}~\bibnamefont {Clausen}},\ and\ \bibinfo
  {author} {\bibfnamefont {P.}~\bibnamefont {Schneeweiss}},\ }\href@noop {}
  {\bibfield  {journal} {\bibinfo  {journal} {Physical Review A}\ }\textbf
  {\bibinfo {volume} {92}},\ \bibinfo {pages} {043819} (\bibinfo {year}
  {2015})}\BibitemShut {NoStop}%
\bibitem [{\citenamefont {Pichler}\ \emph {et~al.}(2015)\citenamefont
  {Pichler}, \citenamefont {Ramos}, \citenamefont {Daley},\ and\ \citenamefont
  {Zoller}}]{Pichler2015}%
  \BibitemOpen
  \bibfield  {author} {\bibinfo {author} {\bibfnamefont {H.}~\bibnamefont
  {Pichler}}, \bibinfo {author} {\bibfnamefont {T.}~\bibnamefont {Ramos}},
  \bibinfo {author} {\bibfnamefont {A.~J.}\ \bibnamefont {Daley}},\ and\
  \bibinfo {author} {\bibfnamefont {P.}~\bibnamefont {Zoller}},\ }\href@noop {}
  {\bibfield  {journal} {\bibinfo  {journal} {Physical Review A}\ }\textbf
  {\bibinfo {volume} {91}},\ \bibinfo {pages} {042116} (\bibinfo {year}
  {2015})}\BibitemShut {NoStop}%
\bibitem [{\citenamefont {Ghatak}\ and\ \citenamefont
  {Das}(2019)}]{ghatak2019new}%
  \BibitemOpen
  \bibfield  {author} {\bibinfo {author} {\bibfnamefont {A.}~\bibnamefont
  {Ghatak}}\ and\ \bibinfo {author} {\bibfnamefont {T.}~\bibnamefont {Das}},\
  }\href@noop {} {\bibfield  {journal} {\bibinfo  {journal} {Journal of
  Physics: Condensed Matter}\ }\textbf {\bibinfo {volume} {31}},\ \bibinfo
  {pages} {263001} (\bibinfo {year} {2019})}\BibitemShut {NoStop}%
\bibitem [{\citenamefont {Ruostekoski}\ \emph {et~al.}(2002)\citenamefont
  {Ruostekoski}, \citenamefont {Dunne},\ and\ \citenamefont
  {Javanainen}}]{ruostekoski2002particle}%
  \BibitemOpen
  \bibfield  {author} {\bibinfo {author} {\bibfnamefont {J.}~\bibnamefont
  {Ruostekoski}}, \bibinfo {author} {\bibfnamefont {G.~V.}\ \bibnamefont
  {Dunne}},\ and\ \bibinfo {author} {\bibfnamefont {J.}~\bibnamefont
  {Javanainen}},\ }\href@noop {} {\bibfield  {journal} {\bibinfo  {journal}
  {Physical Review Letters}\ }\textbf {\bibinfo {volume} {88}},\ \bibinfo
  {pages} {180401} (\bibinfo {year} {2002})}\BibitemShut {NoStop}%
\bibitem [{\citenamefont {Schomerus}(2013)}]{schomerus2013topologically}%
  \BibitemOpen
  \bibfield  {author} {\bibinfo {author} {\bibfnamefont {H.}~\bibnamefont
  {Schomerus}},\ }\href@noop {} {\bibfield  {journal} {\bibinfo  {journal}
  {Optics Letters}\ }\textbf {\bibinfo {volume} {38}},\ \bibinfo {pages} {1912}
  (\bibinfo {year} {2013})}\BibitemShut {NoStop}%
\bibitem [{\citenamefont {Poli}\ \emph {et~al.}(2015)\citenamefont {Poli},
  \citenamefont {Bellec}, \citenamefont {Kuhl}, \citenamefont {Mortessagne},\
  and\ \citenamefont {Schomerus}}]{poli2015selective}%
  \BibitemOpen
  \bibfield  {author} {\bibinfo {author} {\bibfnamefont {C.}~\bibnamefont
  {Poli}}, \bibinfo {author} {\bibfnamefont {M.}~\bibnamefont {Bellec}},
  \bibinfo {author} {\bibfnamefont {U.}~\bibnamefont {Kuhl}}, \bibinfo {author}
  {\bibfnamefont {F.}~\bibnamefont {Mortessagne}},\ and\ \bibinfo {author}
  {\bibfnamefont {H.}~\bibnamefont {Schomerus}},\ }\href@noop {} {\bibfield
  {journal} {\bibinfo  {journal} {Nature Communications}\ }\textbf {\bibinfo
  {volume} {6}},\ \bibinfo {pages} {1} (\bibinfo {year} {2015})}\BibitemShut
  {NoStop}%
\bibitem [{\citenamefont {Zeuner}\ \emph {et~al.}(2015)\citenamefont {Zeuner},
  \citenamefont {Rechtsman}, \citenamefont {Plotnik}, \citenamefont {Lumer},
  \citenamefont {Nolte}, \citenamefont {Rudner}, \citenamefont {Segev},\ and\
  \citenamefont {Szameit}}]{zeuner2015observation}%
  \BibitemOpen
  \bibfield  {author} {\bibinfo {author} {\bibfnamefont {J.~M.}\ \bibnamefont
  {Zeuner}}, \bibinfo {author} {\bibfnamefont {M.~C.}\ \bibnamefont
  {Rechtsman}}, \bibinfo {author} {\bibfnamefont {Y.}~\bibnamefont {Plotnik}},
  \bibinfo {author} {\bibfnamefont {Y.}~\bibnamefont {Lumer}}, \bibinfo
  {author} {\bibfnamefont {S.}~\bibnamefont {Nolte}}, \bibinfo {author}
  {\bibfnamefont {M.~S.}\ \bibnamefont {Rudner}}, \bibinfo {author}
  {\bibfnamefont {M.}~\bibnamefont {Segev}},\ and\ \bibinfo {author}
  {\bibfnamefont {A.}~\bibnamefont {Szameit}},\ }\href@noop {} {\bibfield
  {journal} {\bibinfo  {journal} {Physical Review Letters}\ }\textbf {\bibinfo
  {volume} {115}},\ \bibinfo {pages} {040402} (\bibinfo {year}
  {2015})}\BibitemShut {NoStop}%
\bibitem [{\citenamefont {Leykam}\ \emph {et~al.}(2017)\citenamefont {Leykam},
  \citenamefont {Bliokh}, \citenamefont {Huang}, \citenamefont {Chong},\ and\
  \citenamefont {Nori}}]{leykam2017edge}%
  \BibitemOpen
  \bibfield  {author} {\bibinfo {author} {\bibfnamefont {D.}~\bibnamefont
  {Leykam}}, \bibinfo {author} {\bibfnamefont {K.~Y.}\ \bibnamefont {Bliokh}},
  \bibinfo {author} {\bibfnamefont {C.}~\bibnamefont {Huang}}, \bibinfo
  {author} {\bibfnamefont {Y.~D.}\ \bibnamefont {Chong}},\ and\ \bibinfo
  {author} {\bibfnamefont {F.}~\bibnamefont {Nori}},\ }\href@noop {} {\bibfield
   {journal} {\bibinfo  {journal} {Physical review letters}\ }\textbf {\bibinfo
  {volume} {118}},\ \bibinfo {pages} {040401} (\bibinfo {year}
  {2017})}\BibitemShut {NoStop}%
\bibitem [{\citenamefont {Shen}\ \emph {et~al.}(2018)\citenamefont {Shen},
  \citenamefont {Zhen},\ and\ \citenamefont {Fu}}]{shen2018topological}%
  \BibitemOpen
  \bibfield  {author} {\bibinfo {author} {\bibfnamefont {H.}~\bibnamefont
  {Shen}}, \bibinfo {author} {\bibfnamefont {B.}~\bibnamefont {Zhen}},\ and\
  \bibinfo {author} {\bibfnamefont {L.}~\bibnamefont {Fu}},\ }\href@noop {}
  {\bibfield  {journal} {\bibinfo  {journal} {Physical Review Letters}\
  }\textbf {\bibinfo {volume} {120}},\ \bibinfo {pages} {146402} (\bibinfo
  {year} {2018})}\BibitemShut {NoStop}%
\bibitem [{\citenamefont {Zhao}\ \emph {et~al.}(2018)\citenamefont {Zhao},
  \citenamefont {Miao}, \citenamefont {Teimourpour}, \citenamefont {Malzard},
  \citenamefont {El-Ganainy}, \citenamefont {Schomerus},\ and\ \citenamefont
  {Feng}}]{zhao2018topological}%
  \BibitemOpen
  \bibfield  {author} {\bibinfo {author} {\bibfnamefont {H.}~\bibnamefont
  {Zhao}}, \bibinfo {author} {\bibfnamefont {P.}~\bibnamefont {Miao}}, \bibinfo
  {author} {\bibfnamefont {M.~H.}\ \bibnamefont {Teimourpour}}, \bibinfo
  {author} {\bibfnamefont {S.}~\bibnamefont {Malzard}}, \bibinfo {author}
  {\bibfnamefont {R.}~\bibnamefont {El-Ganainy}}, \bibinfo {author}
  {\bibfnamefont {H.}~\bibnamefont {Schomerus}},\ and\ \bibinfo {author}
  {\bibfnamefont {L.}~\bibnamefont {Feng}},\ }\href@noop {} {\bibfield
  {journal} {\bibinfo  {journal} {Nature Communications}\ }\textbf {\bibinfo
  {volume} {9}},\ \bibinfo {pages} {1} (\bibinfo {year} {2018})}\BibitemShut
  {NoStop}%
\bibitem [{\citenamefont {Parto}\ \emph {et~al.}(2018)\citenamefont {Parto},
  \citenamefont {Wittek}, \citenamefont {Hodaei}, \citenamefont {Harari},
  \citenamefont {Bandres}, \citenamefont {Ren}, \citenamefont {Rechtsman},
  \citenamefont {Segev}, \citenamefont {Christodoulides},\ and\ \citenamefont
  {Khajavikhan}}]{parto2018edge}%
  \BibitemOpen
  \bibfield  {author} {\bibinfo {author} {\bibfnamefont {M.}~\bibnamefont
  {Parto}}, \bibinfo {author} {\bibfnamefont {S.}~\bibnamefont {Wittek}},
  \bibinfo {author} {\bibfnamefont {H.}~\bibnamefont {Hodaei}}, \bibinfo
  {author} {\bibfnamefont {G.}~\bibnamefont {Harari}}, \bibinfo {author}
  {\bibfnamefont {M.~A.}\ \bibnamefont {Bandres}}, \bibinfo {author}
  {\bibfnamefont {J.}~\bibnamefont {Ren}}, \bibinfo {author} {\bibfnamefont
  {M.~C.}\ \bibnamefont {Rechtsman}}, \bibinfo {author} {\bibfnamefont
  {M.}~\bibnamefont {Segev}}, \bibinfo {author} {\bibfnamefont {D.~N.}\
  \bibnamefont {Christodoulides}},\ and\ \bibinfo {author} {\bibfnamefont
  {M.}~\bibnamefont {Khajavikhan}},\ }\href@noop {} {\bibfield  {journal}
  {\bibinfo  {journal} {Physical Review Letters}\ }\textbf {\bibinfo {volume}
  {120}},\ \bibinfo {pages} {113901} (\bibinfo {year} {2018})}\BibitemShut
  {NoStop}%
\bibitem [{\citenamefont {Kawabata}\ \emph {et~al.}(2019)\citenamefont
  {Kawabata}, \citenamefont {Shiozaki}, \citenamefont {Ueda},\ and\
  \citenamefont {Sato}}]{kawabata2019symmetry}%
  \BibitemOpen
  \bibfield  {author} {\bibinfo {author} {\bibfnamefont {K.}~\bibnamefont
  {Kawabata}}, \bibinfo {author} {\bibfnamefont {K.}~\bibnamefont {Shiozaki}},
  \bibinfo {author} {\bibfnamefont {M.}~\bibnamefont {Ueda}},\ and\ \bibinfo
  {author} {\bibfnamefont {M.}~\bibnamefont {Sato}},\ }\href@noop {} {\bibfield
   {journal} {\bibinfo  {journal} {Physical Review X}\ }\textbf {\bibinfo
  {volume} {9}},\ \bibinfo {pages} {041015} (\bibinfo {year}
  {2019})}\BibitemShut {NoStop}%
\bibitem [{\citenamefont {Li}\ \emph {et~al.}(2020{\natexlab{b}})\citenamefont
  {Li}, \citenamefont {Lee},\ and\ \citenamefont {Gong}}]{li2020topological}%
  \BibitemOpen
  \bibfield  {author} {\bibinfo {author} {\bibfnamefont {L.}~\bibnamefont
  {Li}}, \bibinfo {author} {\bibfnamefont {C.~H.}\ \bibnamefont {Lee}},\ and\
  \bibinfo {author} {\bibfnamefont {J.}~\bibnamefont {Gong}},\ }\href@noop {}
  {\bibfield  {journal} {\bibinfo  {journal} {Physical Review Letters}\
  }\textbf {\bibinfo {volume} {124}},\ \bibinfo {pages} {250402} (\bibinfo
  {year} {2020}{\natexlab{b}})}\BibitemShut {NoStop}%
\bibitem [{\citenamefont {Yokomizo}\ and\ \citenamefont
  {Murakami}(2021)}]{yokomizo2021non}%
  \BibitemOpen
  \bibfield  {author} {\bibinfo {author} {\bibfnamefont {K.}~\bibnamefont
  {Yokomizo}}\ and\ \bibinfo {author} {\bibfnamefont {S.}~\bibnamefont
  {Murakami}},\ }\href@noop {} {\bibfield  {journal} {\bibinfo  {journal}
  {Physical Review B}\ }\textbf {\bibinfo {volume} {103}},\ \bibinfo {pages}
  {165123} (\bibinfo {year} {2021})}\BibitemShut {NoStop}%
\bibitem [{\citenamefont {Goban}\ \emph {et~al.}(2012)\citenamefont {Goban},
  \citenamefont {Choi}, \citenamefont {Alton}, \citenamefont {Ding},
  \citenamefont {Lacro{\^u}te}, \citenamefont {Pototschnig}, \citenamefont
  {Thiele}, \citenamefont {Stern},\ and\ \citenamefont
  {Kimble}}]{goban2012demonstration}%
  \BibitemOpen
  \bibfield  {author} {\bibinfo {author} {\bibfnamefont {A.}~\bibnamefont
  {Goban}}, \bibinfo {author} {\bibfnamefont {K.}~\bibnamefont {Choi}},
  \bibinfo {author} {\bibfnamefont {D.}~\bibnamefont {Alton}}, \bibinfo
  {author} {\bibfnamefont {D.}~\bibnamefont {Ding}}, \bibinfo {author}
  {\bibfnamefont {C.}~\bibnamefont {Lacro{\^u}te}}, \bibinfo {author}
  {\bibfnamefont {M.}~\bibnamefont {Pototschnig}}, \bibinfo {author}
  {\bibfnamefont {T.}~\bibnamefont {Thiele}}, \bibinfo {author} {\bibfnamefont
  {N.}~\bibnamefont {Stern}},\ and\ \bibinfo {author} {\bibfnamefont
  {H.}~\bibnamefont {Kimble}},\ }\href@noop {} {\bibfield  {journal} {\bibinfo
  {journal} {Physical Review Letters}\ }\textbf {\bibinfo {volume} {109}},\
  \bibinfo {pages} {033603} (\bibinfo {year} {2012})}\BibitemShut {NoStop}%
\bibitem [{\citenamefont {Reitz}\ \emph {et~al.}(2013)\citenamefont {Reitz},
  \citenamefont {Sayrin}, \citenamefont {Mitsch}, \citenamefont {Schneeweiss},\
  and\ \citenamefont {Rauschenbeutel}}]{reitz2013coherence}%
  \BibitemOpen
  \bibfield  {author} {\bibinfo {author} {\bibfnamefont {D.}~\bibnamefont
  {Reitz}}, \bibinfo {author} {\bibfnamefont {C.}~\bibnamefont {Sayrin}},
  \bibinfo {author} {\bibfnamefont {R.}~\bibnamefont {Mitsch}}, \bibinfo
  {author} {\bibfnamefont {P.}~\bibnamefont {Schneeweiss}},\ and\ \bibinfo
  {author} {\bibfnamefont {A.}~\bibnamefont {Rauschenbeutel}},\ }\href@noop {}
  {\bibfield  {journal} {\bibinfo  {journal} {Physical Review Letters}\
  }\textbf {\bibinfo {volume} {110}},\ \bibinfo {pages} {243603} (\bibinfo
  {year} {2013})}\BibitemShut {NoStop}%
\bibitem [{\citenamefont {Le~Kien}\ and\ \citenamefont
  {Rauschenbeutel}(2014)}]{le2014propagation}%
  \BibitemOpen
  \bibfield  {author} {\bibinfo {author} {\bibfnamefont {F.}~\bibnamefont
  {Le~Kien}}\ and\ \bibinfo {author} {\bibfnamefont {A.}~\bibnamefont
  {Rauschenbeutel}},\ }\href@noop {} {\bibfield  {journal} {\bibinfo  {journal}
  {Physical Review A}\ }\textbf {\bibinfo {volume} {90}},\ \bibinfo {pages}
  {063816} (\bibinfo {year} {2014})}\BibitemShut {NoStop}%
\bibitem [{\citenamefont {Sayrin}\ \emph {et~al.}(2015)\citenamefont {Sayrin},
  \citenamefont {Clausen}, \citenamefont {Albrecht}, \citenamefont
  {Schneeweiss},\ and\ \citenamefont {Rauschenbeutel}}]{sayrin2015storage}%
  \BibitemOpen
  \bibfield  {author} {\bibinfo {author} {\bibfnamefont {C.}~\bibnamefont
  {Sayrin}}, \bibinfo {author} {\bibfnamefont {C.}~\bibnamefont {Clausen}},
  \bibinfo {author} {\bibfnamefont {B.}~\bibnamefont {Albrecht}}, \bibinfo
  {author} {\bibfnamefont {P.}~\bibnamefont {Schneeweiss}},\ and\ \bibinfo
  {author} {\bibfnamefont {A.}~\bibnamefont {Rauschenbeutel}},\ }\href@noop {}
  {\bibfield  {journal} {\bibinfo  {journal} {Optica}\ }\textbf {\bibinfo
  {volume} {2}},\ \bibinfo {pages} {353} (\bibinfo {year} {2015})}\BibitemShut
  {NoStop}%
\bibitem [{\citenamefont {Tiecke}\ \emph {et~al.}(2015)\citenamefont {Tiecke},
  \citenamefont {Nayak}, \citenamefont {Thompson}, \citenamefont {Peyronel},
  \citenamefont {de~Leon}, \citenamefont {Vuleti{\'c}},\ and\ \citenamefont
  {Lukin}}]{tiecke2015efficient}%
  \BibitemOpen
  \bibfield  {author} {\bibinfo {author} {\bibfnamefont {T.}~\bibnamefont
  {Tiecke}}, \bibinfo {author} {\bibfnamefont {K.}~\bibnamefont {Nayak}},
  \bibinfo {author} {\bibfnamefont {J.~D.}\ \bibnamefont {Thompson}}, \bibinfo
  {author} {\bibfnamefont {T.}~\bibnamefont {Peyronel}}, \bibinfo {author}
  {\bibfnamefont {N.~P.}\ \bibnamefont {de~Leon}}, \bibinfo {author}
  {\bibfnamefont {V.}~\bibnamefont {Vuleti{\'c}}},\ and\ \bibinfo {author}
  {\bibfnamefont {M.}~\bibnamefont {Lukin}},\ }\href@noop {} {\bibfield
  {journal} {\bibinfo  {journal} {Optica}\ }\textbf {\bibinfo {volume} {2}},\
  \bibinfo {pages} {70} (\bibinfo {year} {2015})}\BibitemShut {NoStop}%
\bibitem [{\citenamefont {Van~Loo}\ \emph {et~al.}(2013)\citenamefont
  {Van~Loo}, \citenamefont {Fedorov}, \citenamefont {Lalumiere}, \citenamefont
  {Sanders}, \citenamefont {Blais},\ and\ \citenamefont
  {Wallraff}}]{van2013photon}%
  \BibitemOpen
  \bibfield  {author} {\bibinfo {author} {\bibfnamefont {A.~F.}\ \bibnamefont
  {Van~Loo}}, \bibinfo {author} {\bibfnamefont {A.}~\bibnamefont {Fedorov}},
  \bibinfo {author} {\bibfnamefont {K.}~\bibnamefont {Lalumiere}}, \bibinfo
  {author} {\bibfnamefont {B.~C.}\ \bibnamefont {Sanders}}, \bibinfo {author}
  {\bibfnamefont {A.}~\bibnamefont {Blais}},\ and\ \bibinfo {author}
  {\bibfnamefont {A.}~\bibnamefont {Wallraff}},\ }\href@noop {} {\bibfield
  {journal} {\bibinfo  {journal} {Science}\ }\textbf {\bibinfo {volume}
  {342}},\ \bibinfo {pages} {1494} (\bibinfo {year} {2013})}\BibitemShut
  {NoStop}%
\bibitem [{\citenamefont {Janvier}\ \emph {et~al.}(2015)\citenamefont
  {Janvier}, \citenamefont {Tosi}, \citenamefont {Bretheau}, \citenamefont
  {Girit}, \citenamefont {Stern}, \citenamefont {Bertet}, \citenamefont
  {Joyez}, \citenamefont {Vion}, \citenamefont {Esteve}, \citenamefont
  {Goffman} \emph {et~al.}}]{janvier2015coherent}%
  \BibitemOpen
  \bibfield  {author} {\bibinfo {author} {\bibfnamefont {C.}~\bibnamefont
  {Janvier}}, \bibinfo {author} {\bibfnamefont {L.}~\bibnamefont {Tosi}},
  \bibinfo {author} {\bibfnamefont {L.}~\bibnamefont {Bretheau}}, \bibinfo
  {author} {\bibfnamefont {{\c{C}}.}~\bibnamefont {Girit}}, \bibinfo {author}
  {\bibfnamefont {M.}~\bibnamefont {Stern}}, \bibinfo {author} {\bibfnamefont
  {P.}~\bibnamefont {Bertet}}, \bibinfo {author} {\bibfnamefont
  {P.}~\bibnamefont {Joyez}}, \bibinfo {author} {\bibfnamefont
  {D.}~\bibnamefont {Vion}}, \bibinfo {author} {\bibfnamefont {D.}~\bibnamefont
  {Esteve}}, \bibinfo {author} {\bibfnamefont {M.}~\bibnamefont {Goffman}},
  \emph {et~al.},\ }\href@noop {} {\bibfield  {journal} {\bibinfo  {journal}
  {Science}\ }\textbf {\bibinfo {volume} {349}},\ \bibinfo {pages} {1199}
  (\bibinfo {year} {2015})}\BibitemShut {NoStop}%
\bibitem [{\citenamefont {Kakuyanagi}\ \emph {et~al.}(2016)\citenamefont
  {Kakuyanagi}, \citenamefont {Matsuzaki}, \citenamefont {D{\'e}prez},
  \citenamefont {Toida}, \citenamefont {Semba}, \citenamefont {Yamaguchi},
  \citenamefont {Munro},\ and\ \citenamefont
  {Saito}}]{kakuyanagi2016observation}%
  \BibitemOpen
  \bibfield  {author} {\bibinfo {author} {\bibfnamefont {K.}~\bibnamefont
  {Kakuyanagi}}, \bibinfo {author} {\bibfnamefont {Y.}~\bibnamefont
  {Matsuzaki}}, \bibinfo {author} {\bibfnamefont {C.}~\bibnamefont
  {D{\'e}prez}}, \bibinfo {author} {\bibfnamefont {H.}~\bibnamefont {Toida}},
  \bibinfo {author} {\bibfnamefont {K.}~\bibnamefont {Semba}}, \bibinfo
  {author} {\bibfnamefont {H.}~\bibnamefont {Yamaguchi}}, \bibinfo {author}
  {\bibfnamefont {W.~J.}\ \bibnamefont {Munro}},\ and\ \bibinfo {author}
  {\bibfnamefont {S.}~\bibnamefont {Saito}},\ }\href@noop {} {\bibfield
  {journal} {\bibinfo  {journal} {Physical Review Letters}\ }\textbf {\bibinfo
  {volume} {117}},\ \bibinfo {pages} {210503} (\bibinfo {year}
  {2016})}\BibitemShut {NoStop}%
\bibitem [{\citenamefont {Wen}\ \emph {et~al.}(2019)\citenamefont {Wen},
  \citenamefont {Lin}, \citenamefont {Kockum}, \citenamefont {Suri},
  \citenamefont {Ian}, \citenamefont {Chen}, \citenamefont {Mao}, \citenamefont
  {Chiu}, \citenamefont {Delsing}, \citenamefont {Nori} \emph
  {et~al.}}]{wen2019large}%
  \BibitemOpen
  \bibfield  {author} {\bibinfo {author} {\bibfnamefont {P.}~\bibnamefont
  {Wen}}, \bibinfo {author} {\bibfnamefont {K.-T.}\ \bibnamefont {Lin}},
  \bibinfo {author} {\bibfnamefont {A.}~\bibnamefont {Kockum}}, \bibinfo
  {author} {\bibfnamefont {B.}~\bibnamefont {Suri}}, \bibinfo {author}
  {\bibfnamefont {H.}~\bibnamefont {Ian}}, \bibinfo {author} {\bibfnamefont
  {J.}~\bibnamefont {Chen}}, \bibinfo {author} {\bibfnamefont {S.}~\bibnamefont
  {Mao}}, \bibinfo {author} {\bibfnamefont {C.}~\bibnamefont {Chiu}}, \bibinfo
  {author} {\bibfnamefont {P.}~\bibnamefont {Delsing}}, \bibinfo {author}
  {\bibfnamefont {F.}~\bibnamefont {Nori}}, \emph {et~al.},\ }\href@noop {}
  {\bibfield  {journal} {\bibinfo  {journal} {Physical Review Letters}\
  }\textbf {\bibinfo {volume} {123}},\ \bibinfo {pages} {233602} (\bibinfo
  {year} {2019})}\BibitemShut {NoStop}%
\bibitem [{\citenamefont {Fedorov}\ \emph {et~al.}(2021)\citenamefont
  {Fedorov}, \citenamefont {Remizov}, \citenamefont {Shapiro}, \citenamefont
  {Pogosov}, \citenamefont {Egorova}, \citenamefont {Tsitsilin}, \citenamefont
  {Andronik}, \citenamefont {Dobronosova}, \citenamefont {Rodionov},
  \citenamefont {Astafiev} \emph {et~al.}}]{fedorov2021photon}%
  \BibitemOpen
  \bibfield  {author} {\bibinfo {author} {\bibfnamefont {G.}~\bibnamefont
  {Fedorov}}, \bibinfo {author} {\bibfnamefont {S.}~\bibnamefont {Remizov}},
  \bibinfo {author} {\bibfnamefont {D.}~\bibnamefont {Shapiro}}, \bibinfo
  {author} {\bibfnamefont {W.}~\bibnamefont {Pogosov}}, \bibinfo {author}
  {\bibfnamefont {E.}~\bibnamefont {Egorova}}, \bibinfo {author} {\bibfnamefont
  {I.}~\bibnamefont {Tsitsilin}}, \bibinfo {author} {\bibfnamefont
  {M.}~\bibnamefont {Andronik}}, \bibinfo {author} {\bibfnamefont
  {A.}~\bibnamefont {Dobronosova}}, \bibinfo {author} {\bibfnamefont
  {I.}~\bibnamefont {Rodionov}}, \bibinfo {author} {\bibfnamefont
  {O.}~\bibnamefont {Astafiev}}, \emph {et~al.},\ }\href@noop {} {\bibfield
  {journal} {\bibinfo  {journal} {Physical Review Letters}\ }\textbf {\bibinfo
  {volume} {126}},\ \bibinfo {pages} {180503} (\bibinfo {year}
  {2021})}\BibitemShut {NoStop}%
\bibitem [{\citenamefont {Vetsch}\ \emph {et~al.}(2010)\citenamefont {Vetsch},
  \citenamefont {Reitz}, \citenamefont {Sagu{\'e}}, \citenamefont {Schmidt},
  \citenamefont {Dawkins},\ and\ \citenamefont
  {Rauschenbeutel}}]{vetsch2010optical}%
  \BibitemOpen
  \bibfield  {author} {\bibinfo {author} {\bibfnamefont {E.}~\bibnamefont
  {Vetsch}}, \bibinfo {author} {\bibfnamefont {D.}~\bibnamefont {Reitz}},
  \bibinfo {author} {\bibfnamefont {G.}~\bibnamefont {Sagu{\'e}}}, \bibinfo
  {author} {\bibfnamefont {R.}~\bibnamefont {Schmidt}}, \bibinfo {author}
  {\bibfnamefont {S.}~\bibnamefont {Dawkins}},\ and\ \bibinfo {author}
  {\bibfnamefont {A.}~\bibnamefont {Rauschenbeutel}},\ }\href@noop {}
  {\bibfield  {journal} {\bibinfo  {journal} {Physical Review Letters}\
  }\textbf {\bibinfo {volume} {104}},\ \bibinfo {pages} {203603} (\bibinfo
  {year} {2010})}\BibitemShut {NoStop}%
\bibitem [{\citenamefont {Yu}\ \emph {et~al.}(2014)\citenamefont {Yu},
  \citenamefont {Hood}, \citenamefont {Muniz}, \citenamefont {Martin},
  \citenamefont {Norte}, \citenamefont {Hung}, \citenamefont {Meenehan},
  \citenamefont {Cohen}, \citenamefont {Painter},\ and\ \citenamefont
  {Kimble}}]{yu2014nanowire}%
  \BibitemOpen
  \bibfield  {author} {\bibinfo {author} {\bibfnamefont {S.-P.}\ \bibnamefont
  {Yu}}, \bibinfo {author} {\bibfnamefont {J.}~\bibnamefont {Hood}}, \bibinfo
  {author} {\bibfnamefont {J.}~\bibnamefont {Muniz}}, \bibinfo {author}
  {\bibfnamefont {M.}~\bibnamefont {Martin}}, \bibinfo {author} {\bibfnamefont
  {R.}~\bibnamefont {Norte}}, \bibinfo {author} {\bibfnamefont {C.-L.}\
  \bibnamefont {Hung}}, \bibinfo {author} {\bibfnamefont {S.~M.}\ \bibnamefont
  {Meenehan}}, \bibinfo {author} {\bibfnamefont {J.~D.}\ \bibnamefont {Cohen}},
  \bibinfo {author} {\bibfnamefont {O.}~\bibnamefont {Painter}},\ and\ \bibinfo
  {author} {\bibfnamefont {H.}~\bibnamefont {Kimble}},\ }\href@noop {}
  {\bibfield  {journal} {\bibinfo  {journal} {Applied Physics Letters}\
  }\textbf {\bibinfo {volume} {104}},\ \bibinfo {pages} {111103} (\bibinfo
  {year} {2014})}\BibitemShut {NoStop}%
\bibitem [{\citenamefont {Douglas}\ \emph {et~al.}(2015)\citenamefont
  {Douglas}, \citenamefont {Habibian}, \citenamefont {Hung}, \citenamefont
  {Gorshkov}, \citenamefont {Kimble},\ and\ \citenamefont
  {Chang}}]{douglas2015quantum}%
  \BibitemOpen
  \bibfield  {author} {\bibinfo {author} {\bibfnamefont {J.~S.}\ \bibnamefont
  {Douglas}}, \bibinfo {author} {\bibfnamefont {H.}~\bibnamefont {Habibian}},
  \bibinfo {author} {\bibfnamefont {C.-L.}\ \bibnamefont {Hung}}, \bibinfo
  {author} {\bibfnamefont {A.~V.}\ \bibnamefont {Gorshkov}}, \bibinfo {author}
  {\bibfnamefont {H.~J.}\ \bibnamefont {Kimble}},\ and\ \bibinfo {author}
  {\bibfnamefont {D.~E.}\ \bibnamefont {Chang}},\ }\href@noop {} {\bibfield
  {journal} {\bibinfo  {journal} {Nature Photonics}\ }\textbf {\bibinfo
  {volume} {9}},\ \bibinfo {pages} {326} (\bibinfo {year} {2015})}\BibitemShut
  {NoStop}%
\bibitem [{\citenamefont {Gonz{\'a}lez-Tudela}\ \emph
  {et~al.}(2015)\citenamefont {Gonz{\'a}lez-Tudela}, \citenamefont {Hung},
  \citenamefont {Chang}, \citenamefont {Cirac},\ and\ \citenamefont
  {Kimble}}]{gonzalez2015subwavelength}%
  \BibitemOpen
  \bibfield  {author} {\bibinfo {author} {\bibfnamefont {A.}~\bibnamefont
  {Gonz{\'a}lez-Tudela}}, \bibinfo {author} {\bibfnamefont {C.-L.}\
  \bibnamefont {Hung}}, \bibinfo {author} {\bibfnamefont {D.~E.}\ \bibnamefont
  {Chang}}, \bibinfo {author} {\bibfnamefont {J.~I.}\ \bibnamefont {Cirac}},\
  and\ \bibinfo {author} {\bibfnamefont {H.}~\bibnamefont {Kimble}},\
  }\href@noop {} {\bibfield  {journal} {\bibinfo  {journal} {Nature Photonics}\
  }\textbf {\bibinfo {volume} {9}},\ \bibinfo {pages} {320} (\bibinfo {year}
  {2015})}\BibitemShut {NoStop}%
\bibitem [{\citenamefont {Yu}\ \emph {et~al.}(2019)\citenamefont {Yu},
  \citenamefont {Muniz}, \citenamefont {Hung},\ and\ \citenamefont
  {Kimble}}]{yu2019two}%
  \BibitemOpen
  \bibfield  {author} {\bibinfo {author} {\bibfnamefont {S.-P.}\ \bibnamefont
  {Yu}}, \bibinfo {author} {\bibfnamefont {J.~A.}\ \bibnamefont {Muniz}},
  \bibinfo {author} {\bibfnamefont {C.-L.}\ \bibnamefont {Hung}},\ and\
  \bibinfo {author} {\bibfnamefont {H.}~\bibnamefont {Kimble}},\ }\href@noop {}
  {\bibfield  {journal} {\bibinfo  {journal} {Proceedings of the National
  Academy of Sciences}\ }\textbf {\bibinfo {volume} {116}},\ \bibinfo {pages}
  {12743} (\bibinfo {year} {2019})}\BibitemShut {NoStop}%
\bibitem [{\citenamefont {Orazbayev}\ \emph {et~al.}(2018)\citenamefont
  {Orazbayev}, \citenamefont {Kaina},\ and\ \citenamefont
  {Fleury}}]{orazbayev2018chiral}%
  \BibitemOpen
  \bibfield  {author} {\bibinfo {author} {\bibfnamefont {B.}~\bibnamefont
  {Orazbayev}}, \bibinfo {author} {\bibfnamefont {N.}~\bibnamefont {Kaina}},\
  and\ \bibinfo {author} {\bibfnamefont {R.}~\bibnamefont {Fleury}},\
  }\href@noop {} {\bibfield  {journal} {\bibinfo  {journal} {Physical Review
  Applied}\ }\textbf {\bibinfo {volume} {10}},\ \bibinfo {pages} {054069}
  (\bibinfo {year} {2018})}\BibitemShut {NoStop}%
\bibitem [{\citenamefont {Asenjo-Garcia}\ \emph {et~al.}(2017)\citenamefont
  {Asenjo-Garcia}, \citenamefont {Hood}, \citenamefont {Chang},\ and\
  \citenamefont {Kimble}}]{asenjo2017atom}%
  \BibitemOpen
  \bibfield  {author} {\bibinfo {author} {\bibfnamefont {A.}~\bibnamefont
  {Asenjo-Garcia}}, \bibinfo {author} {\bibfnamefont {J.}~\bibnamefont {Hood}},
  \bibinfo {author} {\bibfnamefont {D.}~\bibnamefont {Chang}},\ and\ \bibinfo
  {author} {\bibfnamefont {H.}~\bibnamefont {Kimble}},\ }\href@noop {}
  {\bibfield  {journal} {\bibinfo  {journal} {Physical Review A}\ }\textbf
  {\bibinfo {volume} {95}},\ \bibinfo {pages} {033818} (\bibinfo {year}
  {2017})}\BibitemShut {NoStop}%
\bibitem [{\citenamefont {Le~Kien}\ \emph {et~al.}(2004)\citenamefont
  {Le~Kien}, \citenamefont {Liang}, \citenamefont {Hakuta},\ and\ \citenamefont
  {Balykin}}]{le2004field}%
  \BibitemOpen
  \bibfield  {author} {\bibinfo {author} {\bibfnamefont {F.}~\bibnamefont
  {Le~Kien}}, \bibinfo {author} {\bibfnamefont {J.}~\bibnamefont {Liang}},
  \bibinfo {author} {\bibfnamefont {K.}~\bibnamefont {Hakuta}},\ and\ \bibinfo
  {author} {\bibfnamefont {V.}~\bibnamefont {Balykin}},\ }\href@noop {}
  {\bibfield  {journal} {\bibinfo  {journal} {Optics Communications}\ }\textbf
  {\bibinfo {volume} {242}},\ \bibinfo {pages} {445} (\bibinfo {year}
  {2004})}\BibitemShut {NoStop}%
\bibitem [{\citenamefont {Tong}\ \emph {et~al.}(2004)\citenamefont {Tong},
  \citenamefont {Lou},\ and\ \citenamefont {Mazur}}]{tong2004single}%
  \BibitemOpen
  \bibfield  {author} {\bibinfo {author} {\bibfnamefont {L.}~\bibnamefont
  {Tong}}, \bibinfo {author} {\bibfnamefont {J.}~\bibnamefont {Lou}},\ and\
  \bibinfo {author} {\bibfnamefont {E.}~\bibnamefont {Mazur}},\ }\href@noop {}
  {\bibfield  {journal} {\bibinfo  {journal} {Optics Express}\ }\textbf
  {\bibinfo {volume} {12}},\ \bibinfo {pages} {1025} (\bibinfo {year}
  {2004})}\BibitemShut {NoStop}%
\bibitem [{\citenamefont {Song}\ \emph {et~al.}(2020)\citenamefont {Song},
  \citenamefont {Liu}, \citenamefont {Zheng}, \citenamefont {Zhang},
  \citenamefont {Wang},\ and\ \citenamefont {Lu}}]{song2020two}%
  \BibitemOpen
  \bibfield  {author} {\bibinfo {author} {\bibfnamefont {Y.}~\bibnamefont
  {Song}}, \bibinfo {author} {\bibfnamefont {W.}~\bibnamefont {Liu}}, \bibinfo
  {author} {\bibfnamefont {L.}~\bibnamefont {Zheng}}, \bibinfo {author}
  {\bibfnamefont {Y.}~\bibnamefont {Zhang}}, \bibinfo {author} {\bibfnamefont
  {B.}~\bibnamefont {Wang}},\ and\ \bibinfo {author} {\bibfnamefont
  {P.}~\bibnamefont {Lu}},\ }\href@noop {} {\bibfield  {journal} {\bibinfo
  {journal} {Physical Review Applied}\ }\textbf {\bibinfo {volume} {14}},\
  \bibinfo {pages} {064076} (\bibinfo {year} {2020})}\BibitemShut {NoStop}%
\end{thebibliography}

\providecommand{\noopsort}[1]{}\providecommand{\singleletter}[1]{#1}%

\end{document}